\documentstyle[epsfig,rotate,12pt]{article}
\topmargin - 1.0 true cm
\def\sss{\scriptscriptstyle}
\def\barp{{\raise.35ex\hbox{${\sss (}$}}---{\raise.35ex\hbox{${\sss )}$}}}
\def\bdbarp{\hbox{$B_d$\kern-1.4em\raise1.4ex\hbox{\barp}}}
\def\bsbarp{\hbox{$B_s$\kern-1.4em\raise1.4ex\hbox{\barp}}}
\def\dbarp{\hbox{$D$\kern-1.1em\raise1.4ex\hbox{\barp}}}
\def\dcp{D^0_{\sss CP}}

\def\rly#1{\mathrel{\raise.3ex\hbox{$#1$\kern-.75em\lower1ex\hbox{$\sim$}}}}
\def\lsim{\rly<}
\newread\epsffilein 
\newif\ifepsffileok 
\newif\ifepsfbbfound 
\newif\ifepsfverbose 
\newdimen\epsfxsize 
\newdimen\epsfysize 
\newdimen\epsftsize 
\newdimen\epsfrsize 
\newdimen\epsftmp  
\newdimen\pspoints  
\def\epsfbox#1{\global\def\epsfllx{72}\global\def\epsflly{72}%
 \global\def\epsfurx{540}\global\def\epsfury{720}%
 \def\lbracket{[}\def\testit{#1}\ifx\testit\lbracket
 \let\next=\epsfgetlitbb\else\let\next=\epsfnormal\fi\next{#1}}%
\def\epsfgetlitbb#1#2 #3 #4 #5]#6{\epsfgrab #2 #3 #4 #5 .\\%
 \epsfsetgraph{#6}}%
\def\epsfnormal#1{\epsfgetbb{#1}\epsfsetgraph{#1}}%
\def\epsfgetbb#1{%
\openin\epsffilein=#1
\ifeof\epsffilein\errmessage{I couldn't open #1, will ignore it}\else
 {\epsffileoktrue \chardef\other=12
 \def\do##1{\catcode`##1=\other}\dospecials \catcode`\ =10
 \loop
  \read\epsffilein to \epsffileline
  \ifeof\epsffilein\epsffileokfalse\else
   \expandafter\epsfaux\epsffileline:. \\%
  \fi
 \ifepsffileok\repeat
 \ifepsfbbfound\else
 \ifepsfverbose\message{No bounding box comment in #1; using defaults}\fi\fi
 }\closein\epsffilein\fi}%
\def\epsfclipstring{}
\def\epsfsetgraph#1{%
 \epsfrsize=\epsfury\pspoints
 \advance\epsfrsize by-\epsflly\pspoints
 \epsftsize=\epsfurx\pspoints
 \advance\epsftsize by-\epsfllx\pspoints
 \epsfxsize\epsfsize\epsftsize\epsfrsize
 \ifnum\epsfxsize=0 \ifnum\epsfysize=0
  \epsfxsize=\epsftsize \epsfysize=\epsfrsize
  \epsfrsize=0pt
arithmetic! 
  \else\epsftmp=\epsftsize \divide\epsftmp\epsfrsize
  \epsfxsize=\epsfysize \multiply\epsfxsize\epsftmp
  \multiply\epsftmp\epsfrsize \advance\epsftsize-\epsftmp
  \epsftmp=\epsfysize
  \loop \advance\epsftsize\epsftsize \divide\epsftmp 2
  \ifnum\epsftmp>0
   \ifnum\epsftsize<\epsfrsize\else
    \advance\epsftsize-\epsfrsize \advance\epsfxsize\epsftmp \fi
  \repeat
  \epsfrsize=0pt
  \fi
 \else \ifnum\epsfysize=0
  \epsftmp=\epsfrsize \divide\epsftmp\epsftsize
  \epsfysize=\epsfxsize \multiply\epsfysize\epsftmp   
  \multiply\epsftmp\epsftsize \advance\epsfrsize-\epsftmp
  \epsftmp=\epsfxsize
  \loop \advance\epsfrsize\epsfrsize \divide\epsftmp 2
  \ifnum\epsftmp>0
  \ifnum\epsfrsize<\epsftsize\else
   \advance\epsfrsize-\epsftsize \advance\epsfysize\epsftmp \fi
  \repeat
  \epsfrsize=0pt
 \else
  \epsfrsize=\epsfysize
 \fi
 \fi
 \ifepsfverbose\message{#1: width=\the\epsfxsize, height=\the\epsfysize}\fi
 \epsftmp=10\epsfxsize \divide\epsftmp\pspoints
 \vbox to\epsfysize{\vfil\hbox to\epsfxsize{%
  \ifnum\epsfrsize=0\relax
  \includegraphics{#1}%
  \else
  \epsfrsize=10\epsfysize \divide\epsfrsize\pspoints  
  \includegraphics{#1}%
  \fi
  \hfil}}%
\global\epsfxsize=0pt\global\epsfysize=0pt}%
 {\catcode`\%=12
\global\let\epsfpercent=
\long\def\epsfaux#1#2:#3\\{\ifx#1\epsfpercent
 \def\testit{#2}\ifx\testit\epsfbblit
  \epsfgrab #3 . . . \\%
  \epsffileokfalse
  \global\epsfbbfoundtrue
 \fi\else\ifx#1\par\else\epsffileokfalse\fi\fi}%
\def\epsfempty{}%
\def\epsfgrab #1 #2 #3 #4 #5\\{%
\global\def\epsfllx{#1}\ifx\epsfllx\epsfempty
  \epsfgrab #2 #3 #4 #5 .\\\else
 \global\def\epsflly{#2}%
 \global\def\epsfurx{#3}\global\def\epsfury{#4}\fi}%
\def\epsfsize#1#2{\epsfxsize}
\let\epsffile=\epsfbox
\def\bsll{$b \rightarrow s \ell^+ \ell^- $ }
\def\bxsll{$B \rightarrow X_s \ell^+ \ell^- $ }
\def\bxsee{B \rightarrow X_s e^+ e^-  }
\def\bxsmm{B \rightarrow X_s \mu^+ \mu^-  }
\def\bxstt{B \rightarrow X_s \tau^+ \tau^- }

\def\bxsg{$B \rightarrow X_s \gamma $ }
\def\s{\hat{s}}

\newcommand{\ra}{\rightarrow}
\newcommand{\bs}{B_s^0}

\def\be{\begin{equation}}
\def\ee{\end{equation}}
\def\g{\gamma}

\newcommand{\M}{{\cal M}}
\def\mt{m_t}
\def\mb{m_b}
\def\mc{m_c}
\def\bra{\langle}
\def\ket{\rangle}
\def\bea{\begin{eqnarray}}
\def\eea{\end{eqnarray}}
\def\be{\begin{equation}}
\def\ee{\end{equation}}
\def\a{\alpha}

\def\g{\gamma}

\def\p{\pi}

\def\l{\lambda}

\def\G{\Gamma}

\def\mt{m_t}

\newcommand{\bgamaxs}{$B \to X _{s} + \gamma$}

\newcommand{\BGAMAXS}{B \ra X _{s} + \gamma}
\newcommand{\BGAMAXD}{B \ra X _{d} + \gamma}
\newcommand{\BBGAMAXS}{{\cal B}(B \ra  X _{s} + \gamma)}
\newcommand{\BBGAMAXD}{{\cal B}(B \ra  X _{d} + \gamma)}

\newcommand{\BGAMAKSTAR}{B \ra  K^{\star} + \gamma}
\newcommand{\GGAMAXD}{\Gamma(B \ra  X _{d} + \gamma)}

\newcommand{\GGAMAXS}{\Gamma (B \ra  X _{s} + \gamma)}

\def\beq{\begin{equation}}
\def\eeq{\end{equation}}
\def\Vcdabs{\vert V_{cd} \vert}
\def\Vus{V_{us}}
\def\Vusabs{\vert V_{us} \vert}

\def\Vcsabs{\vert V_{cs} \vert}
\def\Vud{V_{ud}}
\def\Vudabs{\vert V_{ud} \vert}
\def\Vbc{V_{cb}}

\def\Vcbabs{\vert V_{cb} \vert}
\def\Vbu{V_{ub}}

\def\Vubabs{\vert V_{ub}\vert}
\def\Vtd{V_{td}}
\def\Vtdabs{\vert V_{td} \vert}
\def\Vts{V_{ts}}
\def\Vtsabs{\vert V_{ts} \vert}

\def\Vtbabs{\vert V_{tb}\vert}
\newcommand{\bd}{B_d^0}

\newcommand{\abseps}{\vert\epsilon\vert}

\def\BZ{B_d^0}
\def\BZB{\bar{B_d^0}}
\def\BS{B_s^0}
\def\BSB{\bar{B_s^0}}

\newcommand{\fbb}{f^2_{B_d}B_{B_d}}
\newcommand{\fbbs}{f^2_{B_s}B_{B_s}}
\newcommand{\fbd}{f_{B_d}}
\newcommand{\fbs}{f_{B_s}}
\newcommand{\go}[1]{\gamma^{#1}}
\newcommand{\gu}[1]{\gamma_{#1}}

\newcommand{\delmd}{\Delta M_d}
\newcommand{\delms}{\Delta M_s}

\def\sw{\sin{^2}\theta _{W}}

\def\qbar{\overline q}

\newcommand{\bu}{B_u^\pm}

\def\q5q{\qbar{{\lambda_a}\over 2} i\gamma_5 q}

\def\gmuu{\gamma^\mu}

\newcommand{\bgamaxd}{$B \to X _{d} + \gamma$}

\def\to{\rightarrow}

\def\mb{m_b}

\def\xs{x_s}
\def\xd{x_d}
\newcommand{\kkbar}{$K^0$-${\overline{K^0}}$}
\newcommand{\bdbdbar}{$B_d^0$-${\overline{B_d^0}}$}
\newcommand{\bsbsbar}{$B_s^0$-${\overline{B_s^0}}$}

\def\as{\alpha _s}

\def\Vbc{V_{cb}}
\def\Vbu{V_{ub}}
\def\Vtd{V_{td}}
\def\Vts{V_{ts}}

\begin{document}
\begin{flushright}
DESY 96-106 \\
June 1996
\end{flushright}
\begin{center}
{\large \bf
\centerline
{$B$ Decays, Flavour Mixings and CP Violation}
\centerline {in the Standard Model
$\footnote{Lectures given at the XX International
Nathiagali Summer College on Physics and Contemporary Needs, Bhurban, 
Pakistan, June 24 - July 13, 1995; to appear in the Proceedings
(Nova Science Publishers, New York), Editors: Riazuddin, K.A. Shoaib et 
al.}$}}
 \vspace*{1.5cm}

{\large A.~Ali}
\vskip0.2cm
  Deutsches Elektronen Synchrotron DESY, Hamburg \\
\vskip0.5cm
{\Large Abstract\\}
\parbox[t]{\textwidth}{
\indent
These lectures review the progress made in our present understanding of 
$B$ decays. The emphasis here is on  applications of
QCD to $B$ decays and the attendant perturbative and non-perturbative 
uncertainties, which limit present theoretical precision in some cases but
the overall picture that emerges is consistent with the standard model (SM). 
This is illustrated by quantitatively analyzing some of the key measurements
in $B$ physics. These lectures are divided in five parts.
In the first part, the Kobayashi-Maskawa generalization of
the Cabibbo-GIM matrix for quark flavour mixing is discussed.
In the second part, the bulk 
properties of $B$ decays, such as the inclusive decay rates, semileptonic
branching ratios, $B$-hadron lifetimes, and the so-called charm counting in
$B$ decays are taken up. The third part is devoted to
theoretical studies of rare $B$ decays, in particular the electromagnetic
penguins involving the decays $B \to K^* + \gamma$ and $B \to X_s + \gamma$. 
The photon energy spectrum and the branching
ratios in the SM are discussed and compared with data, enabling a
determination of the CKM matrix element $\Vtsabs$, the $b$-quark mass, and
the kinetic energy of the $b$-quark in the $B$ meson.
The CKM-suppressed inclusive decay $B \to X_d + \gamma$, and the exclusive 
decays $B \to (\rho, \omega) + \gamma$, are discussed in the SM using QCD 
sum rules for the latter. The importance of these decays in determining the
CKM parameters is emphasized. This part ends with a discussion of the
decay rates and distributions in  $B \to X_s \ell^+ \ell^-$, including
the long-distance effects, and 
 estimates of a large
number of other rare $B$ decays, including $B \to X \nu \bar{\nu}$, and the  
two-body decays $B^0 \to \ell^+ \ell^-$ and $B^0 \to \gamma \gamma$. 
 The fourth part is devoted to reviewing the
present estimates of the CKM matrix elements from $B$ decays and
$B^0$ - $\overline{B^0}$ mixings, which determine five of the nine elements
in this matrix. This is combined with our present knowledge of the other 
four CKM matrix elements and  a quantitative test of the unitarity of the CKM
matrix is presented.
 This information is then combined with the constraints from the
CP-violation parameter $\abseps$ in order to provide a profile of the CKM
unitarity triangle and CP-violating asymmetries in $B$ decays.
These aspects are discussed in the fifth part of these lectures.}

\end{center}
\thispagestyle{empty}
\newpage
\setcounter{page}{1}
\textheight 23.0 true cm
\newpage
\vspace*{2cm}
\begin{center}
 {\bf  TABLE OF CONTENTS}
\end{center}
\vspace{6mm}
{\bf 1. Weak-coupling Lagrangian for the {\bf $b$} Quark
and the CKM Matrix \dotfill {} 2 }\\
\qquad 1.1 {\bf Some popular representations of the CKM Matrix}{\dotfill {} 
8}\\ \qquad 1.2 {\bf The CKM unitarity triangles}{\dotfill {} 10}\\
{\bf 2. Dominant $B$ Decays in the Standard Model \dotfill {} 12}\\
\qquad 2.1 {\bf Inclusive semileptonic decay rates of the $B$ 
hadrons}{\dotfill {} 14}\\
\qquad 2.2 {\bf Inclusive non-leptonic decay rates of the $B$ hadrons}
{\dotfill {} 15}\\
\qquad 2.3 {\bf Power corrections in $\Gamma_{SL }(B)$ and $\Gamma_{NL}(B)$}
{\dotfill {} 21}\\
\qquad 2.4 {\bf Numerical estimates of ${\cal B}_{SL }(B)$ and $\langle n_c
\rangle$}{\dotfill {} 23}\\
\qquad 2.5 {\bf $B$-Hadron Lifetimes in the Standard Model}{\dotfill {} 
25}\\ 
\qquad 2.6 {\bf Determination of $\Vcbabs$ and $\Vubabs$}{\dotfill {} 26}\\
{\bf 3. Electromagnetic Penguins and Rare $B$ Decays in the Standard Model
\dotfill {} 29}\\
\qquad 3.1 {\bf The effective Hamiltonian for $\BGAMAXS$}{\dotfill {} 30}\\
\qquad 3.2 {\bf Real and virtual $O(\a_s)$ corrections for the matrix 
element} {\dotfill {} 34}\\
\qquad 3.3 {\bf Estimates of long-distance effects in $\BGAMAXS$}{\dotfill
{} 38}\\
\qquad 3.4 {\bf Estimates of $\BBGAMAXS$ in the Standard Model}{\dotfill {}
39}\\
\qquad 3.5 {\bf Photon energy spectrum in $\BGAMAXS$}{\dotfill {} 41}\\
\qquad 3.6 {\bf Inclusive radiative decays \bgamaxd }{\dotfill {} 43}\\
\qquad 3.7 {\bf  Estimates of ${\cal B}(B \to V + \gamma )$
 and constraints on the CKM parameters}{\dotfill {} 46}\\
\qquad 3.8 {\bf Inclusive rare decays $B \to X_s \ell^+ \ell^-$ in the SM}
{\dotfill {} 52}\\
\qquad 3.9 {\bf Summary and overview of rare $B$ decays}{\dotfill {} 58}\\
{\bf 4. An Update of the CKM Matrix \dotfill {} 59}\\
\qquad 4.1 {\bf The present profile of the CKM matrix}{\dotfill {} 62}\\
\qquad 4.2 {\bf $x_s$ and the unitarity triangle}{\dotfill {} 65}\\
{\bf 5. CP Violation in the $B$ System \dotfill {} 67}\\
\qquad 5.1. {\bf Summary of the CKM fits and CP asymmetries in $B$ decays 
\dotfill {} 71}\\
 {\bf 6. Acknowledgements  \dotfill {} 73}\\
{\bf 7. References  \dotfill {} 74}\\
\newpage

\noindent
\section{Weak-coupling Lagrangian for the {\bf $b$} Quark
 and the CKM Matrix}

\par
The Lagrangian density of the standard model
for electroweak interactions (henceforth called SM)
can be symbolically written as \cite{GSW}:
\begin{equation}
 {\cal L} = {\cal L} (f, W, B) + {\cal L} (f, \Phi)
+ {\cal L} (W, B, \Phi) - V(\Phi)
\end{equation}
where the symbols $f, W, B$ and $\Phi$
represent fermions (leptons and quarks),
$SU(2)_L$ gauge bosons, $W_\mu^i$, $U(1)_Y$
gauge boson, $B_\mu$ and the Higgs doublet field,
 respectively.
  The SM particle content
together with the (weak) isospin properties of the basic quanta are
given below, where we have assumed that there are
three families of leptons and quarks. \\ \\

\noindent
\underline{Fermions}   \\
\[ \mbox{ Leptons:} \
   \left( \begin{array}{c} \nu_e \\ e \end{array} \right)_L ,\
   \left( \begin{array}{c} \nu_\mu \\ \mu \end{array} \right)_L ,\
   \left( \begin{array}{c} \nu_\tau \\ \tau \end{array} \right)_L ;\
e_R, \mu_R, \tau_R \] \\
\[ \mbox{Quarks:} \
   \left( \begin{array}{c} u \\ d \end{array} \right)_L ,\
   \left( \begin{array}{c} c \\ s \end{array} \right)_L ,\
   \left( \begin{array}{c} t \\ b \end{array} \right)_L ;\
u_R, d_R, c_R, ... \] \\
\underline{Gauge bosons} \\
\[ \left( \begin{array}{c} {W_\mu}^1  \\ {W_\mu}^2 \\ {W_\mu}^3
\end{array} \right) \;\;; ~~B_\mu \] \\
\underline{Scalars}\\
\begin{equation}
 \Phi = \left(\begin{array} {c} \phi^+ \\ \phi^0
\end{array} \right) \,\, , ~~\Phi^{\dag} =
\left( \phi^- , \bar{\phi^0} \right).
\end{equation}
The various terms in the SM Lagrangian
can be written by demanding $SU(2)_L\otimes
U(1)$ gauge invariance, lepton-quark
universality, and family-independence of the
electroweak interactions. Since the
SM Lagrangian is given in any standard textbook
on electroweak interactions \cite{Okun} we shall
not reproduce it here. The main interest in $B$ physics lies
in the study of the first two terms
in (1) involving fermions, gauge bosons and the
Higgs fields, in particular flavour-changing transitions.
In order to appreciate how flavour-changing transitions
emerge in the SM, it is worth while to write
the first two terms in (1) explicitly. The interaction
between the fermions and gauge bosons has the form:
\begin{equation}
{\cal L} (f,W,B) = \sum_{j=1}^{3}\{\bar{l}_L^j
D\hspace{-2.8mm}/\hspace{1.3mm} l_L^j + \bar{l}_R^j
D\hspace{-2.8mm}/\hspace{1.3mm} 'l_R^j + \bar{q}^j_L
D \hspace{-2.8mm}/\hspace{1.3mm} q_L^j\} + \sum_{i=1}^6
\bar{q}^i_R D \hspace{-2.8mm}/\hspace{1.3mm} '
q^i_R ,
\end{equation}
where $j$ is the family index,
\[ l_L^1 =\left( \begin{array}{c} \nu_e \\ e \end{array}
\right)_L \,\,,
~l_R^1 = e_R\,\,,  ~q_L^1 =
\left( \begin{array}{c} u \\ d \end{array}
\right)_L\,\,, q_R^1 = u_R \,, ~ q^2_R = d_R,  ... \]
and the two covariant
derivatives are defined as
\begin{eqnarray}
D \hspace{-2.8mm}/\hspace{1.3mm} \equiv D_\mu \gamma^\mu\,,\,\,
D \hspace{-2.8mm}/\hspace{1.3mm} ' \equiv D_\mu' \gamma^\mu,
\nonumber\\
D_\mu = \partial_\mu - i g_2 \left( \vec{W}_\mu
\cdot \frac{\vec{\sigma}}{2} \right)
- ig_1 \frac{Y}{2} B_\mu ,\nonumber\\
D'_\mu = \partial_\mu - i g_1 \frac{Y}{2} B_\mu ,
\end{eqnarray}
where $g_1$ and $g_2$ are, respectively, the
$U(1)_Y$ and $SU(2)_L$ coupling constants,
$\sigma^a$ $(a = 1,2,3)$ are the (weak) isospin
Pauli matrices, and the Gell-Mann -- Nishijima formula
$Q = I_3 + Y $ defines the (weak) hypercharge
of the quarks and leptons.
The interaction term ${\cal L} (f, \phi)$ involving
the fermions and the Higgs fields has the Yukawa form
\begin{eqnarray}
{\cal L} (f,\Phi) &=& \sum^3_{j=1} \left\{
\left( h_l \right)_j \bar{l}^j_L \Phi
\bar{l}_R^j \right\} \nonumber\\
&+& \sum^3_{j,k=1} \bigg\{ \left( h'_q \right)_{j k}
\bar{q}_L^j \Phi u_R^k
+ (h_q)_{jk}  \bar{q}_L^j \Phi^c d_R^k \bigg\},
\end{eqnarray}
where
 \[ \Phi^c = i \sigma_2 \phi^* =
\left( \begin{array}{c} {\phi^0}^*\\
- \phi^-  \end{array} \right) , \,\, \]
with both $\Phi$ and $\Phi^c$ transforming
as a (weak) isospin doublet. The hypercharge
of the Higgs fields can be written by inspection.
The Yukawa coupling constants
$(h_l)_j$, $(h_q)_{jk}$ and $(h'_q)_{jk}$ are
arbitrary complex numbers and each term in
(5) is independently $SU(2)_L \otimes U(1)_Y$
invariant due to the fact that the $SU(2)_L$
acts only on $\bar{l}_L$ and $\bar{q}_L$ and on the
Higgs doublet $\Phi$ and $\Phi^c$, whose products are $SU(2)$ scalars.
\par
After spontaneous symmetry breaking $SU(2)_L
\otimes U(1)_Y \longrightarrow U(1)_{EM}$,
which preserves the symmetry under $U(1)$ electromagnetism,
the gauge bosons, fermions and the neutral
scalar field $\phi$ acquire non-zero masses
through the Higgs mechanism \\
\begin{equation}
        V(\phi) =  \mu^2 | \phi |^2
+  \lambda | \phi |^4 \,\,;
~~\mu^2 < 0 \,\,, \lambda > 0 ,\nonumber
\end{equation}
with $ V(\phi)_{\rm min} $ at $ |\phi| =
v / \sqrt{2} = \sqrt{-\mu^2 / 2\lambda}$,
                  where $v$ is the neutral-Higgs vacuum expectation
value.
Making the Higgs transformation $\phi \longrightarrow
\phi + v$ in (5), one finds $( \phi$ is the physical Higgs scalar
with $m_\phi^2=2 \lambda v^2$):
\begin{eqnarray}
{\cal L} (f,\phi)^{\rm SSB} &=& \sum^3_{j=1}
(m_j)_l \bar{l}_L^j l_R^j
\left( 1 + \frac{1}{v} \phi \right)\nonumber\\
&-& \sum^3_{j,k=1} \left\{
\left(m_{jk} \right)_U \bar{u}_L^j u_R^k +
\left( m_{jk} \right)_D \bar{d}_L^j
d_R^k \right\}
   \left( 1+ \frac{1}{v} \phi \right) + \mbox{h. c.},
\end{eqnarray}
where
\begin{eqnarray}
\left( m_j \right)_l &=& \left(h_j \right)_l
\frac{v}{\sqrt{2}} ,\nonumber\\
\left( m_{jk} \right)_U &=&
- \left( h_q \right)_{jk} \frac{v}{\sqrt{2}},
\nonumber\\
\left( m_{jk} \right)_D &=& -
\left( h'_q \right)_{jk} \frac{v}{\sqrt{2}}.
\end{eqnarray}
Since in the SM there are no right-handed fields $\nu_R^i$
$(i = 1,2,3)$, the neutrinos $\nu^i$ remain
massless and the charged lepton mass matrix ${(m_j)}_l$ is
diagonal. Hence, in the SM
 there are no family-changing leptonic interactions -- an aspect that
will soon come under experimental scrutiny from the ongoing and planned
experiments dedicated to the neutrino mass and oscillation measurements.
In (8),
${(m_{jk})}_U $ and ${(m_{jk})}_D$ are the $(3 \times 3)$
quark mass matrices for the up- and down-type quarks, respectively.
In order to write the Lagrangian in terms of the quark
mass eigenstates one has to diagonalize the mass
matrices $ \left( m_{jk} \right)_U ~\mbox{and}  ~\left( m_{jk}
\right)_D$. This can be done
with the help of two unitary matrices. It is customary to denote them
by $V_L^{up} \mbox{and} ~{V_R^{up}}^{\dag}$ (likewise for the down
type quarks):
\begin{eqnarray}
V^{\rm up}_L m_U V^{{\rm up}^{\dag}}_R
& \equiv &  \left( m_{\rm diag.} \right)_U
\equiv {\rm Diag.} \left( m_u\,\,, m_c \,\,,
m_t \right) ,\\
V_L^{\rm down} m_D  V_R^{{\rm down} \dag}
 & \equiv & \left( m_{\rm diag.}
\right)_D \equiv {\rm Diag.} \left(
m_d\,\,, m_s\,\,, m_b \right), \nonumber
\end{eqnarray}
with ${V_L^{\rm up}}^{\dag} V_L^{\rm up} = 1$, etc.
Concentrating on the up-type quarks in (7)
one can do the following manipulation :
\begin{eqnarray}
\bar{u}_L m_U u_R &=& \bar{u}_L V_L^{{\rm up}^{\dag}} V_L^{\rm up}
m_U V^{{\rm up} \dag}_R V_R^{\rm up} u_R \\
&=& \overline{V_L^{\rm up} u_L}
\left( m_{\rm diag.} \right)_U
\left( V_R^{\rm up} u_R \right), \nonumber
\end{eqnarray}
which shows that the physical quark fields (mass eigenstates)
are:
\begin{eqnarray}
u_L^{i\rm (Phys)} &=& \left(V_L^{\rm up} u_L\right)^{i} ,\nonumber \\
d_L^{i\rm (Phys)} &=& \left(V^{\rm down}_L d_L\right)^{i} .
\end{eqnarray}
Likewise,
$u_R^{i\rm (Phys)}  =  \left(V_R^{\rm up} u_R \right)^{i},
~d_R^{i\rm (Phys)} =  \left(V_R^{\rm down} d_R\right)^{i} $.
One can now rewrite the term ${\cal L} (f, \Phi)$
in the SM Lagrangian in terms of $(u^i_L)^{\rm Phys}$ and
$(d^i_L)^{\rm Phys}$, obtaining
\begin{equation}
{\cal L} (f, \Phi)^{\rm SSB} =
-\left( 1+ \frac{\phi}{v} \right)
\bigg\{ \sum^6_{i = 1}
m_{q_i} \bar{q}_i q_i
+ \sum^3_{j=1} m_{l_i}
\bar{l}_j l_j \bigg\},
\end{equation}
where it is now understood that
$q_1 = \frac{1}{2} \left( {u^1}_L^{\rm Phys} +
{u^1}_R^{\rm Phys} \right) = u \,\, $ etc., and we have dropped the
superscript on the quark fields.
The identification of the parameters $m_{l_i}$,
$m_{q_i}$ with the lepton and quark masses is
now evident. In addition, the SM  Lagrangian has
specified the Higgs-fermion $(\phi f \bar{f})$ Yukawa couplings and
their Lorentz structure, as well as their C, P, and
CP properties. All these symmetries are
separately conserved in ${\cal L} (f, \phi)^{\rm SSB}$
and the Higgs-fermion Yukawa couplings are manifestly
diagonal in flavour space. It should be emphasized that this is a
consequence of the choice made in the SM of a single doublet Higgs field 
since, otherwise, flavour-changing neutral-current
(FCNC) transitions in the Higgs sector would be allowed in general.
\par
Finally, one can carry through the transformation (10) in the part of
the SM Lagrangian describing the fermion-gauge boson couplings,
 ${\cal L} (f,W,B)$.
Written in terms of the physical boson
$(W_\mu^{\pm}, ~Z_\mu^0, ~A_\mu )$ and fermion fields,
it is easy to show that the neutral current
(NC) part of $ {\cal L} (f,W,B)$ is manifestly flavour-diagonal.
Thus, all flavour-changing transitions in the SM are confined to the
the charged current (CC) sector.
Denoting the quarks and leptons by $f_i (i=1...6)$,
the neutral current in the SM is given by:
\begin{equation}
J_\mu^{NC} = \sum_i \bar{f}_i
\bigg[ \frac{e}{\sin \theta_W \cos \theta_W}
Z_\mu \left( {I_3}_L - Q \sw \right)_i
 +  e A_\mu Q_i \bigg] f_i,
\label{neutralc}
\end{equation}
where $(I_3)_L=(1-\gamma_5)/2 (I_3)$ with
 $I_3=+1/2$ for $u_i$ and $\nu_i$ and $-1/2$ for $d_i$ and $Q_i$ is 
the electric charge of the fermion $f_i$ in units of the electron charge,
i.e., $Q_e=+1$. The electroweak mixing
angle in $J_\mu^{NC}$, denoted by $\theta_W$, has its origin in the
diagonalization of the gauge boson mass matrix, and it has 
the usual definition $\cos \theta_W = g_2/\sqrt{g_1^2
+g_2^2}$, with the electric charge defined as
$e \equiv g_2 \sin \theta_W$.
It is easy to check that the neutral current interaction
induced by the $Z$ exchange violates P and C
but conserves CP. The electromagnetic interaction
conserves, of course, all three   C, P, and CP
separately.
\par
 The neutral current couplings specified in
$J_\mu^{NC}$ have been measured in $e^+e^-$ 
annihilation 
experiments, in particular at LEP and SLC, and elsewhere. A
comparison of these couplings with data can be seen in the 
Particle Data Group (PDG) review of particle properties \cite{PDG94}. The 
present situation can be summarized by the statement that
the SM is consistent with the vast majority of these measurements. 
Quantitatively, this can be expressed in terms of the following
values of the basic parameters of the SM, obtained by averaging the
LEP, SLD, $p\bar{p}$ and $\nu N$ data \cite{PDG94}\footnote{The numbers are
taken from P. Langacker's 1995 updated review available through the
World Wide Web.}:
\bea\label{SMparameters}
m_Z ~\mbox{(GeV)} &=& 91.184 \pm 0.0022, \nonumber\\
m_W ~\mbox{(GeV)} &=& 80.26 \pm 0.16  \nonumber\\
\sin^2\hat{\theta}_{W}(m_Z) &=& 0.2315 \pm 0.0002 \pm 0.0003~,
\nonumber\\
m_t ~\mbox{(GeV)} &=& 180 \pm 7 ^{+12}_{-13} \nonumber\\
\as(m_Z) &=& 0.123 \pm 0.004 \pm 0.002 ~,
\eea 
where  
$\sin^2\hat{\theta}_{W}(m_Z)$ is defined in the modified minimal
subtraction scheme $(\overline{\mbox{MS}})$ at the scale $m_Z$, and 
the particular definition of 
$\sin^2\hat{\theta}_W$ used in the analysis can be seen in Langacker's 
review in
\cite{PDG94}. In addition, $N_\nu = 2.991 \pm 0.016$ \cite{LEPEWGroup95},
where $N_\nu$ is the number of light neutrini.
 The fitted value of $\mt$ from the electroweak 
data 
is in excellent agreement with the direct measurements of the same at the
Fermilab experiments CDF and DO \cite{top95},  
$\mt=180 \pm 12$ GeV. However, precision  measurements at the $Z^0$ also
yield decay rates $\Gamma(Z \to b\bar{b})$ and (to a smaller extent) 
 $\Gamma(Z\to c\bar{c})$, which are on their face value at variance
with those predicted in the SM, taken together with the measurement of the 
top quark mass. The experimental situation at the end of 1995 has been
summarized in \cite{PDG94,LEPEWGroup95}:
\bea\label{gambbcc}
R_{b} \equiv \frac{\Gamma(Z^0 \to b\bar{b})}{\Gamma(Z^0 \to \mbox{hadrons})}
&=& 0.2219 \pm 0.0017 ~~~[R_{b}(\mbox{SM})=0.2156], \nonumber\\
R_{c} \equiv \frac{\Gamma(Z^0 \to c\bar{c})}{\Gamma(Z^0 \to \mbox{hadrons})}
&=& 0.1543 \pm 0.0074 ~~~[R_{c}(\mbox{SM})=0.1724]. 
\eea
These measurements constitute a pull factor of $3.7$ and $-2.5$  on the
SM values of $R_b$ and $R_c$, respectively
\footnote{This is defined as
 $ P = (O(\rm{expt}) - O(\rm{th}))/\sigma(O)(\rm{exp})$.}.
We refer the interested reader to \cite{PDG94,LEPEWGroup95} for further
details of the data and a recent 
comprehensive theoretical analysis of the same in \cite{Londonetal96}, in
which a number of non-SM Ans\"atze are put forward to explain the 
present experimental anomalies. In the meanwhile, the value of the
top quark mass has come down marginally; the present
world average  $\mt=175 \pm 9.0$ 
GeV \cite{DPG96} has slightly eased the situation for the SM.

\par
At this particular junction of experiments and SM, it is a fair question
to ask: Quo Vadis SM? In our view,
it is perhaps still too premature to argue persuasively that the 
SM and experiments have parted with each other. This is a tenable
point of view as the case against the SM  is at best
circumstantial and by no means impeccable. It should be stressed that
a good fraction of the LEP data remains to be analyzed. Likewise, lot more
data will be collected at the SLC collider. The experimental
jury is, therefore, still out on $R_b$ and $R_c$ and the final verdict on 
the validity of the SM is yet to be spoken. Improved measurements of $m_W$ at
LEP200, projected to yield a precision of $\Delta m_W \leq 50$ MeV
 \cite{LEP200WG}, 
 and a projected precision of $\Delta m_t=\pm 4$ GeV on the top quark mass at
the Tevatron collider \cite{DPG96}, likewise, will also have a direct 
bearing on the issues being discussed.

\par 
We shall concentrate in these lectures on the physics of the charged
current processes in $B$ decays, which is not effected directly by the
possible modifications of $J_\mu^{NC}$. However, the additional NC 
couplings, if present, may lead to modifications of some FCNC rare $B$ decay
rates and distributions,
such as in $B \to X \ell^+ \ell^-$ and $B \to X \nu \bar{\nu}$,
in which such couplings do   
play a role. Since, in absolute terms, the possible deviation of the 
experimental widths in question from their SM value is small,
we do not expect that such possible modifications in $J_\mu^{NC}$ will   
significantly alter the SM-based profile
of FCNC $B$ decays.
 The implications of a modified $J_\mu^{NC}$ in $B$ decays 
must be investigated quantitatively, in case $Z^0$ data 
force such a modification.
In the CC sector itself, despite suggestions to the contrary 
\cite{Shifmannp}, we shall argue in these lectures that there
is no evidence for new physics so far, though this may change as 
data and theory of weak decays get precise and compelling.
\par
       We now proceed to discuss the charged current, $J_\mu^{CC}$,
which is to be derived from the ${\cal L} (f,W,B)$ part of
the SM Lagrangian using the mass eigenstates.
The CC couplings in the SM involve only the left-handed
fermions $q_L^i$ and $l_L^i$.
Concentrating on the hadronic (quark) part of ${\cal L}^{CC}$, one can
now do the following manipulations:
\begin{eqnarray}
{\cal L}^{CC} &=& \frac{e}{ \sqrt{2} \sin \theta_W}
\sum^3_{i=1} \bar{u}^i_L \gmuu W_\mu^+ d_L^i + \mbox{h.c.}
\nonumber\\
&=& \frac{e}{ \sqrt{2} \sin \theta_W }
\sum^3_{i=1} \bar{u}^{i}_L V_L^{{\rm up}\dag}
V^{\rm up}_L  \gmuu W_\mu^+
V_L^{{\rm down}\dag} V_L^{\rm down} d_L^i +\mbox{h.c.}
\nonumber\\
&=& \frac{e}{ \sqrt{2} \sin \theta_W }
\sum^3_{i=1} \left( \bar{u}_L^{\rm Phys} \right)^i
\gmuu W_\mu^+ \left( V_L^{\rm up} V_L^{{\rm down}\dag} \right)_{ij}
\left(d_L^{\rm Phys}\right)^j + \mbox{h.c.}
\end{eqnarray}
Thus, the charged current $J_\mu^{CC}$, which couples
to the $W^{\pm}$, is
\begin{equation}
J_\mu^{CC} =  \frac{e}{ \sqrt{2} \sin \theta_W }
 \left( \bar{u}, \bar{c}, \bar{t} \right)_L
\gamma_\mu V_{\mbox{\footnotesize CKM}}
\left( \begin{array}{c} d\\ s \\ b \end{array} \right)_L ,
\end{equation}
where we have again dropped the superscript and
      $V_{\mbox{\footnotesize CKM}} \equiv V_L^{\rm up} V_L^{{\rm down}\dag}$
is a $(3 \times 3)$ unitary matrix in flavour space, first
written down by Kobayashi and Maskawa in 1973 \cite{KM}. It is a
generalization of the Cabibbo rotation \cite{Cab} for the
three-quark-flavour $(u, d, s)$ case, invented to keep the universality
of weak interactions, which took the form of a
                                           $(2\times 2)$ matrix by the
inclusion of $c$-quark with the GIM construction \cite{GIM}, and is
called the Cabibbo-Kobayashi-Maskawa (CKM)
matrix. The charged current Lagrangian has the
property that it has a $(V-A)$ structure, hence it violates
P and C maximally, conserves the electric charge and the lepton-
and baryon-number
separately, but otherwise there are no restrictions
on it except that $V_{\mbox{\footnotesize CKM}}^{\dag}
 V_{\mbox{\footnotesize CKM}} = 1$.
                                                In general,
${\cal L}^{CC}$ violates CP due to the
possibility of a non-trivial phase in $V_{\mbox{\footnotesize CKM}}$.\\
\par
From this discussion
it is clear that the flavour-changing transitions in the SM
are confined to the CC sector. They emerge
in the process of diagonalization of the quark mass matrices after
spontaneous symmetry breaking, which in turn depend on
the Higgs-fermion Yukawa couplings. As stated 
already, the Yukawa couplings in the standard model are arbitrary complex
numbers. Their adhoc nature in
the standard model is  in all likelihood pointing to
the physics beyond the SM,
 which may help in understanding the observed
pattern of the fermion masses and mixing angles. How such extensions will 
look like is, however, neither obvious nor 
the subject of these lectures. Here, we have a
restricted mandate, namely we will investigate the question if the SM, which 
comes together with these masses and mixings, is a consistent and complete
description of data or not. 

\par
The matrix elements of $V_{\mbox{\footnotesize CKM}}$ are determined
by the charged current coupling to the
$W^{\pm}$ bosons. Symbolically this matrix can be written as:
\begin{equation}
V_{\mbox{\footnotesize CKM}} \equiv \left( \begin{array}{lll}
V_{ud} & V_{us} & V_{ub}\\
V_{cd} & V_{cs} & V_{cb}\\
V_{td} & V_{ts} & V_{tb}
\end{array} \right).
\end{equation}
 Of the nine
elements of $V_{\mbox{\footnotesize CKM}}$, the four involving the $u,d,c,s$ 
quarks and present in the upper left $2 \times 2$ block,
are studied in decays which are not our principal concern here. Their current
values can be seen in the PDG
review, where also references to the original literature can be 
found \cite{PDG94}.

\par
 We shall concentrate here on the remaining five matrix elements
in which quarks in the third family are involved. Two of the 
matrix elements involving the $b$ quark, $V_{ub}$ and
$V_{cb}$, have been measured in
direct decays of the $B$ hadrons in experiments
at the $e^+ e^-$ storage rings CESR, DORIS and LEP.
The theory and phenomenology behind their determination will be discussed
at some length here. The remaining
three elements $V_{td}$, $V_{ts}$ and $V_{tb}$ can, in principle, be 
directly measured in the decays of the top quark $t \to b W^+, ~t \to s 
W^+$ and $t \to d W^+$, respectively.
 First measurements of 
 $V_{tb}$ have been reported by the CDF collaboration \cite{CDFvtb},
through the measurement of the ratio $R_{tb}$,
\be\label{rtbcdf}
 R_{tb} \equiv \frac{{\cal B}(t \to
bW)}{\sum_{q=d,s,b} {\cal B}(t \to q W)}=
 0.87^{+0.13 ~+0.13}_{-0.30 ~-0.11} ~,
\ee 
 which is consistent with unity but
within experimental errors also consistent with a value of $V_{tb}$ which is 
considerably less than that. Apart from establishing the dominance of 
$V_{tb}$ (over the other two matrix elements) by improved measurements of 
the ratio $R_{tb}$ \footnote{A precision of $\Vtbabs =1.0 \pm 12\%$ is, for 
example,
projected at the Fermilab Tevatron with an integrated luminosity of
$2 (fb)^{-1}$, expected to be collected at the turn of this century
\cite{DPG96}.} it will be difficult in the foreseeable 
future to get quantitative information on  
 $V_{td}$ and $V_{ts}$ from decays of top quarks,
both due to the scarcity of data involving top quark production  
and, more importantly, the issues having to do with
efficient tagging of light-quark jets.

\par
 Fortunately, the matrix elements $V_{ti}$  are also
accessible in $B$ and $K$ decays through virtual transitions
involving the couplings $Wt\bar{b}$, $Wt\bar{s}$ and
$Wt\bar{d}$. Examples of these transitions in the $B$ system are the 
$|\Delta B| =2$
processes, $B^0$ - $\bar{B}^0$ mixings, and $|\Delta B|=1$ FCNC processes,
 such as rare $B$ decays
$b \to (s,d) + \gamma$ and $b \to (s,d) + l^+ l^-$. We discuss the theory
and phenomenology of these processes here.
Precision experiments involving $B$ decays and mixings will completely
determine the matrix $V_{\mbox{\footnotesize CKM}}$,
including the CP-violating phase, establishing either that the SM provides a
consistent theoretical framework for describing flavour physics or else
that the charged current $J_\mu^{cc}$  must be modified.
\subsection{Some popular representations of the CKM Matrix}

\par
  We discuss a couple of popular representations of the CKM matrix.
          The original parametrization due to
Kobayashi and Maskawa \cite{KM}
                      was constructed from the
rotation matrices in the flavour space
involving the angles $\theta_i ~(i = 1,2,3)$ and a
phase $\delta$,
\begin{equation}
V_{\mbox{\footnotesize KM}} = R_{23} ( \theta_3, \delta) R_{12} (\theta_1, 0)
R_{23} (\theta_2,0),
\end{equation}
where $0 \le \theta_i \le \pi/2,\,\, 0 \le \delta \le 2 \pi$, and
$R_{ij}(\theta, \phi)$ denotes a unitary rotation in the $(i,j)$
plane by the angle $\theta$ and the phase $\phi$.
The resulting representation is:
\begin{equation}
V_{\mbox{\footnotesize KM}} = \left( \begin{array}{lll}
c_1 & -s_1c_3 & -s_1 s_3 \\
s_1 c_2 & c_1c_2c_3-s_2s_3e^{i \delta} & c_1c_2s_3+
s_2c_3e^{i\delta} \\
s_1s_2 & c_1 s_2c_3+c_2s_3 e^{i \delta} & c_1s_2s_3-c_2c_3
e^{i\delta}
\end{array} \right),
\end{equation}
with $ c_i = \cos \theta_i, s_i = \sin \theta_i $.
This reduces to the usual Cabibbo form for $ \theta_2 = \theta_3 = 0$,
with the angle $\theta_1$, identified
(up to a sign) with the Cabibbo angle. In the PDG review \cite{PDG94}, 
however, another parametrization is advocated which differs
from $V_{\mbox{\footnotesize KM}}$ in assigning the
complex phases (dominantly) to the (1,3) and (3,1)
matrix elements of $V_{\mbox{\footnotesize CKM}}$.
We shall not write the PDG representation
of the CKM matrix but give instead the simpler approximate form,
which follows from the {\it a posteriori }realization that
      $c_{13} \simeq 1 $  to a very high accuracy,
due to the measurement of
$|V_{ub} | = s_{13} \simeq 0.003$--$0.006$ (see below).
\begin{equation}
V_{\mbox{\footnotesize PDG}} \simeq \left( \begin{array}
{lll} c_{12} c_{13} &  s_{12} c_{13} &
s_{13} e^{-i\delta_{13}}\\
-s_{12}c_{23} & c_{12}c_{23} & s_{23} c_{13} \\
s_{12} s_{23} - c_{12} c_{23} s_{13} e^{i \delta_{13}}
& -c_{12} s_{23}  & c_{23}c_{13}
\end{array} \right).
\end{equation}
The three angles called $\theta_{ij} , ~i \neq j$,
                                                 can be chosen to lie in
the range $0 \le \theta_{ij} \le \pi/2$, making all the sines and
cosines $c_{ij}$ and $ s_{ij}$ real and positive. The KM phase,
called for obvious reasons
 $\delta_{13}$, lies in the range $0 \le \delta_{13}
\le 2 \pi$. In the limit $\theta_{13} = \theta_{23} = 0$,
the third generation decouples; identifying the
Cabibbo angle with $\theta_{12}$, one recovers the Cabibbo form. To
an excellent accuracy, which we will discuss, one could set $c_{13}=1$ and
$c_{23}=1$. In that case,
for the PDG parametrization,
\begin{equation}
| V_{us} | =  | s_{12} c_{13} |  \simeq | s_{12} |;
~~~| V_{ub} | = | s_{13} |;
~~~| V_{cb} | = | s_{23} c_{13}| \simeq |s_{23}|.
\end{equation}
These can then be taken as three independent parameters,
measured directly in decays, together with the
phase $\delta_{13}$.
\par
Another approximate but very useful form of the matrix
$V_{\mbox{\footnotesize CKM}}$ is due to Wolfenstein \cite{Wolfenstein},
who made the observation that (empirically) the following
pattern for the $V_{\mbox{\footnotesize CKM}}$ matrix elements is
suggested by data:
\begin{eqnarray}
|V_{ii}| & \simeq & 1, ~i=1...3,  \nonumber\\
|V_{12} | & \simeq & |V_{21}| \sim \lambda, \nonumber\\
|V_{23}| & \simeq & |V_{32}| \sim \lambda^2 \nonumber\\
|V_{13}| &\sim & \lambda^3, ~~|V_{23}| \sim \lambda^3,
\end{eqnarray}
with $\lambda \equiv \sin \theta_c \simeq 0.221$.
Thus, it is useful to write a perturbative form (in $\lambda$) for
$V_{\mbox{\footnotesize CKM}}$. Denoting this by $V_{\footnotesize{\small 
\mbox{Wolfenstein}}}$,
 \begin{equation}\label{Vwolf}
V_{\mbox{\footnotesize Wolfenstein}} = \left( \begin{array}{lll}
1-\frac{1}{2} \lambda^2 & \lambda &
A\lambda^3 (\rho - i \eta) \\
- \lambda & 1-\frac{1}{2} \lambda^2 
 & A \lambda^2 \\
A\lambda^3 (1-\rho-i \eta) & -A \lambda^2 & 1
\end{array} \right).
\label{Vwolfenstein}
\end{equation}
Like the previous representations, $V_{\footnotesize{\mbox{Wolfenstein}}}$
 has also three
real parameters called $A$, $\lambda$ and $\rho$, and a phase $\eta$.
Since we shall be making extensive use of this parametrization, we
write some relations involving
the matrix elements of interest in this representation:
\begin{equation}
\frac{|\Vbu|}{|\Vbc|} = \lambda \sqrt{\rho^2 + \eta^2},
~~~\frac{|\Vtd|}{|\Vbc|} = \lambda \sqrt{(1-\rho)^2 + \eta^2},
\nonumber \\
\end{equation}
\begin{equation}
\frac{|\Vtd|}{|\Vbu|} =  \sqrt{\frac{(1-\rho)^2 + \eta^2}{\rho^2+ \eta^2}},
~~~\frac{|\Vts|}{|\Vbc|} = 1 ~,
\end{equation}
and the dominant phases are:
\begin{equation}
\Im (\Vbu ) = \Im (\Vtd ) = - A \lambda^3 \eta .
\end{equation}
 It should be recalled  
that the Wolfenstein parameterization given in Eq.~(\ref{Vwolfenstein})
is an approximation and in certain 
situations in the future it may become mandatory to specify  the matrix 
by taking into account the dropped terms in $O(\lambda^4)$ 
in $V_{\footnotesize{\mbox{Wolfenstein}}}$. For the present experimental and 
theoretical accuracy, the representation (\ref{Vwolf}) is
entirely adequate and we shall restrict ourselves to this form.
Further discussions on this point and suggestions on improved 
treatment to include higher order terms in $\lambda$
 can be seen in \cite{BuBu94}. \\ 

  The four CKM parameters and the six quark masses together with the masses
of the three charged leptons make thirteen of the nineteen parameters of the
standard model; the remaining six can be taken as the three gauge coupling
constants, the Higgs boson mass, the mass of the $W$ boson, and the 
parameters $\theta_{vac}(QCD)$
and $\theta_{vac}(EW)$, which are related to the instanton
sectors of QCD and the electroweak theory, respectively.
 Being fundamental
constants of nature, it is utmost important to measure them as precisely
as possible and the  main goal of flavour physics is to pin down at
least the first thirteen of them.    

\subsection{The CKM unitarity triangles}
\par
\indent
The CKM matrix elements obey unitarity constraints, which state that
any pair of rows, or any pair of columns, of the CKM matrix are
orthogonal. This leads to six orthogonality conditions. The three
involving the orthogonality of columns
  are listed
below, with the quark pair in the parenthesis $(jk)$ representing the
product of the $j$'th and $k$'th columns:
\begin{eqnarray}
  (ds): ~~~\sum_{i}  V_{id}V_{is}^* &=& 0, \nonumber\\
  (sb): ~~~\sum_{i}  V_{is}V_{ib}^* &=& 0,          \\
  (db): ~~~\sum_{i}  V_{id}V_{ib}^* &=& 0. \nonumber
\label{dsb}
\end{eqnarray}
Similarly, there are three more such orthogonality conditions on the
rows:
\begin{eqnarray}
  (uc): ~~~\sum_{j}  V_{uj}V_{cj}^* &=& 0, \nonumber\\
  (ct): ~~~\sum_{j}  V_{cj}V_{tj}^* &=& 0,          \\
  (ut): ~~~\sum_{j}  V_{uj}V_{tj}^* &=&0. \nonumber
\label{uct}
\end{eqnarray}
The six orthogonality conditions can be
depicted as six triangles in the complex plane of the CKM parameter space
\cite{AKL94}. 
The FCNC transitions in which the pair of quarks depicted
participate test these triangular constraints directly. Thus,
 the triangle labeled  $(ds)$ represents the unitarity
constraints on the transition $s \to d$, which, for example, one
encounters in
the $K^0$--$\overline{K^0}$ transition and in rare $K$ decays such as
$K_L \to \pi^0 \ell^+ \ell^-$. The two others in this group,
namely $(sb)$ and $(db)$, are encountered in the FCNC transitions
in $B$ decays, such as particle-antiparticle mixings  $\BZ$ - $\BZB$ and 
$\BS$ - $\BSB$ and rare $B$ decays $B \to K^* + \gamma$ and
$B \to \omega + \gamma$, respectively. The unitarity conditions on the
rows will be difficult to test in FCNC
transitions shown, as the rates of such transitions are enormously 
suppressed in the SM, and we shall not discuss this set in these lectures.	
\par
\begin{figure}
\epsfig{file=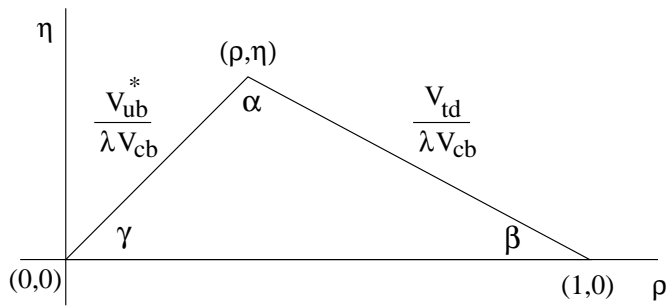,bbllx=30pt,bblly=285pt,bburx=390pt,bbury=494pt,
width=10cm}

\caption{The unitarity triangle. The angles $\alpha$, $\beta$ and $\gamma$
can be measured via CP violation in the $B$ system and the sides
from the CC- and FCNC-induced $B$ decays.}
\label{triangle}
\end{figure}

 The constraint 
stemming from the orthogonality condition on the first and third
row of $V_{\mbox{\footnotesize CKM}}$,
\begin{equation}
V_{ud} V_{td}^* + V_{us} V_{ts}^* + V_{ub} V_{tb}^* = 0 
\label{tdunit}
\end{equation}
 has received considerable attention. Since
$V_{ud} \simeq  V_{tb} \simeq 1$ and $V_{ts}^* \simeq - V_{cb}$,
the unitarity relation (\ref{tdunit}) simplifies:
\begin{equation}
V_{ub} + V_{td}^* = V_{us} V_{cb},
\end{equation}
which can be conveniently depicted as a triangle relation in
the complex plane, as shown in Fig. ~\ref{triangle}. Thus, knowing the
sides of the CKM-unitarity triangle, the three
angles  of the triangle $\alpha, \beta$ and $\gamma$ are determined.
These angles are all related to the Kobayashi-Maskawa phase
$\delta$ (equivalently the phase $\delta_{13}$
 in $V_{\mbox{\footnotesize PDG}}$ or the
phase $\eta$ in $V_{\mbox{\footnotesize Wolfenstein}}$),
 and they can, in principle, be
independently measured in various
CP-violating $B$ decays.
As we shall discuss below, the matrix elements
$V_{cb}$ and $V_{ub}$ are already known from the  CC $B$ decays.
         With more data from $B$ decays and an improved theory
one would be able to determine them rather
precisely. The matrix element
           $V_{td}$ can, in principle, be determined from the
rare decays $b \to d + ~\gamma$, $b \to d + ~l^+ l^-$,
 $ b \to d + \nu \bar{\nu}$
(and some selected exclusive decays),
 and $B_d^0$--$B_d^0$
mixing, which already provides a first measurement of $\Vtdabs$ which
is, however, not very precise.
This set of experiments
then provides another way of determining the triangle, namely by measuring
its sides. The unitarity triangle 
represents an important testing ground for the SM flavour
physics, in particular in $B$ and $K$ decays.

\par
It has been pointed out by Jarlskog \cite{Jarlskog},
that there exists a large number of 
different parametrizations of the CKM matrix.
However, since the phases of the quark fields are unphysical
quantities, the different parametrizations, emerging from
specific choices of these phases, must all be equivalent.
The parametrization independent quantities are the
absolute values of the matrix elements $|V_{ij}|$ (hence also the angles 
of the unitarity triangles) and the area of the unitarity triangles, which is
the same for all six triangles and is an invariant measure of  CP violation.
This can be expressed as
 \begin{eqnarray}
 \mbox{area}[\Delta \mbox{(CKM)}] & = &
  \frac{1}{2}s_{12}s_{23}s_{13} \sin{\delta}_{13}
  ~~~~\mbox{[PDG]},\nonumber \\
     & = & \frac{1}{2}A\lambda^{6} \eta
  ~~~~~~~~~~~~~~~~\mbox{[\rm Wolfenstein]},
\end{eqnarray}
for the PDG and
Wolfenstein parametrizations of $ V_{\mbox{\footnotesize CKM}} $.
The Jarlskog invariant denoted by the symbol $J$ \cite{Jarlskog} is 
twice this area, which  
in the standard model is typically of $O(10^{-5})$. It is being 
debated if the intrinsic smallness of $J$ in the standard model is a 
serious problem in 
explaining the measured baryon asymmetry of the universe (BAU),
whose quantitative measure is the ratio of the baryon number density to
entropy density,
\begin{equation}
  \Delta_B= \frac{\rho(B)}{s},
\end{equation}
and its present value is $\Delta_B=(4 - 6) \times 10^{-11}$ \cite{PDG94}.
 Electroweak baryogenesis is a
subject of great theoretical interest and one which is not quite under 
quantitative control as it requires a deeper
understanding of the dynamics of the electroweak theory
at high energies and temperatures. We refer to a recent review
on this subject  by Rubakov 
and Shaposhnikov \cite{RS96}, where several of the key theoretical
concepts and calculational strategies are
discussed. There is one input to the electroweak baryogenesis issue that can 
possibly come from $B$ decays, namely these decays will probe  
if nature admits more than one mechanism of CP
violation. Along the same lines, new sources of CP violation, i.e., 
additional phases, such as are present in multi-Higgs models, may uncover
themselves by inducing electric dipole moments for the neutron and 
the electron at a level just above their present upper bounds \cite{PDG94},
which for sure will not be accommodated by the phase in the CKM matrix.
%
\section{Dominant $B$ Decays in the Standard Model}

\par
 With this introduction, we now turn to study questions which are 
specific to $B$ physics. We start by
 discussing the dominant decay rates 
which determine the lifetimes of the $B$ hadrons, $\tau_B$,
 their semileptonic branching
ratios ${\cal B}_{SL }$ and the charm quark multiplicity in $B$ decays
$\langle n_c\rangle$, a quantity which has become an important ingredient
in understanding the semileptonic branching ratio in the standard model.
 
 The effective lowest-order
weak interaction Hamiltonian can be expressed in terms of
 $J_{\mu}^{CC}$, introduced earlier,\\
\begin{equation}
{\cal{H}}_{W}
      = \frac{G_{F}}{2 \sqrt{2}} \left( J_{\mu}^{CC}
J^{\mu\dag CC} + \mbox{h.c.} \right) ,
\end{equation}
where $G_{F}$ is the Fermi coupling constant. The calculational
framework that will be used is QCD and we concentrate first on
perturbative QCD improvements of the decay rates and distributions
in $B$ decays. The leading order (in $\as$)
perturbative QCD improvements using ${\cal{H}_W}$ have been worked out in
semileptonic processes in \cite{CM} - \cite{JK89},
 which are modeled on the electromagnetic radiative
corrections in the decay of the $\mu$-lepton \cite{Behrends}. This 
argument rests on the property of asymptotic freedom enabling one to
use the parameter $\as(\mb)/\pi$ to do a perturbation expansion of the
decay amplitudes. There still remains the question of how to relate
the perturbative calculations to the actual decays of $B$ hadrons. This
involves some kind of parton-hadron duality that will be discussed later.
For the non-leptonic decays, such perturbative corrections are
calculated 
using the renormalization group techniques \cite{AM74}$ - $\cite{BW90}.
We shall discuss some of these techniques here. The 
underlying theoretical framework and its numerous 
applications in weak decays of the $K$ and $B$ mesons have been recently 
reviewed in a comprehensive paper by
Buchalla, Buras and Lautenbacher \cite{BBL95}, to which we refer for 
details.

The quantitative description of the
 physical decay processes, with hadrons in the initial
and (in most cases of interest) also final states, requires the knowledge of
the wave functions (using the quark-parton model language) or
hadronic form factors and structure functions. The primary task of the 
theory therefore is to evaluate these non-perturbative functions which
depend on the strong interaction dynamics.
 The methods that we shall be using here in studying weak decays are 
based 
on the lattice QCD framework, QCD sum 
rules, and the heavy quark limit of QCD which allows one to
do a systematic expansion of decay amplitudes in $1/m_Q$, where
$m_Q \gg \Lambda_{QCD}$, and $\Lambda_{QCD}$ is the QCD scale parameter
which is typically of $O(200$ MeV) \cite{PDG94}. 
 The lattice QCD framework aims at calculating Green's functions 
and their S-matrix elements from  first principles, i.e. QCD. However, 
in practice,  predictions are hampered by 
the limitations having to do with the computing power
and/or appropriate lattice formulations. Unquenching lattice QCD is one
of the foremost technical problems.
 In addition, simulating $B$ systems directly on the lattice with
present day technology introduces finite-size effects. In view of this,
present lattice-QCD estimates in $B$ systems are to be taken with some 
caution. For results on topics relevant for these lectures, we refer to
recent reviews \cite{CMichel,Soni95,Shigemitsu} from where we shall
be drawing heavily.
 The approach, involving the QCD sum rules \cite{SVZ},
 allows one to make predictions about
current correlation functions which are calculated using the operator
product expansion (OPE). The results allow themselves to be expressed
in terms of a limited number of non-perturbative parameters. To extract
physical quantities, the notion of quark-hadron duality is invoked which
enables one to compare suitably weighted quantities. While the method is
 theoretically well-founded, 
reliable calculations can only be made if the higher twist- and higher
order QCD corrections have been calculated. This is not a limitation in
principle but again in practice a trustworthy theoretical treatment is
available only in a limited number of cases. For a recent review of
the applications of the QCD sum rules in $B$ decays 
and discussions of some of the
inherent uncertainties, see, for example, the review by Braun 
\cite{Braunbrussels}.

 Recently, remarkable progress has been made in the formulation of QCD as
an effective theory in the heavy quark limit, in which  the
resulting theory shows symmetry properties not present in the original QCD 
Lagrangian. These symmetries enable one to make model-independent predictions
for hadronic transition form factors involving  some exclusive $B \to 
(D,D^*,...)$ decays at kinematic points where such symmetry relations hold.
Some of the pioneering work in this direction can be seen in
\cite{HQET,IW}. 
In particular, these methods have enabled us to determine $\Vcbabs$ with
controlled theoretical errors and we shall discuss this application here.
Along the same lines, equally interesting are the related
techniques which involve a systematic expansion
of the inclusive decay amplitudes in the inverse heavy
quark mass \cite{Chayetal} - \cite{Bigi}. A very satisfying
feature of this framework is that the
parton model for heavy quark decays emerges as the leading term in a
systematic expansion of the decay amplitudes.
Power corrections to some of the inclusive and exclusive decay processes
have been calculated in terms of a limited number of non-perturbative
parameters. These methods, like the QCD sum rules, use operator
product expansions which are well defined in Euclidean space.
To make predictions in time-like regions, some notion of quark-hadron 
duality is again required. Theoretically, this aspect of the effective
theories is not quite understood.
 We shall not 
enter into the technical details surrounding these developments
and refer to the original literature and some excellent reviews on
this subject \cite{Georgi92} - \cite{Neubert95}. However, we shall discuss
several illustrative applications of this method in these lectures.

\subsection{Inclusive semileptonic decay rates of the $B$ hadrons}
\par
With this theoretical prelude, we start with the assumption
that the inclusive decays of $B$ hadrons can
be modeled on the QCD-improved quark model decays. More specifically,
 while calculating rates, we shall be equating the partial and total decay
rates of the
$B$ hadrons to the corresponding expressions obtained in the parton model,
relying on the heavy quark expansion \cite{Chayetal} - \cite{MW94}:
\begin{equation}
\Gamma(B \to X) = \Gamma (b \to x) + O(1/m_b^2) ~, 
\end{equation}
 For $b$ quark semileptonic
decays involving CC interactions, one has two partonic transitions:
\begin{eqnarray}
  b & \longrightarrow & c\ell^{-}\bar{\nu}_{\ell},  \\
    & \longrightarrow & u\ell^{-}\bar{\nu}_{\ell}. \nonumber
\label{bsemilept}
\end{eqnarray}   
There exists a  close analogy between the $b$ quark decays
  and $\mu$ decay,
$\mu^{-}\longrightarrow e^{-}\bar{\nu}_{e}\nu_{\mu}$,
with the identification:\\
\begin{equation}
[b,(c,u),\bar{\nu}_{\ell},\ell^{-}]\leftrightarrow [\mu^{-},e^{-},
\bar{\nu}_{e},\nu_{\mu}].
\end{equation}
 This analogy holds in general; in particular,
       at the one loop
level $O(\alpha$) QED corrections to $\mu^{-}$ decay and
$O(\alpha_{s}$) QCD corrections to $b$ semileptonic
 decays are related by simply
replacing \cite{CM,Suzuki,Alipiet}\\
\begin{equation}
\alpha\longrightarrow\  \frac{1}{3}\alpha_{s} Tr\sum_{i=1}^{8}
\lambda_{i}\lambda_{i}   = \frac{4}{3}\alpha_{s},
\end{equation}
where $\lambda_{i}$ are the Gell-Mann $SU(3)$ matrices, and
$\as$ is the lowest order QCD effective coupling constant,
\begin{equation}
\alpha_{s} = \frac{12\pi}{(33-2n_{f})\ln(  \frac{m_{b}^{2}}
{ \Lambda_{\mbox{ \small QCD}}^{2} }  )} \;\;,
\end{equation}
where $n_f$ is the number of effective quarks.
                     The semileptonic decay rates can then be
read off the expression for the $O(\alpha$) radiatively
corrected $\mu$-decay rate \cite{Behrends}.
The rates for $b\longrightarrow (u,c)\ell\nu_{\ell}$ decays,
setting $m_{\ell}=m_{\nu_{\ell}}=0$, are given by the expression: 
\begin{equation}
\Gamma_{SL }(b\longrightarrow (u,c)\ell\nu_{\ell})=
\Gamma_{0} f(r_{i})\left[1-\frac{2}{3}\frac{\alpha_{s} (m_{b}^{2})}
{\pi} g(r_{i})\right],
\end{equation}
with $\Gamma_{0}$ being the normalization factor in the lowest-order
rate
 \begin{equation}
\Gamma_{0} = \frac{G_{F}^{2}}{192\pi^{3}}
\mid V_{ib}\mid^{2} m_{b}^{5},
\end{equation}
 $r_{i} = m_{i}/m_{b} ~(i=u,b)$, and
\begin{equation}
f(r) = 1-8r^{2}+8r^{6}-r^{8}-24r^{4} \ln r .
\end{equation}
The function $g(r)$
              has the normalization $g(0)=\pi^{2}-\frac{25}{4}$,
and numerically $g(0.3)\simeq 2.51$, relevant for the $b \to u $ and $b 
\to c$ transitions, respectively \cite{CM,Suzuki,Alipiet}.
With $\Lambda_{\mbox{\small QCD}} \simeq 200$ MeV and $n_f=5$, this gives 
about 
$(15)\%$ corrections to the semileptonic decay widths involving $\ell =e,
\mu$, reducing $\Gamma_{SL }$ compared to the lowest order result
$ \Gamma_{SL }^{(0)} =\Gamma_{0} f(r)$.
The corresponding decrease in the decay width for the semileptonic
decay $b \to c \tau \nu_\tau$ is obtained by an expression very  
similar to the above one in which the $\tau$-mass effects are included
in the phase space and in the QCD corrections.
 \begin{equation}\label{bcln}
\Gamma(b\to c \tau \nu_\tau)=\Gamma_0 P(x_c,x_\tau,0)\left[
 1+\frac{2\as(\mu)}{3\pi} g(x_c,x_\tau,0)\right]
\end{equation}
where $P(x_1,x_2,x_3)$ is the well known three-body phase space factor
given for arbitrary masses $x_i=m_i/m_b$ by \cite{BKbook}:
\begin{equation}\label{px123}
P(x_1,x_2,x_3)=12\int\limits_{(x_2+x_3)^2}^{(1-x_1)^2} \frac{ds}{s}
(s-x^2_2-x^2_3)(1+x^2_1-s) w(s,x^2_2,x^2_3) w(s,x^2_1,1) ~,
\end{equation}
\begin{equation}\label{wabc}
w(a,b,c)=(a^2+b^2+c^2-2ab-2ac-2bc)^{1/2}
\end{equation}
The function $g(x_1,x_2,x_3)$ has been calculated for arbitrary arguments
in \cite{PHam83} in terms of a one-dimensional integral.
The functions $P(x_1, 0, 0)$ and $g(x_c, 0,0)$ go over to the
functions $f(r)$ and $(-)g(r)$, respectively, given above for the massless
lepton case.  The numerical
values for $g(x_c, x_\tau,0)$ and $g(x_c,0, 0)$ are tabulated in
\cite{Bagan94}. For the default value $x_c=0.3$, one has $g(x_c,
x_\tau,0) = -2.08$, yielding about a 12 \% decrease in $\Gamma (b \to c
\tau \nu_\tau)$ compared to $\Gamma_{SL }^{(0)} (b \to c \tau \nu_\tau)$
as a result of the leading order QCD corrections \cite{PHam83}. For 
more modern calculations of the decay rate
$\Gamma_{SL } (b \to c \tau \nu_\tau)$, see \cite{FCz95}.

\subsection{Inclusive non-leptonic decay rates of the $B$ hadrons}
\par
  The dominant CC-induced non-leptonic and semileptonic decays of $B$
hadrons are governed by the effective Lagrangian,
\begin{eqnarray}\label{Leff1}
{\cal L}_{eff} = -4\frac{G_F}{\sqrt{2}}\Vud^*\Vbc \left[ C_1(\mu){\cal 
O}_1(\mu)
 + C_2(\mu){\cal O}_2(\mu)\right] \nonumber\\
 -4\frac{G_F}{\sqrt{2}}\Vus^*\Vbc \left[ C_1(\mu){\cal O}_1^\prime(\mu)
+ C_2(\mu){\cal O}_2^\prime(\mu)\right] \nonumber\\
-4\frac{G_F}{\sqrt{2}}\Vbc \left[\sum_{\ell=e,\mu,\tau} \bar{\ell}_L 
\gamma_\mu \nu_\ell \bar{c}_L\gamma^\mu b_L \right] + h.c.~,
\end{eqnarray}
and we have just discussed the $O(\as)$ renormalization effects
to the matrix elements of the semileptonic piece in ${\cal L}$.
Here ${\cal O}_1$ and ${\cal O}_2$ are the colour-octet and colour-singlet
four-Fermi operators, respectively (here $\alpha$ and $\beta$ are colour 
indices),
 \begin{eqnarray}
{\cal O}_1 &=& (\bar{d}_\alpha u_\beta)_L(\bar{c}_\beta b_\alpha)_L, 
\nonumber\\
{\cal O}_2 &=& (\bar{d}_\alpha u_\alpha)_L(\bar{c}_\beta b_\beta)_L,
\end{eqnarray}
and $q_L=1/2(1-\gamma_5)$ denotes a left-handed quark field. The operators
${\cal O}_{i}^\prime$ are related to the corresponding fields ${\cal O}_i$
by the relacement $\bar{d} \to \bar{s}$. The octet-octet $({\cal O}_1)$
and singlet-singlet $({\cal O}_2)$ operators emerge due to a single
gluon exchange between the weak current lines (quark fields) and follow from
the colour charge matrix $(T^{a}_{ij})$ algebra:
\begin{equation}
T_{ik}^{a} T_{jl}^{a} = \frac{1}{2N_c} \delta_{ik} \delta_{jl} + \frac{1}{2}
 \delta_{il} \delta_{jk}~.
\end{equation}
Here, $N_c=3$ for QCD. The Wilson coefficients 
$C_i(\mu)$ are calculated at the scale $\mu =m_W$ and then scaled down 
to the scale typical for $B$ decays, $\mu =O(m_b)$, using the renormalization
group equations, which brings to the fore the influence of strong 
interactions
on the dynamics of weak non-leptonic decays. Without QCD corrections, the
two Wilson coefficients 
have the values $C_1(m_W)=0, ~C_2(m_W)=1$.
 Since the operators
${\cal O}_1$ and ${\cal O}_2$ mix under QCD renormalization, it is convenient
to introduce the operators ${\cal O}_\pm \equiv ({\cal O}_2 \pm {\cal 
O}_1)/2$ having the Wilson coefficients $C_\pm$
 which renormalize multiplicatively \cite{AM74}. The results
are now known to two-loop accuracy \cite{BW90}:
\begin{equation}\label{Cpm2loop}
C_\pm (\mu) = L_\pm(\mu) \left[1 + \frac{\as(m_W) - \as(\mu)}{4 \pi}
\frac{\gamma_\pm^{(0)}}{2 \beta_0} 
\bigg(\frac{\gamma_\pm^{(1)}}{\gamma_\pm^{(0)}}
-\frac{\beta_1}{\beta_0}\bigg) + \frac{\as(m_W)}{4 \pi} B_\pm \right],
\end{equation}
where the multiplicative factor in this expression represents the
solution of the RG equations in the leading 
order QCD \cite{AM74},
\begin{equation}\label{lpmmu}
L_\pm(\mu)=\left[\frac{\as(M_W)}{\as(\mu)} 
\right]^{d_\pm}~,
\end{equation}
and the exponents have the values $d_+=\gamma_+^{(0)}/(2\beta_0)$, 
$d_-=\gamma_-^{(0)}/(2 \beta_0)$. The
quantities $\gamma_\pm^{(i)}$ are the coefficients of the anomalous 
dimensions involving the operators ${\cal O}_\pm$ (and ${\cal 
O}_\pm^\prime$),
\begin{equation}
\gamma_\pm =\gamma_\pm^{(0)} \frac{\as}{4 \pi} + \gamma_\pm 
^{(1)}(\frac{\as}{4\pi})^2 + O(\as^3) ,
\end{equation}
 with
\begin{equation}
\gamma_+^{(0)}=4, ~~\gamma_-^{(0)}=-8, ~~\gamma_+^{(1)}=-7+\frac{4}{9}n_f,
~~\gamma_-^{(1)}=-14 -\frac{8}{9} n_f ,
\end{equation}
in the naive dimensional regularization (NDR) scheme, i.e., with 
anticommuting $\gamma_5$.
 The $\beta_i$ are the
first two coefficients of the QCD $\beta$-function, and they have the values
\begin{equation}
\beta_0=11 -\frac{2}{3} n_f, ~~~\beta_1=102-\frac{38}{3}n_f~.
\end{equation}
Finally, the functions $B_\pm$ are the matching conditions obtained by
demanding the equality of the matrix elements of the effective Lagrangian
calculated at the scale $\mu=m_W$ and in the full theory (i.e., SM)
up to terms of $O(\as(m_W^2))$. They
have the values:
\begin{equation}
B_\pm=\pm B \frac{N_c\mp 1}{2N_c},
\end{equation}
 The constant $B$ and the two-loop
anomalous dimension $\gamma_\pm^{(1)}$ are both regularization-scheme 
dependent. In the NDR scheme one has $B=11$. Following 
\cite{BW90}, we define a scheme-independent quantity $R_\pm$,
\begin{equation}
R_\pm = B_\pm + \frac{\gamma_\pm^{(0)}}{2 \beta_o}
\bigg(\frac{\gamma_\pm^{(1)}}{\gamma_\pm^{(0)}}-\frac{\beta_1}
{\beta_0}\bigg),
\end{equation}
in terms of which the Wilson coefficients read
\begin{equation}
C_\pm (\mu) = L_\pm(\mu) \left[1 + \frac{\as(m_W) - \as(\mu)}{4 \pi}
 R_\pm + \frac{\as(\mu)}{4 \pi} B_\pm \right].
\end{equation}
In this form all the scheme-dependence resides in the coefficients 
$B_\pm$ which is to be cancelled by the scheme-dependence of the matrix
elements of the corresponding operators.

  In addition to the decays $b \to c + \bar{u}d, ~b\to c + 
\bar{u}s$ and
$b \to c + \ell \nu_\ell$, which are described by the 
effective Lagrangian (\ref{Leff1}), there are other decays
involving the CC transitions $b \to u X$, $b \to (c,u) + 
\bar{c} s$ and $b \to (c,u) + \bar{c}d$,
 which are not included in this 
Lagrangian. In a systematic treatment involving QCD renormalization, 
one has to enlarge the operator basis to include these transitions and the 
so-called
penguin operators. We shall return to a discussion of this part of
the Lagrangian later in these lectures as we discuss rare $B$-decays,
where the operator basis
will be enlarged and the corresponding Wilson coefficients calculated
in the leading logarithmic approximation.

\par
   We now discuss the semileptonic branching ratio
 ${\cal B}_{\mbox{\small SL }}$ for the
$B$ mesons and to be specific will consider the case $\ell=e, \mu$. This
branching ratio is to a large extent free of the CKM matrix element
uncertainties but requires a QCD-improved calculation of the inclusive
decay rates, $\Gamma_{\mbox{SL }}$, discussed above, and 
$\Gamma_{\mbox{tot}}$,
\begin{equation}\label{bslr}
{\cal B}_{\mbox{SL }} \equiv \frac{\Gamma (B \to X e 
\nu_e)}{\Gamma_{\mbox{tot}}(B)}, \end{equation}
with 
\bea\label{gamatot}
\Gamma_{\mbox{tot}} (B) &=& \sum_{\ell =e, \mu, \tau} \Gamma (B \to X \ell 
\nu_\ell) + \Gamma(B \to X_cX)  + \Gamma(B \to 
X_{c\bar{c}}X) \nonumber\\
 &+& \Gamma (B \to X_uX) +  \Gamma (B) (\mbox{Penguins})~.
\eea
In the spirit of the parton model, we shall equate $ \Gamma(B \to X_cX) =
\Gamma (b \to c \bar{u} d) +~\Gamma (b \to c \bar{u} s)$, noting that
the so-called $W$-annihilation and $W$-exchange two-body decays are
expected to be small in inclusive $B$ decays. This will be quantified 
later as we discuss the lifetime differences among $B$ hadrons which arise
from the matrix elements of the operators representing these 
contributions.
 The corrections for the  decay widths
 $\Gamma (b \to c \bar{u} d)$ and $\Gamma (b \to c \bar{u} s)$
 are identical neglecting
$m_u$ and $m_s$, and so their contributions can be described by
similar functions. The
resulting next-to-leading order QCD corrected sum can be expressed as:
\begin{eqnarray}\label{bcud}
\Gamma(b\to c\bar ud) + \Gamma(b \to c\bar us )&=&\Gamma_0 
P(x_c,0,0)\nonumber\\
&\times&\left[2L(\mu)^2_++L(\mu)^2_-+
 \frac{\as(M_W)-\as(\mu)}{2\pi}(2L(\mu)^2_+ R_++L(\mu)^2_- R_-)\right.  
\nonumber\\
 &+&\frac{2\as(\mu)}{3\pi}\left(\frac{3}{4}(L(\mu)_+-L(\mu)_-)^2
c_{11}(x_c)+\frac{3}{4}(L(\mu)_++L(\mu)_-)^2c_{22}(x_c)\right.  
\nonumber\\
&&+\left.\left.\frac{1}{2}
(L(\mu)^2_+-L(\mu)^2_-)(c_{12}(x_c,\mu)-12 \ln\frac{\mu}{m_b})\right)\right]
\nonumber\\
&&\equiv 3\Gamma_0 \eta(\mu) J(x_c,\mu) ~,
\end{eqnarray}
with $\eta(\mu)$ representing the leading 
order QCD corrections.
 The scheme independent $R_\pm$ come from the NLO   
renormalization group evolution and are given by \cite{BW90}
\begin{eqnarray}\label{R+-}
R_{+} &=& \frac{10863 -1278n_f +80n_f^2}{6(33-2n_f)^2} , \nonumber\\
R_{-} & =& -\frac{15021 -1530 n_f + 80 n_f^2}{3(33-2n_f)^2}
\end{eqnarray}
For $n_f=5$,  $R_+=6473/3174$,
$R_-=-9371/1587$.  Note that the leading dependence of $L(\mu)_\pm$ on the
renormalization scale $\mu$ is canceled to ${\cal O}(\as)$ by the
explicit $\mu$-dependence in the $\as$-correction terms.  Virtual
gluon and Bremsstrahlung corrections to the matrix elements of four
fermion operators are contained in the mass dependent functions $c_{ij}(x)$.
 The analytic expressions for the functions $c_{11}(x), 
c_{12}(x), c_{22}(x)$ are given in \cite{Bagan94} where also their
numerical values are tabulated. For our default value $x_c=0.3$, these
coefficients have the values:
\begin{equation}\label{cijnum}
c_{11}(0.3)\simeq 2 \qquad
c_{12}(0.3,\mu=\mb)\simeq -10
\qquad c_{22}(0.3)\simeq -1~.
\end{equation}
Lumping together all the perturbative and finite charm quark corrections in
a multiplicative factor $\Delta_{c}(m_b, x_c, \as(m_Z))$, the 
perturbatively corrected decay width can be expressed as:
\begin{equation}
\Gamma(b\to c\bar{u}d) + \Gamma(b\to c\bar{u}s)
=3  \Gamma_0 P(x_c,0,0)\left[1 +\Delta_{c}(x_c,
m_b, \as (m_Z)) \right].
\end{equation}
 For the central values of the parameters used here
($m_b=4.8$ GeV, $x_c=0.3$, $\mu=\mb$ and $\as(m_Z)=0.117$), the QCD 
corrections lead to an enhancement \cite{Bagan94}:
\begin{equation}
\Delta_{c}(m_b, x_c, \as (m_Z)) = 0.17 .
\end{equation}
Out of this, the bulk is contributed by the leading log factor
\begin{equation}
\eta(\mu)-1=\frac{1}{3} \left(2 L_{+}^2 + L_{-}^2 \right) -1= 0.10 ~.
\end{equation}

Next, we equate
 $\Gamma (B \to X_{c\bar{c}}) =\Gamma (b \to c \bar{c} s) + \Gamma 
(b \to c \bar{c} d)$ and 
 discuss the perturbative QCD corrections to the decay width
$\Gamma (b \to c \bar{c} s)$ and $\Gamma (b \to c \bar{c} d)$. Neglecting
$m_d$ and $m_s$, an assumption which has been found to be valid to a high
accuracy in \cite{Bagan95a}, the corrections in the two decay widths are
identical and the result can be written in close analogy with the ones
for the decay widths  $\Gamma (b \to c \bar{u} s)$ discussed above.  
\begin{eqnarray}\label{bccs}
\Gamma(b\to c\bar cs) + \Gamma(b\to c\bar cd)
&=&\Gamma_0 P(x_c,x_c,x_s) \nonumber\\
&\times&\left[2L(\mu)^2_++L(\mu)^2_-+   
 \frac{\as(M_W)-\as(\mu)}{2\pi}(2L(\mu)^2_+ R_++L(\mu)^2_- R_-)\right.
\nonumber\\
 &+&\frac{2\as(\mu)}{3\pi}\left(\frac{3}{4}(L(\mu)_+-L(\mu)_-)^2
k_{11}(x_c,\mu)+\frac{3}{4}(L_++L_-)^2k_{22}(x_c)\right.
\nonumber\\
&&+\left.\left.\frac{1}{2}
(L^2_+-L^2_-)(k_{12}(x_c)-12 \ln\frac{\mu}{m_b})\right)\right].
\end{eqnarray}
The functions $k_{ij}(m_b,x_c,\as (m_Z))$ have been calculated and 
their numerical values are tabulated in \cite{Bagan95b}. For $m_b=4.8$ 
GeV, $x_c=0.3$ and $\as (m_Z) =0.117$ and $\mu=m_b$, they have the values:
\begin{equation}\label{kijnum}
k_{11}(0.3)=6.44 \qquad
k_{12}(0.3,\mu=\mb)=0.82
\qquad k_{22}(0.3)=2.99 ~,
\end{equation}
showing that these coefficients are rather sensitive functions of $x_c$.
The corresponding numbers for $k_{ij}$ for the choice $m_s/m_b=0.04$ are
given in \cite{Bagan95b}. Again,
lumping together all the perturbative and finite charm quark corrections in
a multiplicative factor $\Delta_{cc}(m_b, x_c, \as(m_Z))$, the
perturbatively corrected decay width can be expressed as:
\begin{equation}\label{Deltac}
\Gamma(b\to c\bar{c}s)=3  \Gamma_0 P(x_c,x_c,x_s)\left[1 
+\Delta_{cc}(x_c, m_b, \as (m_Z)) \right].
\end{equation}
 For the values of the parameters used here
($m_b=4.8$ GeV, $x_c=0.3$ and $\as(m_Z)=0.117$), the QCD corrections lead
to an enhancement \cite{Bagan95a,Bagan95b}:
\begin{equation}
\Delta_{cc}(m_b, x_c, \as (m_Z)) = 0.37 .
\end{equation}
Out of this, the bulk is contributed by the next to leading order
correction.
This is by far the largest correction to the inclusive rates we have
discussed so far. Using pole quark masses and the renormalization scale
$\mu=m_b$, one gets \cite{Bagan95a}:
\begin{equation}
\frac{\Gamma (b \to c\bar{c}s)(NLO)}{\Gamma (b \to c \bar{c}s)(LO)}= 1.32 
\pm 0.07 ~.
 \end{equation}
The NLO corrections go in the right direction in bringing theoretical
estimates closer to the experimental value for the semileptonic branching
ratio. However, this will also lead to enhanced charmed quark multiplicity
$\langle n_c \rangle$ in $B$ decays. We shall return to a quantitative 
analysis of ${\cal B}_{\mbox{\small SL }}(B)$ and $\langle n_c \rangle$ at 
the end of this section.

   We now turn to a discussion of the CKM-suppressed and penguin transitions
contributing at a smaller rate to $\Gamma_{\mbox{tot}}(B)$. They are of
two kinds:
\begin{itemize}
\item $\Gamma(B \to X_u+X)$,
 which is suppressed due to the CKM matrix
element $\vert V_{ub} \vert$,
 with the rate depending on $\vert V_{ub} \vert^2$, and
\item $\Gamma(B)(\mbox{Penguin})$: The so-called penguin transitions $b 
\to s +X$, where $X=c\bar{c}$
and $X=g$ (QCD penguins), $X=\gamma$ (electromagnetic penguins),
$X=\ell^+ \ell^-, \nu\bar{\nu}$ (electroweak penguins).
\end{itemize}
 There are even smaller transitions involving $b \to d +X$, as well as
a host of other rare decays but we shall neglect them in estimating the
total decay width for obvious reasons. We list below the numerical
contributions where in all entries involving $b \to u +X$ transitions,
we have set $\vert V_{ub} \vert /\vert V_{cb}\vert =0.08$, corresponding
to the present central value of this ratio \cite{Tomasz95}.    
The contribution of $b\to u\bar ud$ is calculated without penguins and 
the contribution to the $b\to c\bar cs$ rate given below is due to 
interference 
of the leading current-current type transitions with the penguin operators.
\begin{equation}\label{bwsmall1}
\Gamma(b\to u\sum_l l\nu)\approx 0.015 \Gamma_0 \qquad
\Gamma(b\to u\bar cs)\approx 0.010 \Gamma_0
\end{equation}
\begin{equation}\label{bwsmall2}
\Gamma(b\to u\bar ud)\approx 0.022 \Gamma_0 \qquad
\Delta\Gamma_{penguin}(b\to c\bar cs)\approx -0.041 \Gamma_0
\end{equation}
\begin{equation}\label{bwsmall3}
\Gamma(b\to s g)\approx 5.0 \times 10^{-3} \Gamma_0 \qquad
\Gamma(b\to s \gamma)\approx 8.0 \times 10^{-4} \Gamma_0 \qquad
\end{equation}
\begin{equation}\label{bwsmall4}
\Gamma(b\to s \nu \bar{\nu})\approx 1.2 \times 10^{-4} \Gamma_0 \qquad
\Gamma(b\to s \ell^+ \ell^-)\approx 5.0 \times 10^{-5} \Gamma_0 \qquad
\end{equation}
Note that the contribution due to the interference with the penguin
transitions in $b\to c\bar cs$ is negative and it
tends to cancel the other small contributions listed above in the total 
non-leptonic width.  The sum of all these contributions add up to
\begin{equation}
\Gamma(B \to X_u+X) + \Gamma(B)(\mbox{Penguins)} \simeq 1.25 \times 
10^{-2} \Gamma_0 ~,
\end{equation}
and hence not of much consequence for
the semileptonic branching ratio or the  $B$ hadron lifetime estimates.
 Rare $B$ decays are
either constrained by direct experimental searches or indirectly through
the measurement of the branching ratio ${\cal B}(\BGAMAXS)$ which excludes
most of the allowed parameter space where these decay modes may have 
appreciably larger branching ratios.
 It may be parenthetically added here
that the chromomagnetic QCD-penguin contribution $b \to s +g$ could 
be significantly large in some extensions of the SM, in particular the 
minimal supersymmetric standard model (MSSM) \cite{Masieroetal,Giudice96}.
This is sometime suggested as a possible source of enhanced non-leptonic
decay width.  
 There exist already an experimental
bound on this decay branching ratio \cite{Tomasz95}, obtained through the
decay chain $b \to s + g \to X_s + \phi$ on the assumption that the $\phi$-
energy spectrum in inclusive $B$ decays is rather hard, 
similar to the photon energy spectrum in electromagnetic penguin decays
$\BGAMAXS$. This quasi two-body $E_\phi$-spectrum is suggested in 
\cite{Deshpandebsg}. 
 However, in our
opinion, such an assumed $E_{\phi}$-spectrum is rather unrealistic
as it is unlikely that the fragmentation of an $O(5$ GeV) $s+g$ 
system will give rise to dominantly two-body or quasi-two body final 
states, hence the 
experimental bound so-obtained \cite{Tomasz95} is not very compelling.
\subsection{Power corrections in $\Gamma_{SL}(B)$ and $\Gamma_{NL}(B)$}
 Before we discuss the numerical results for ${\cal 
B}_{\mbox{\small SL}}$, we include the
$O(1/m_b^2)$ power corrections in the inclusive partonic decay widths, 
which have been calculated using the operator product expansion techniques 
\cite{Chayetal}- \cite{Bigi} and they constitute  
the first non-trivial corrections to the parton model results.
 Taking the typical
hadronic scale to be $O(1)$ GeV, one expects
as a ball-park estimate $O(5) \%$ corrections in 
the inclusive rates. This is now discussed more quantitatively.

\par
 In HQET, the $b$-quark field is represented by a 
four-velocity-dependent field, denoted by $b_v(x)$. To first 
order in $1/m_b$, the $b$-quark field in  QCD $b(x)$
 and the HQET-field
$b_v(x)$ are related through:
\begin{equation}\label{hqetb}
b(x) = e ^{-im_b v.x} \left[ 1 + i\frac{\not\!\! D}{2m_b} \right] b_v(x)
\end{equation}
The QCD Lagrangian for the $b$ quark in HQET in this order is:
\begin{equation}\label{hqetlang}
 {\cal L}^{\mbox{\small HQET}} = \bar{b}_v iv.\not\!\! D b_v + \bar{b}_v
 \frac{i(\not\!\! D)^2}{2m_b} b_v
   -Z_b \bar{b}_v\frac{gG_{\alpha\beta}\sigma^{\alpha\beta}}{4 m_b} b_v
+ O\left[ \frac{1}{m_b^2} \right],
\end{equation}  
where $Z_b$ is a renormalization factor, with $Z_b(\mu=m_b)=1$
and $\not\!\! D = D_\mu \gamma^{\mu}$, with $D_\mu$ being the covariant
derivative. The 
operator
$\bar{b}_v(i\not\!\! D)^2b_v/2m_b$ is not renormalized due to the 
symmetries of
HQET. (In technical jargon, this is termed as a consequence of the
reparametrization invariance of ${\cal L}^{\mbox{\small HQET}}$.)
 With this Lagrangian, it has been shown in \cite{Chayetal} -
\cite{Bigietal} that in
the heavy quark expansion in order $(1/m_b^2)$, the
hadronic corrections can be expressed in terms of two matrix elements
\begin{equation}\label{lambda1}
\langle B^{(*)} \vert \bar{b}_v (iD)^2 b_v \vert B^{(*)}\rangle = 2 
m_{B^{(*)}} \lambda_1 , \nonumber 
\end{equation}
\begin{equation}\label{lambda2} 
\langle B^{(*)} \vert \bar{b}_v \frac{g}{2}\sigma_{\mu \nu} F^{\mu \nu} b_v 
\vert B^{(*)}\rangle = 2 d_{B^{(*)}} m_{B^{(*)}} \lambda_2 ,
\end{equation}
where $F^{\mu \nu}$ is the gluonic field strength tensor,
and the constants $d_{B^{(*)}}$ have the value 3 and $-1$ for $B$ and $B^*$,
respectively. The
constant $\lambda_2$ can be related to the hyperfine splitting in the $B$ 
mesons, which gives:
\begin{equation}
\lambda_2 \simeq \frac{1}{4} (m_{B^*}^2 - m_B^2) = 0.12 ~\mbox{GeV}^2.
\end{equation}
The other quantity $\lambda_1$ is just the average kinetic energy of the
$b$ quark inside a $B$ meson and has been estimated in the QCD sum rule 
approach \cite{kinetic}
yielding  $\lambda_1= -(0.5 \pm 0.1)$ GeV$^2$ \cite{BB94}. As we
shall discuss later, this parameter influences the lepton- and
photon-energy spectrum in the decays $B \to X \ell \nu_\ell$ and $B \to X_s + 
\gamma$, and has also been estimated using these 
spectra \cite{ag95,GKLW96}. Taking into account these corrections, the 
semileptonic
and non-leptonic decay rates of a $B$ meson $B \to X\ell \nu_\ell$ and
$B \to X_cX$ can be written as \cite{Bigietal,MW94}:
\begin{eqnarray}\label{pcsld}
\Gamma(B\longrightarrow X_c\ell\nu_{\ell}) &=&
\Gamma^{(0)} f(r_{c})\bigg[\left(1-\frac{2}{3}\frac{\alpha_{s} (m_{b}^{2})}
{\pi} g(r_{c})\right) \left( 1 + \frac{\lambda_1}{2 m_b^2}
+ \frac{3 \lambda_2}{2m_b^2} - 
\frac{6(1-r_c)^4}{f(r_c)}\frac{\lambda_2}{m_b^2}\right) \nonumber\\
& & \mbox{}
+ O(\as^2,\frac{\as}{m_b^2},\frac{1}{m_b^3})\bigg],
\end{eqnarray}
and
\begin{eqnarray}\label{pcnld}
\Gamma(B\longrightarrow X_cX) &=&
3 \Gamma^{(0)} \bigg[\eta(\mu)J(\mu)
 \left( 1 + \frac{\lambda_1}{2 m_b^2}
+ \frac{3 \lambda_2}{2m_b^2} - 
\frac{6(1-r_c)^4}{f(r_c)}\frac{\lambda_2}{m_b^2}\right) \nonumber\\
& & \mbox{}
-\left(L_+(\mu)^2 -L_-(\mu)^2\right) \frac{4(1-r_c)^3}{f(r_c)} 
\frac{\lambda_2}{m_b^2} +
 O(\as^2,\frac{\as}{m_b^2},\frac{1}{m_b^3})\bigg],
\end{eqnarray}
where the product $\eta(\mu)J(\mu)$ denotes the NLO corrected result for
the partonic decay discussed above in (\ref{bcud}), to which Eq. 
(\ref{pcnld}) reduces in the limit $\lambda_1=\lambda_2=0$.

\par
The decay rates depend on the quark masses, which unlike lepton masses,
do not appear as poles in the $S$-matrix nor do the quarks  
exist as asymptotic states. They are parameters of an interacting theory 
and hence subject to renormalization effects. Consequently, they require 
a regularization scheme, such as the $\overline{MS}$ scheme, and a scale,
where they are normalized, to become well-defined quantities.
 For example, the quark masses
in the so-called $\overline{MS}$ scheme and the pole masses (OS scheme) are 
related in the leading order \cite{mtmsbar},
\begin{equation}\label{polmsbar}
\overline{m}_Q(m_Q)=m_Q\left[1- 4 \frac{\as (m_Q)}{(3\pi)} +...\right] ~.
\end{equation}

 In HQET, quark masses can be expressed in terms of the heavy meson 
masses $m_M$ and the parameters $\lambda_1, ~\lambda_2$ and a quantity 
called $\bar{\Lambda}$, where
\begin{equation}\label{biglambda}
m_M=m_Q + \bar{\Lambda} - \frac{\lambda_1 + d_M \lambda_2}{2m_Q}+...
\end{equation}
With this the decay rates then depend on the QCD-related parameters, 
$\Lambda_{QCD}, \lambda_1, \lambda_2$ and $\bar{\Lambda}$.
 Since
 quark masses are scale- and 
(regularization) scheme-dependent quantities, this  also holds for the
parameter $\bar{\Lambda}$.
As a consequence of this, the decay rates
become scale- and scheme-dependent. This will reflect itself through
increased theoretical dispersion on various physical quantities.
More precise predictions for
the decay rates and related quantities (such as ${\cal B}_{SL}$ and
 $\langle n_c \rangle$) 
require
knowledge of higher order corrections, which are unfortunately not known.
In the context of the heavy quark expansion, one can relate the parameters
$\lambda_1$ and $\lambda_2$ to the quark and hadron masses:
\begin{equation}
m_b-m_c=m_B-m_D + \frac{\lambda_1 + 3 \lambda_2}{2} (\frac{1}{m_b} - 
\frac{1}{m_c}) + O(\frac{1}{m^2}) ,
\end{equation}
and the quark mass differences can then be calculated knowing $\lambda_1$ 
and $\lambda_2$, giving $(m_b - m_c)= (3.4 \pm 0.03 \pm 0.03)$ GeV 
\cite{Neubert95}.
 This difference, which determines the inclusive
rates and shape of the lepton energy spectrum in semileptonic decays,  
has also been determined from an analysis of the experimental lepton energy 
spectrum in $B$ decays, yielding  $(m_b-m_c)=3.39\pm 0.01$ GeV
for the pole masses \cite{GKLW96}, in excellent agreement with the QCD 
sum rule based estimates. The quark masses themselves 
have considerable uncertainties. We shall not discuss them here.
For a comprehensive discussion of the quark masses and their
current values, we refer to the reviews by Leutwyler \cite{Leutwyler96}
and by Manohar in the PDG review of particle properties \cite{PDG94}.

\subsection{Numerical estimates of ${\cal B}_{SL}(B)$ and $\langle n_c 
\rangle$}
\par
The theoretical framework described in the previous section
can now be used to predict two important quantities
in $B$ decays ${\cal B}_{SL}(B)$ and
$\langle n_c \rangle $, which have been measured. Concerning 
${\cal B}_{SL}(B)$, there is some discrepancy between the two set of 
experiments
performed at the $\Upsilon(4S)$ and at the $Z^0$ resonance, although it
must be stressed that these experiments measure a different mixture
of $B$ hadrons. The 
present measurements are reviewed recently in \cite{Tomasz95}:
\begin{eqnarray}\label{slbr95}
{\cal B}_{SL}(B) &=& (10.56 \pm 0.17 \pm 0.33)\% ~~~\mbox{at} ~\Upsilon(4S) 
\nonumber\\
{\cal B}_{SL}(B) &=& (10.89 \pm 0.18 \pm 0.24)\% ~~~\mbox{at} ~Z^0 
\nonumber\\
\langle n_c \rangle &<& 1.16 \pm 0.05 ~~~\mbox{at} ~\Upsilon(4S).
\end{eqnarray}
The first error quoted for ${\cal B}_{SL}$ is experimental and the second 
is an estimate
of the model dependence. The number for $\langle n_c \rangle$ is obtained
by summing over various bound and unbound charmed hadron states in
$B$ decays, and the
inequality is to due to the present upper limit on the inclusive
branching ratio ${\cal B}(B\to \eta_c X) < 0.018$. 

\begin{figure}[htb]
\vspace{0.10in}
\centerline{
\epsfig{file=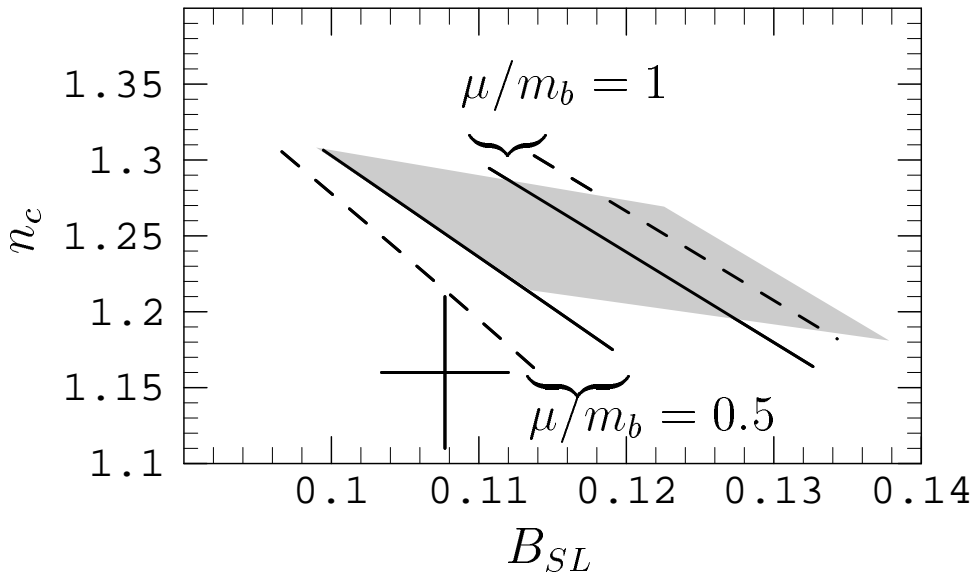,height=3in,angle=0}
}
\vspace{0.08in}
\caption[]{The charm content $n_c$ vs. $B_{\small SL}$ in $B$ decays.
Solid lines: theoretical predictions in the OS scheme for
$0.23 < \mc/\mb < 0.33$; dashed lines: the same in the $\overline{MS}$
scheme for $0.18 < \bar{m}_c(\bar{m}_c)/\bar{m}_b(\bar{m}_b) < 0.28$.
Shaded area: theoretical predictions in the OS scheme
by varying over the parameters as discussed in
the text. The data point is from \protect\cite{Tomasz95} (Figure taken
from \protect\cite{Bagan95a}).
\label{ncslBB}}
\end{figure}

   The theoretical predictions for these quantities 
at present are somewhat fuzzy due to the
uncertainties in the input parameters. We shall rely here on a recent
theoretical update by Bagan et al. \cite{Bagan95a}
 [see also the recent analysis by Neubert and Sachrajda \cite{NS96}],
 where the following ranges of parameters have 
been used:
\begin{equation}\label{BBparameters1}
m_b(\mbox{pole})=4.8 \pm 0.2 ~\mbox{GeV}; ~~~\as (m_Z)=0.117 \pm 0.007,
~~~m_b/2 < \mu < 2 m_b,\nonumber\\
\end{equation}
\begin{equation}\label{BBparameters2}
\lambda_1=-(0.5 \pm 0.1) ~\mbox{GeV}^2; ~~~\lambda_2 =0.12 ~\mbox{GeV}^2
\end{equation}
Their analysis leads to the following values \cite{Bagan95a}:
\begin{equation}\label{bslrpole}
{\cal B}_{SL} = (12.0 \pm 0.7 \pm 0.5 \pm 0.2 ^{+0.9}_{-1.2})\%,
\end{equation}
and
\begin{equation}\label{bslrmsbar}
\overline{{\cal B}}_{SL} = (11.3 \pm 0.6 \pm 0.7\pm 0.2^{+0.9}_{-1.7})\% ,
\end{equation}
using the pole (also called on shell OS)
 and $\overline{MS}$ masses, respectively. The errors are from 
$\Delta(m_b), ~\Delta \as (m_Z), ~\Delta (\lambda_1)$, $\Delta (\Gamma(B \to
X_{cc})$ and from the scale$(=\mu)$ variation, respectively.
 An estimate of the present theoretical uncertainty can be obtained
by adding these errors in quadrature (a reasonable procedure since most
errors given above are independent), yielding
\begin{eqnarray}\label{bslrpole2}
{\cal B}_{SL} &=& (12.0 \pm 1.4)\%, \nonumber\\
\overline{{\cal B}}_{SL} &=& (11.3 \pm 1.6)\%~. 
\end{eqnarray}
This shows that this quantity has still significant theoretical spread:
 \begin{equation}
\delta {\cal B}_{SL}(B) \simeq \pm 0.15 {\cal B}_{SL}(B) \simeq \pm 0.018 ~.
\end{equation}
The corresponding estimates for $\langle n_c \rangle$ and
 $\langle \bar{n}_c\rangle$ in the OS and $\overline{MS}$ schemes,
respectively, are \cite{Bagan95a}:
\begin{eqnarray}\label{ncav}
\langle n_{c} \rangle = (1.24 \pm 0.05 \pm 0.01)
\qquad \mbox{and} \qquad
\langle \overline{n_{c}}\rangle = (1.30 \pm 0.03 \pm 0.04), 
\end{eqnarray}
where the first error in both the schemes is due to $\Delta \mb$, and the
second error combines the uncertainties in the rest of the
parameters. The estimates (\ref{bslrpole2}) and (\ref{ncav}) should be
compared with the experimental measurements given in Eq.~(\ref{slbr95}).
A comparison of these theoretical estimates (eqs. (\ref{bslrpole}), 
(\ref{bslrmsbar}) and (\ref{ncav})) and data on
 $\langle n_c\rangle $ and ${\cal B}_{SL}$ is shown in Fig. ~\ref{ncslBB}.
%
%
It should be pointed out that in the analysis of Bagan et al.
~\cite{Bagan95a} for
${\cal B}_{SL}$ and $\langle n_c \rangle$, possible contributions from the
so-called spectator effects ($W$-annihilation and $W$-exchange) have
been neglected. Such effects depend on the wave-functions at the origin,
using the quark model language,
and are proportional to $f_B^2$, which could lead to some $O(5) \%$
effects, well within the present theoretical noise $\delta({\cal B}_{SL})$
discussed above. All in all, it is fair to conclude that
within existing uncertainties, the current theoretical estimates
for ${\cal B}_{SL}$ and $\overline{n_c}$ do not disagree
significantly with the experimental values, though there is a tendency
in  these estimates to yield somewhat larger value for $\overline{n_c}$.
More work is needed to make these comparisons precise. While still on
the same subject, we note that
the pattern of power corrections predicted by the
short-distance expansion of the heavy quark effective theory in
inclusive non-leptonic decays has been questioned in \cite{AMPR96}.
It is argued on phenomenological grounds
 that data on $b$- and $c$-decays are better accommodated if
some account is taken of the linear $(1/m_Q)$ corrections,
which are absent in the HQET approach. 
Naive identification of the $m_Q^5$ factor in the inclusive decay widths
with  $m_H^5$, with $m_H$ being the heavy hadron mass, brings data and 
theoretical estimates in 
better agreement! The presence of $1/m_Q$ corrections could be related to 
the breakdown of local 
duality in non-leptonic decays. The real question is if one could
derive this suggested pattern of power corrections in QCD.   

\par
While the suggestion of Altarelli et al. \cite{AMPR96} is interesting 
and it raises new theoretical 
issues as to the reliability of $\Gamma_{\mbox{\small NL}}$
calculated in the HQET approach, their arguments are purely
phenomenological.
 It is fair to say that
a big step in reducing the present theoretical uncertainties will be the 
completion of all the perturbative 
 NLL corrections to the dominant decays discussed above. Parts
of them, the so-called gluon bubble graphs having an arbitrary number of
fermion loops, 
which take into account the
effects of the running of $\as$ on ${\cal B}_{SL}$ in all orders in 
perturbation theory, are available in the literature 
\cite{Lukeetal95,Balletal95}. This is related to the issue of
defining a factorization scale, discussed earlier in a more general context
by Brodsky et al. \cite{BLM83}, which at present is a big uncertainty in
this approach, as can be seen in Fig. ~\ref{ncslBB}.
%
\subsection{ $B$-Hadron Lifetimes in the Standard Model}

\par
   A matter closely related to the semileptonic branching ratios is that 
of the
individual $B$ hadron lifetimes. The QCD-improved spectator model gives
almost equal lifetimes. Power corrections will split the
$B$-baryon lifetime from those of $B_d, B^{\pm}$ and $B_s$. However,
first estimates of these differences 
are at the few per cent level \cite{Bigietal}.
 The experimental situation has been 
summarized as of  summer 1995 in \cite{Kroll95}:
\begin{equation}
\frac{\tau(B^-)}{\tau(B_d)}=1.02 \pm 0.04;
~~\frac{\tau(B_s)}{\tau(B_d)}=1.01 \pm 0.07; 
~~\frac{\tau(\Lambda_b)}{\tau(B_d)}=0.76 \pm 0.05 ~.
\end{equation}
Since then, a new result for the lifetime of the $\Lambda_b$ baryon has been
reported by the CDF collaboration,  $\tau(\Lambda_b)/\tau(B_d)
=0.85 \pm 0.10 \pm 0.05$  \cite{CDFlambda} , which is still less than the 
other two ratios involving the $B$-meson lifetimes but reduces the gap in
the $\Lambda_b$ and $B$-meson lifetimes.

    This subject has received renewed theoretical attention lately
 \cite{NS96,Uraltsev96,Rosner96}, in which the possibly enhanced roles of 
the four-Fermion operators between baryonic states has been
studied. We recall that such operators enter at 
$O(1/m_b^3)$ in the heavy quark expansion discussed above \cite{Bigietal}. 
In this order, there are
four such operators, which using the notation of \cite{NS96}, can be
expressed as:
\begin{eqnarray}\label{4quarkNS}
{\cal O}^q_{V-A} &=& (\bar{b}_L\gamma_\mu q_L)(\bar{q}_L\gamma^{\mu}b_L),
\nonumber\\
{\cal O}^q_{S-P} &=& (\bar{b}_R q_L)(\bar{q}_L b_R),
\nonumber\\
{\cal T}^q_{V-A} &=& (\bar{b}_L\gamma_\mu t_a 
q_L)(\bar{q}_L\gamma^{\mu}t_a b_L), \nonumber\\
{\cal T}^q_{S-P} &=& (\bar{b}_Rt_a q_L)(\bar{q}_Lt_a b_R),
\end{eqnarray}
where $t_a$ are generators of colour $SU(3)$. The matrix elements of these
operators between various $B$-meson and $\Lambda_b$-baryons
are in general different and this contribution will thus split the decay 
widths
of the various $B$ hadrons. In general, the operators
(\ref{4quarkNS}) introduce eight new 
parameters corresponding to the matrix elements of these operators.
In the large-$N_c$ limit, however, it has been
argued in \cite{NS96} that the $B$-mesonic matrix elements of the operators
$\langle B_q \vert {\cal O}^q_{V-A} \vert B_q \rangle$ and
$\langle B_q \vert {\cal O}^q_{S-P} \vert B_q \rangle$ are the dominant 
ones. While accurate numerical estimates require a precise knowledge of
these matrix elements, one expects that they give rise 
typically 
to  the spectator-type effects (using the parton model language):
\begin{equation}
\frac{\Gamma_{\mbox{spec}}}{\Gamma_{\mbox{tot}}} \simeq (\frac{2 \pi f_B}
{m_B})^2 \simeq 5\% ~,
\end{equation}
with $f_B$ of order 200 MeV.
In the case of $\Lambda_b$ baryons, one can use the heavy quark spin symmetry
to derive two relations among the operators between the $\Lambda_b$ states.
The problem is then reduced to the estimate of two matrix elements
which in \cite{NS96} are taken to be the following:
\begin{equation}
 \frac{1}{2m_{\Lambda_b}}\langle \Lambda_b \vert {\cal O}^q_{V-A} \vert 
\Lambda_b \rangle \equiv -\frac{f_B^2 m_B}{48} r(\frac{\Lambda_b}{B_q}) ,
\end{equation}
and
\begin{equation}
\langle  \Lambda_b \vert \tilde{ {\cal O}}^q_{V-A} 
\vert \Lambda_b \rangle = -\tilde{B}
\langle \Lambda_b \vert  {\cal O}^q_{V-A}
\vert \Lambda_b \rangle ~,  
\end{equation}
The operator $\tilde{\cal O}_{V-A}$ is a linear combination of the
operators ${\cal T}_{V-A}$ and ${\cal O}_{V-A}$ introduced earlier,
$\tilde{{\cal O}}_{V-A}= 2{\cal T}_{V-A} + 3 {\cal O}_{V-A}$, following from
colour matrix algebra \cite{NS96}, and
$r(\Lambda_b/B_q)$ is the ratio of the squares of the wave functions  
which can be expressed in terms of the probability of finding a light quark
at the location of a $b$ quark inside $\Lambda_b$ baryon and the $B$ meson,
i.e.
\begin{equation}
r(\frac{\Lambda_b}{B_q})  = \frac{\vert \Psi^{\Lambda_b}_{bq} \vert^2}
        {\vert \Psi^{B_q}_{b\bar{q}} \vert^2} ~.
\end{equation}
 One expects $\tilde{B} =1$ in the valence-quark approximation.
However, the ratio $r(\Lambda_b/B_q)$ has a large uncertainty on it, ranging 
from $r(\Lambda_b/B_q) \simeq 0.5$ in the non-relativistic quark model 
\cite{Guberinaetal79}
to $r(\Lambda_b/B_q) =1.8 \pm 0.5$ if one uses the ratio of the spin 
splittings 
between $\Sigma_b$ and $\Sigma_b^*$ baryons and $B$ and $B^*$ mesons, as
advocated by Rosner \cite{Rosner96} and using the preliminary data 
from DELPHI, $m(\Sigma_b^*) - m(\Sigma_b) = (56 \pm 16)$ MeV 
\cite{DELPHISIGMA}.

Using the ball-park estimates that $\tilde{B} $ and
$r(\Lambda_b/B_q) $ are both of order unity yields
for the lifetime ratio  $\tau(\Lambda_b)/\tau(B_d) > 0.9$ \cite{NS96},
significantly larger than the present world average.
 Reliable estimates of these constants can be got, in principle, using 
lattice-QCD and QCD sum rules. Very recently, QCD sum rules have been used
to estimate
$\langle  \Lambda_b \vert \tilde{ {\cal O}}^q_{V-A}
\vert \Lambda_b \rangle$ and $\tilde{B}$, yielding 
$\langle  \Lambda_b \vert \tilde{ {\cal O}}^q_{V-A}
\vert \Lambda_b \rangle= (0.4 - 1.2) \times 10^{-3} ~\mbox{GeV}^3$
and $\tilde{B}=1.0$ \cite{CD96}. This corresponds to the parameter
$r(\Lambda_b/B_q)$ having a value in the range
$r(\Lambda_b/B_q) \simeq 0.1 - 0.3$, much too small to explain the
observed lifetime difference. One must conclude that the lifetime ratio
$\tau(\Lambda_b)/\tau(B_d)$ remains a puzzle. New and improved measurements
are needed, which we trust will be forthcoming from HERA-B and Tevatron 
experiments in not-too-distant a future.
  
\subsection{Determination of $\Vcbabs$ and $\Vubabs$}
\par
 The CKM matrix element $V_{cb}$ can
be obtained from semileptonic decays of $B$ mesons. We shall restrict
ourselves to the methods based on HQET to calculate the exclusive 
semileptonic decay rates and use the heavy quark expansion to estimate the
inclusive rates. Concerning exclusive decays, we recall that
 in the heavy quark limit $(\mb \to \infty)$, it has been
observed that all hadronic form factors in the semileptonic decays $B \to
(D,D^*) \ell \nu_\ell$ can be expressed in terms of a single function, the
Isgur-Wise function \cite{IW}. It has been shown that the HQET-based
method works best for $B\to D^*l\nu$ decays, since these are unaffected by
$1/m_Q$ corrections \cite{Luke,Boyd,Neubert}. Since the rate is zero at
the kinematic point $\omega=1$, one uses data for $\omega >1$ and an 
extrapolation procedure (discussed below)  to determine $\xi (1)
\Vcbabs$ and the slope of the Isgur-Wise function $\hat{\rho}^2$.
      
Using HQET, the differential decay rate in $B \to D^* \ell \nu_\ell$ is
\begin{eqnarray}
\frac{d\Gamma (B \to D^* \ell \bar{\nu})}{d\omega }
&=& \frac{G_F^2}{48 \pi^3} (m_B-m_{D^*})^2 m_{D^*}^3 \eta_{A}^2
 \sqrt{\omega^2-1} (\omega + 1)^2 \\ \nonumber
&~& ~~~~~~~~~~~~~~\times [ 1+ \frac{4 \omega}{\omega + 1}
 \frac{1-2\omega r + r^2}{(1-r)^2}] \Vcbabs ^2 \xi^2(\omega) ~,
\label{bdstara1}
\end{eqnarray}
where $r=m_{D^*}/m_B$, $\omega=v\cdot v'$ ($v$ and $v'$ are the
four-velocities of the $B$ and $D^*$ meson, respectively), and $\eta_{A}$
is the short-distance correction to the axial vector form factor. In the
leading logarithmic approximation, this was calculated by Shifman and
Voloshin some time ago -- the so-called hybrid anomalous dimension
\cite{hybrid}. In the absence of any power corrections, $\xi (\omega=1)=1$.
The size of the $O(1/\mb^2)$ and $O(1/\mc^2)$ corrections to the Isgur-Wise
function, $\xi (\omega)$, and partial next-to-leading order
corrections to $\eta_A$ have received a great deal of theoretical
attention, and the state of the art has been
summarized  by Neubert \cite{neubert95} and Shifman
 \cite{shifmanpascos}. Following \cite{neubert95}, we take:
\begin{eqnarray}
\label{neubertxiold}
\xi (1) &=& 1+ \delta (1/m^2)= 0.945 \pm 0.025 ~, \nonumber \\
 \eta_{A} &=& 0.965 \pm 0.020 ~.
\end{eqnarray}
This gives the range \cite{neubert95}:
\be
{\cal F}(1)=\xi \cdot \eta_{A}=0.91 \pm 0.04~.
\label{alxi}
\ee
Recently, the quantity $\eta_{A}$, and its counterpart for the vector
current matrix element renormalization, $\eta_{V}$, have been calculated
in the complete next-to-leading order by Czarnecki \cite{Cz96}, getting
\begin{eqnarray}
\label{Czetas}
\eta_{A} &=& 0.960 \pm 0.007~, \nonumber \\
\eta_{V} &=& 1.022 \pm 0.004 ~.
\end{eqnarray}
The NLO central value for $\eta_{A}$ is in agreement with the estimate 
of the same given in eq.~(\ref{neubertxiold}) but the error on it is now 
reduced by a factor of 3. So, the error on ${\cal F}(1)$ is now completely 
dominated by the power corrections in $\xi (1)$.

Since the range of accessible energies in the decay
$B \to D^* \ell \bar{\nu}$ is rather small\\ $(1 ~< \omega < ~1.5)$, the
extrapolation to the symmetry point can be done using an expansion
around $\omega =1$,
\be
{\cal F}(\omega) = {\cal F}(1) \left[1- \hat{\rho}^2(\omega -1) + \hat{c}
          (\omega -1)^2+...\right].
\ee
 It is usual to use a linear form for
extrapolation with the slope $\hat{\rho}^2$ left as a free parameter, as
the effect of a curvature term is small \cite{Stone}.
The present experimental input from the exclusive
semileptonic channels is based on the data by
 CLEO, ALEPH, ARGUS, and DELPHI,
which is summarized in \cite{alpisa95} to which we refer for details
and references to the experimental analysis.
 For the updated ARGUS numbers, see \cite{ARGUS96}.
 The statistically weighted average used in the analysis \cite{alpisa95} is:
\be
  \vert V_{cb}\vert \cdot {\cal F}(1) = 0.0353 \pm 0.0018 ~.
\ee
This agrees with the numbers presented by Skwarnicki   last summer
\cite{Tomasz95},
\bea
  \vert V_{cb}\vert \cdot {\cal F}(1) &=& 0.0351 \pm 0.0017 ~,\nonumber\\
  \hat{\rho}^2 &=& 0.87 \pm 0.10 ~.
\eea  
Using ${\cal F}(1)$ from Eq.~(\ref{alxi}), gives the following
value:
 \be
  \vert V_{cb} \vert= 0.0388 \pm 0.0019 ~(\rm{expt}) \pm 0.0017 ~(\rm{th}).
\label{Vcbhqet95}
\ee
Combining the errors linearly gives $\vert V_{cb} \vert = 0.0388 \pm
0.0036$. This is in good agreement with the value $\Vcbabs = 0.037
^{+0.003}_{-0.002}$ obtained from the exclusive decay $B \to D^* \ell
\nu_\ell$, using a dispersion relation approach \cite{BGL95}.

\par
 The value of $\vert V_{cb}
\vert$ obtained from the inclusive semileptonic $B$
 decays using heavy quark expansion is quite compatible
with the above determination of the same. The inclusive analysis has the 
advantage of having very small statistical error.
However, as discussed previously,  there is
about $2\sigma$ discrepancy between the semileptonic branching ratios
at the $\Upsilon(4S)$ and in $Z^0$ decays.   
 Using an averaged value for the semileptonic
decay width from these two sets of measurements gives: $\Gamma(b \to c \ell
\nu_\ell)={\cal B}_{SL}/\langle \tau_B \rangle =(67.3\pm 2.7) ~{\mbox 
ns}^{(-1)}$ \cite{Tomasz95}, where 
 $\langle \tau_B \rangle =(\tau_{B^-} + \tau_{B^0})/2=1.60 \pm 0.03 
~\mbox{ps}$ and
 $\langle \tau_B \rangle =1.55 \pm 0.02 ~\mbox{ps}$ have been used for the
$\Upsilon(4S)$ and $Z^0$ decays, respectively, and the 
error has been inflated to take into account the disagreement
mentioned. This leads to a value \cite{Tomasz95}:
\be
 \vert V_{cb} \vert = 0.0398 \pm 0.0008 ~\mbox{ (expt)} \pm 0.004 
~\mbox{ (th)} ~. \ee
where the theoretical error estimate ($\pm 10 \%)$ has been taken from  
Neubert \cite{neubert95}. While still on the same quantity, it should be 
noted that Vainshtein has quoted a much smaller theoretical error, 
\cite{Vainshtein95}:
\be
\vert V_{cb} \vert = 0.0408 \left[\frac{{\cal B}(B \to X_c\ell \nu_\ell)}
{0.105}\right]^{1/2} \left[\frac{1.6 \mbox{ps}}{\tau_B}\right]^{1/2} (1.0 
\pm 0.03\mbox{(th)}) ~.
 \ee
For further discussion of these matters we refer to
\cite{shifmanpascos,Vainshtein95,neubert95}.
 We shall use  the following values
for $\Vcbabs$ and the Wolfenstein parameter $A$:
\be
 \vert V_{cb} \vert = 0.0388 \pm 0.0036 \Longrightarrow ~A = 0.80 \pm 0.075~.
\label{Avalueal}
\ee
\par
Up to recently, $\vert V_{ub}/V_{cb}\vert$ was obtained by looking at the
endpoint of the inclusive lepton spectrum in semileptonic $B$ decays.
Unfortunately, there still exists quite a bit of model dependence in the
interpretation of the inclusive data by themselves
yielding $\Vubabs/\Vcbabs = 0.08 \pm 0.03$ \cite{Bartelt93,ARGUS96}.
A recent new input to this quantity is provided by the measurements of the
exclusive semileptonic decays $B \to (\pi, \rho) \ell \nu_\ell$
\cite{Thorndike95,Tomasz95}. The extracted branching ratios depend on
the model used to correct for the experimental acceptance
but they do provide some discrimination among the various models.
 In particular, models such as that
of Isgur et al.~\cite{ISGW}, which give values well in excess of 3 for the
ratio of the decay widths $\Gamma (B^0 \to \rho^-\ell^+ \nu)/\Gamma (B^0
\to \pi^- \ell^+ \nu)$, are disfavoured by the CLEO data
which yield typically $1.7 \pm 1.0$ for the same quantity,
with some marginal model dependence. The models with larger values for
the ratio $\Gamma (B^0 \to \rho^-\ell^+ \nu)/\Gamma (B^0
\to \pi^- \ell^+ \nu)$ also lead to softer lepton energy spectrum in
inclusive $B$ decays, requiring larger values of $\Vubabs/\Vcbabs$. 
 A recent
estimate of this ratio in the light-cone QCD sum rule approach gives
$2.4 \pm 0.8$  \cite{BB96}, and yields a branching ratio
${\cal B}(B \to \rho \ell \nu_\ell) = (19 \pm 7) \Vubabs^2$, giving
a value of $(1.95 \pm 0.72) \times 10^{-4}$, for $\Vubabs =0.08$, in 
agreement with the experimental numbers, which, however, have 
significant errors. 
 Excluding  models which fail to reproduce the exclusive data 
from further consideration, measurements in both the inclusive and
exclusive modes are compatible with
\be
\left\vert \frac{V_{ub}}{V_{cb}} \right\vert = 0.08\pm 0.02~,
\label{vubvcbn}
\ee
 which is also the value adopted by the PDG Review \cite{PDG94}. This gives
a constraint on the Wolfenstein parameters $\rho$ and $\eta$:
\be
\sqrt{\rho^2 + \eta^2} = 0.36 \pm 0.08~.
\ee
With the measurements of the form factors in semileptonic decays $B \to   
(\pi,\rho,\omega) \ell \nu_\ell$, one should be able to further constrain
the models, thereby reducing the present theoretical uncertainty on this
quantity.

\par
   We summarize this section by observing that the bulk properties of
$B$ decays are largely accounted for in the standard model. On the
theoretical front, parton model estimates of the earlier epoch have been
replaced by theoretically better founded calculations with controlled
errors, though this point of view has not found universal acceptance
\cite{AMPR96}!
 In particular, methods based on  HQET and heavy
quark expansion have led to a quantitative determination of $\Vcbabs$ at
$\pm 10 \%$ accuracy, which makes it after $\Vudabs$ and $\Vusabs$, the
third best measured CKM matrix element. The 
matrix element $\Vubabs$ has still 
large uncertainties ($\pm 25 \%)$ and there is every need to reduce this,
as it is one of the principal handicaps at present in testing the unitarity 
of the CKM matrix precisely (more on this later).
The quantities ${\cal B}_{SL}$, $\langle n_c \rangle$, and the individual
$B$-hadron lifetimes are not in perfect agreement with data but the
present theoretical estimates have still large uncertainties to abandon
the SM. A
completely quantitative comparison requires the missing NLL corrections and
in the case of lifetime differences better evaluations of the matrix
elements of four-quark operators, which we hope will be forthcoming.
Finally, we stress that it will be very helpful to measure the semileptonic
branching ratios $B_{SL}$ for the $\Lambda_b$ baryons. With the lifetimes
of the $B$ hadrons now well measured, such a measurement would allow to
compare $\Gamma_{SL}(B_d), \Gamma_{SL} (B^\pm)$ and 
$\Gamma_{SL}(\Lambda_b)$, to check the pattern of power corrections 
in semileptonic decays.
%
%
 \section{Electromagnetic Penguins and Rare $B$ Decays in the Standard Model}

\par
  The SM does not admit FCNC transitions in the Born approximation,
which is obvious from the Lagrangian given at the very outset of
these lectures. However,
they are induced through the exchange of $W^\pm$ bosons
in loop diagrams. We shall discuss  representative examples
from several such transitions involving $B$ decays,
 starting with the decay $\BGAMAXS$, which has been measured by CLEO
 \cite{CLEOrare2}. This was preceded by the measurement of the exclusive 
decay $\BGAMAKSTAR$ \cite{CLEOrare1}:
\begin{eqnarray}
{\cal B}(\BGAMAXS) &=& (2.32\pm 0.57\pm 0.35)\times 10^{-4} ~,\\
{\cal B}(\BGAMAKSTAR) &=& (4.5\pm 1.0\pm 0.6)\times 10^{-5}~,
\end{eqnarray}
yielding an exclusive-to-inclusive ratio:
\begin{equation}
R_{K^*} = \frac{\Gamma(\BGAMAKSTAR)}{\Gamma(\BGAMAXS)}=(19\pm 6\pm 4)\% ~.
\end{equation}
These decay rates test the SM and the models for decay form factors
and we shall study them quantitatively.

 The leading contribution to $b \to s +\gamma$ arises
at one-loop from the so-called penguin diagrams and
the matrix element in the lowest order can be written as:
\begin{equation} \label{e1}
  {\cal M} (b \to s ~+\gamma)
    = \frac{G_F}{\sqrt{2}} \,\frac{e}{2 \pi^2}
 \sum_{i} V_{ib} V_{is}^*
      F_2 (x_i)\, q^\mu \epsilon^\nu \bar{s} \sigma_{\mu \nu}
      (m_bR ~+ ~m_sL)b ~,
 \end{equation}
where $x_i= ~m_i^2/m_W^2$,
$q_\mu$  and $\epsilon_\mu$ are, respectively, the photon four-momentum
and polarization vector,
the sum is over the quarks, $u, ~c$, and $t$, and $V_{ij}$ are the
CKM matrix elements.
The (modified) Inami-Lim function $F_2(x_i)$ derived from the (1-loop) 
penguin diagrams is given by \cite{InamiLim}:
\begin{equation}
F_{2}(x) = \frac{x}{24 (x-1)^{4}} \ \left[6 x (3 x -2 )
\log x - (x-1) (8 x^{2} +5 x -7 ) \right], \nonumber \\
\end{equation}
where in writing the expression for $F_2(x_i)$ above we have
left out a constant from the function derived by Inami and Lim, since on 
using the unitarity constraint these sum to zero.
It is instructive to write the unitarity constraint for the decays
$\BGAMAXS$ in full:
\begin{equation} \label{e4}
 V_{tb} V_{ts}^* + V_{cb}V_{cs}^* + V_{ub}V_{us}^* =0 ~.
 \end{equation}
Now, since the last term in this sum is completely negligible compared to 
the others
(by direct experimental measurements), one could set it to zero enabling
us to express the one-loop electromagnetic penguin amplitude as follows:
\begin{equation} \label{e2}
 {\cal M }(b \to s ~+\gamma)
    = \frac{G_F}{\sqrt{2}}\,\frac{e}{2 \pi^2} \,\lambda_{t}
   \,(F_2 (x_t)-F_2(x_c))\, q^\mu \epsilon^\nu \bar{s} \sigma_{\mu \nu}
      (m_bR ~+ ~m_sL)b ~.
 \end{equation}
The GIM mechanism \cite{GIM} is manifest in this amplitude and the
CKM-matrix element dependence is factorized in $\lambda_t\equiv V_{tb} 
V_{ts}^*$. The measurement of the branching ratio for $\BGAMAXS$ can then be 
readily interpreted in terms of the CKM-matrix element product
$\lambda_t/\Vcbabs$ or equivalently $\Vtsabs/\Vcbabs$. In the approximation
we are using (i.e., setting $\lambda_u=0$), this is equivalent to
measuring $\Vcsabs$.
For a quantitative determination
of $\Vtsabs/\Vcbabs$, however,  QCD radiative
corrections have to be computed and
the contribution  of the so-called long-distance effects estimated.
 We proceed to
discuss them below.
 \subsection{The effective Hamiltonian for $B \to X_s \gamma$}
\par
  The appropriate framework to incorporate
QCD corrections is that of an effective theory obtained by integrating 
out the
heavy degrees of freedom, which in the present context are the top quark 
and $W^\pm$ bosons. This effective theory
is an expansion in $1/m_W^2$ and  involves a tower of
increasing higher dimensional operators
built from the  quark fields $(u,d,s,c,b)$, photon, gluons and leptons. The 
presence
of the top quark and of the $W^\pm$ bosons is reflected through the
effective coefficients of these operators which become functions of
their masses. The operator basis depends on the underlying theory and 
in these lectures we shall concentrate on the standard model. The
basis that we shall use is restricted to dimension-6 operators and it closes
under QCD renormalization. The operators which vanish on using the
equations of motion are not included.
The effective Hamiltonian ${\cal H}_{eff}$ given below
 covers not only the decay $b \to s + \gamma$, in which we are
principally interested in this section, but also other processes
such as $b \to s + g$ and $b \to s + q \bar{q}$.
				 
 It is to be expected in general that due to
 QCD corrections, which induce
 operator-mixing, 
 additional contributions with  different CKM pre-factors have to be
included in the amplitudes.
Thus, QCD effects alter the CKM-matrix element dependence of the
decay rates for both $\BGAMAXS$ and (more importantly) $\BGAMAXD$. 
 However, with the help of the unitarity condition given above,
the CKM matrix
dependence in the effective Hamiltonian incorporating the QCD
corrections for the decays $\BGAMAXS$ factorizes, and
one can write this Hamiltonian as \footnote{Note
that in addition to the penguins with the $u$-quark intermediate state
there are also non-factorizing contributions due to the
operators $ (\bar{u}_{L \alpha} \go{\mu} b_{L \alpha})
(\bar{s}_{L \beta} \gu{\mu} u_{L \beta})$, which like the $u$-quark
contribution to the 1-loop electromagnetic penguins are proportional
to the CKM-factor $\lambda_u \equiv V_{us} V_{ub}^*$,
                              and hence are consistently set to zero.}:
\begin{equation}\label{heffbsg}
{\cal H}_{eff}(b \to s +\gamma) = - \frac{4 G_F}{\sqrt{2}} V_{ts}^* V_{tb}
        \sum_{i=1}^{8} C_i (\mu) {\cal O}_i (\mu) ,
\end{equation}
where the operator basis is chosen to be (here $\mu$ and $\nu$ are 
Lorentz indices and $\alpha$ and $\beta$ are colour indices)
\begin{eqnarray}
{\cal O}_1 &=& (\bar{s}_{L \alpha} \gamma_\mu b_{L \alpha})
               (\bar{c}_{L \beta} \gamma^\mu c_{L \beta}),    \\
{\cal O}_2 &=& (\bar{s}_{L \alpha} \gamma_\mu b_{L \beta})
               (\bar{c}_{L \beta} \gamma^\mu c_{L \alpha}),    \\
{\cal O}_3 &=& (\bar{s}_{L \alpha} \gamma_\mu b_{L \alpha})
               \sum_{q=u,d,s,c,b}
               (\bar{q}_{L \beta} \gamma^\mu q_{L \beta}),    \\
{\cal O}_4 &=& (\bar{s}_{L \alpha} \gamma_\mu b_{L \beta})
                \sum_{q=u,d,s,c,b}
               (\bar{q}_{L \beta} \gamma^\mu q_{L \alpha}),    \\
{\cal O}_5 &=& (\bar{s}_{L \alpha} \gamma_\mu b_{L \alpha})
               \sum_{q=u,d,s,c,b}
               (\bar{q}_{R \beta} \gamma^\mu q_{R \beta}),    \\
{\cal O}_6 &=& (\bar{s}_{L \alpha} \gamma_\mu b_{L \beta})
                \sum_{q=u,d,s,c,b}
               (\bar{q}_{R \beta} \gamma^\mu q_{R \alpha}),    \\
{\cal O}_7 &=& \frac{e}{16 \pi^2} m_b
               (\bar{s}_{L \alpha} \sigma_{\mu \nu} b_{R \alpha})
                F^{\mu \nu},                                   \\
{\cal O}_7^\prime &=& \frac{e}{16 \pi^2} m_s
               (\bar{s}_{R \alpha} \sigma_{\mu \nu} b_{L \alpha})
                F^{\mu \nu},                                    \\
{\cal O}_8 &=& \frac{g}{16 \pi^2} m_b
(\bar{s}_{L \alpha} T_{\alpha \beta}^a \sigma_{\mu \nu} b_{R \beta})
                G^{a \mu \nu},                                   \\
{\cal O}_8^\prime &=& \frac{g}{16 \pi^2} m_s
(\bar{s}_{R \alpha} T_{\alpha \beta}^a \sigma_{\mu \nu} b_{L \beta})
                G^{a \mu \nu},                                   
\end{eqnarray}
where $e$ and $g_s$ are the electromagnetic and the strong
coupling constants, and
 $F_{\mu \nu}$ and $G^A_{\mu \nu}$
denote the electromagnetic and the gluonic field strength
tensors, respectively.
We call attention to  the explicit mass factors in
 ${\cal O}_7 ({\cal O}_7^\prime)$
and ${\cal O}_8({\cal O}_8^\prime)$, which will undergo renormalization 
just as the Wilson coefficients.
                The dominant contributions
in the radiative decays $\BGAMAXS$
arise from the operators  ${\cal O}_2$, ${\cal O}_7$ and ${\cal O}_8$,
whereas the operators ${\cal O}_3,..., {\cal O}_6$
get coefficients through operator mixing only, which numerically
are negligible. Historically, the anomalous
dimension matrix was calculated in a truncated basis \cite{BSGAM}
 and this basis is
still often used for the sake of ease 
 in calculating the real and virtual corrections, though
as we discuss below, now the complete anomalous dimension matrix is
available \cite{Ciuchini}.

  The perturbative QCD corrections to the decay rate $\GGAMAXS$ have two 
distinct contributions:
\begin{itemize}
\item Corrections to the Wilson coefficients
$C_i(\mu)$, calculated with the help of the 
 renormalization group equation, whose solution requires the
knowledge of the anomalous dimension matrix in a given order in $\as$.

\item Corrections to the matrix elements of the operators
${\cal O}_i$ entering through the effective Hamiltonian
at the scale $\mu=O(m_b)$.
\end{itemize}
The anomalous dimension matrix is needed in order 
to use the renormalization group and sum up large logarithms, 
i.e., terms 
like $\as^{n}(m_W)\log^{m}(m_b/M)$, where $M=m_t$ or $ m_W$ and $m\leq n$
(with $n=0,1,2,...)$. At present only the leading logarithmic corrections 
$(m=n)$ have been calculated systematically in the complete basis given 
above \cite{Ciuchini}.
 
Next-to-leading 
order corrections to the matrix elements are now available completely. 
They are of two kinds: 
 \begin{itemize}
 \item QCD Bremsstrahlung corrections $b \to s \gamma + g$, which are
needed both to cancel the infrared divergences
 in the decay rate for
$\BGAMAXS$ and in obtaining a non-trivial QCD contribution to the
photon energy spectrum in the inclusive decay $\BGAMAXS$.
\item Next-to-leading order virtual corrections to the matrix elements
in the decay $b \to s +\gamma$. 
\end{itemize}
The Bremsstrahlung corrections were calculated in \cite{ag1} - 
\cite{ag3} in the truncated basis and last year also in the complete 
operator basis \cite{ag95}, which have been checked recently in 
\cite{Pott95}\footnote {This paper as it stands has some errors
which we expect to be rectified!}.
The higher order matching conditions, i.e., $C_i(m_W)$, are known up to
the desired accuracy, i.e., up to $O(\as(M_W)$ terms \cite{Yao94}. 
 Very recently, the next-to-leading order virtual
corrections have also been calculated \cite{GHW96}. What still remains to 
be done  is the calculation
of $\gamma^{(1)}$, the next-to-leading order anomalous dimension matrix,
which is the hardest part of this computation.
This is clearly needed to get theoretical estimates of the branching ratio
${\cal B}(\BGAMAXS )$ in the complete next-to-leading order. 
We discuss the presently available pieces to estimate $\BBGAMAXS$.

\par
We recall that the Wilson coefficients obey the 
renormalization group equation
\begin{equation} \label{RGE}
\left[\mu \frac{\partial}{\partial \mu}
+ \beta(g) \frac{\partial}{\partial g} \right]
C_i \left(\frac{M^2_W}{\mu^2},g \right)
= \hat{\gamma}_{ji} (g) C_j \left(\frac{M^2_W}{\mu^2},g \right) .
\end{equation}
The  QCD beta function $\beta(g)$ has been defined earlier
and $\hat{\gamma}(g)$ is the anomalous dimension matrix, which,
to leading logarithmic accuracy, is given by
\begin{equation}  \label{gam}
\hat{\gamma}(g) =  \gamma_0 \frac{g^2}{16 \pi^2} .
\end{equation}
Here $\gamma_0$ is a $8 \times 8$ matrix given by \cite{Ciuchini,Buras94}
{\footnotesize 
\begin{eqnarray*}
&& \gamma_0 = \\
&& \!\!\!\!\!\!\!\!\!\!\!\!\!\!\!\!\! \left[ \begin{array}{cccccccc}
\vspace{0.2cm}
    -2        &         6        &        0        &         0     &
     0        &         0        &        0        &
     3        \\
\vspace{0.2cm}
     6        &        -2        &-\frac{2}{9}     &  \frac{2}{3}  &
 -\frac{2}{9} &  \frac{2}{3}     & \frac{464}{81}  &
\frac{76}{27} \\
\vspace{0.2cm}
     0        &         0        &-\frac{22}{9}    & \frac{22}{3}  &
 -\frac{4}{9} &  \frac{4}{3}     &-\frac{368}{81}  &
\frac{152}{27}+ 3 f \\
\vspace{0.2cm}
     0        &         0        &  6-\frac{2}{9}f &-2+\frac{2}{3}f&
-\frac{2}{9}f &  \frac{2}{3}f    &\frac{464}{81}u-\frac{184}{81}d  &
6+\frac{76}{27}f  \\
\vspace{0.2cm}
     0        &         0        &          0     &      0     &
     2        &        -6        &  -\frac{80}{9}  &
\frac{8}{3}- 3 f    \\
\vspace{0.2cm}
     0        &         0        &-\frac{2}{9}f   & \frac{2}{3}f   &
-\frac{2}{9}f &-16+\frac{2}{3}f  &-\frac{400}{81}u+\frac{248}{81}d-
\frac{80}{3}  & -4-\frac{113}{27}f     \\
\vspace{0.2cm}
     0        &         0        &          0     &         0     &
     0        &         0        &   \frac{32}{3} &
     0        \\
\vspace{0.2cm}
     0        &         0        &          0     &          0    &
     0        &         0        & -\frac{32}{9}  &
 \frac{28}{3} 
\end{array} \right] .
\end{eqnarray*}}
where $f$ is the number of active flavours, and $u$ ($d$) is the
number of up (down) flavours in the effective theory. For the case
at hand we have $u = 2$, $d = 3$ and $f = u+d = 5$. This matrix is
given in the NDR dimensional regularization scheme which is also
used in the calculations of the matrix elements of the operators
discussed below. The difference between the NDR and the HV schemes lies
in the seventh and eighth columns and the corresponding matrix in the
HV scheme can be seen in \cite{AGM94}.

 The solution of the
renormalization group flow is obtained as
\begin{equation}
\vec{C} (\mu) = T^{-1} D (\mu, m_W) T \vec{C} (m_W)
\end{equation}
where $T$ diagonalizes the anomalous dimension matrix and $D$ is a
diagonal matrix, which is given for the relevant case by
$$
D = \mbox{Diag}[\eta^{0.1456}, \eta^{-0.8994}, \eta^{16/23}, \eta^{14/23},
                \eta^{-12/23}, \eta^{-0.4230},
                \eta^{0.4086}, \eta^{6/23}]
$$
where
$$
\eta = \frac{\alpha_s (M_W)}{\alpha_s (\mu)} .
$$

The transformation of the anomalous dimension matrix into a
diagonal form can then be  performed. Assuming
that the coefficients $C_3 \cdots C_6$ vanish at the matching
scale $\mu = M_W$, the result for the coefficients at the scale $\mu$
reads
\begin{eqnarray}
C_1 (\mu ) &=& \frac{1}{2} C_2 (M_W)
               \left( \eta^{6/23} -  \eta^{-12/23} \right),
\\
C_2 (\mu ) &=& \frac{1}{2} C_2 (M_W)
               \left( \eta^{6/23} +  \eta^{-12/23} \right),
\\
C_3 (\mu ) &=& C_2 (M_W) \left(-0.0112 \eta^{-0.8994}
                         + \frac{1}{6} \eta^{-12/23}
                         - 0.1403 \eta^{-0.4230}
                         + 0.0054 \eta^{0.1456} \right.
\\ \nonumber
           && \qquad \qquad  \left. \vphantom{\frac{1}{6}}
                         - 0.0714 \eta^{6/23}
                         + 0.0509 \eta^{0.4086} \right),
\\
C_4 (\mu) &=& C_2 (M_W) \left(0.0156 \eta^{-0.8994}
                         - \frac{1}{6} \eta^{-12/23}
                         + 0.1214 \eta^{-0.4230}
                         + 0.0026 \eta^{0.1456} \right.
\\ \nonumber
           && \qquad \qquad  \left. \vphantom{\frac{1}{6}}
                         - 0.0714 \eta^{6/23}
                         + 0.0984 \eta^{0.4086} \right),
\\
C_5 (\mu) &=& C_2 (M_W) \left(-0.0025 \eta^{-0.8994}
                         + 0.0117 \eta^{-0.4230}
                         + 0.0304 \eta^{0.1456}
                         - 0.0397 \eta^{0.4086} \right), \nonumber\\
&&
\\
C_6 (\mu) &=& C_2 (M_W) \left(-0.0462 \eta^{-0.8994}
                         + 0.0239 \eta^{-0.4230}
                         - 0.0112 \eta^{0.1456}
                         + 0.0335 \eta^{0.4086} \right), \nonumber\\
&&
\\
C_7 (\mu) &=& C_7 (M_W) \eta^{16/23} +
              C_8 (M_W) \frac{8}{3}
                        \left( \eta^{14/23} - \eta^{16/23} \right),
\\ \nonumber
          && + C_2 (M_W) \left(- 0.0185 \eta^{-0.8994}
                               - 0.0714 \eta^{-12/23}
                               - 0.0380 \eta^{-0.4230}
                               - 0.0057 \eta^{0.1456} \right.
\\ \nonumber
           && \qquad \qquad  \left.
                               - 0.4286 \eta^{6/23}
                               - 0.6494 \eta^{0.4086}
                               + 2.2996 \eta^{14/23}
                               - 1.0880 \eta^{16/23} \right),
\\
C_8 (\mu) &=& C_8 (M_W) \eta^{14/23}
\\ \nonumber
          && + C_2 (M_W) \left(- 0.0571 \eta^{-0.8994}
                               + 0.0873 \eta^{-0.4230}
                               + 0.0209 \eta^{0.1456} \right.
\\ \nonumber
           && \qquad \qquad  \left.
                               - 0.9135 \eta^{0.4086}
                               + 0.8623 \eta^{14/23} \right).
\end{eqnarray}

The non-zero initial conditions in the SM are given at the scale $M_W$ and 
read \cite{InamiLim}
\begin{eqnarray}
C_2 &=& 1 \\
C_7 (M_W) &=& -\frac{1}{2} x \left[ \frac{2x^2/3 + 5x/12-7/12}{(x-1)^3} -
                  \frac{3x^2/2-x}{(x-1)^4} \ln x \right],
\\
C_8(M_W) &=& -\frac{1}{2} x \left[ \frac{x^2/4 - 5x/4-1/2}{(x-1)^3} +
                  \frac{3x/2}{(x-1)^4} \ln x \right],
\end{eqnarray}
and
$x = m^2_t / M_W^2$. The numerical values for the Wilson coefficients
at the scale $\mu=M_W$ (``Matching Conditions") and at three other scales
$\mu= 10.0$ GeV, $5.0$ GeV and $10.0$ GeV are given in Table \ref{wcmudep}. 
Also, for subsequent discussion it is useful to define two
effective Wilson coefficients $C_7^{\mathit{eff}}(\mu)$ and
 $C_8^{\mathit{eff}}(\mu)$ 
\cite{Buras94}:
 \begin{eqnarray}
\label{C78eff}
C_7^{\mathit{eff}} &\equiv & C_7 - \frac{C_5}{3} -  C_6 \quad , \nonumber\\
C_8^{\mathit{eff}} &\equiv & C_8 + C_5 \quad .
\end{eqnarray}
Their values are also given in Table 1.  

\begin{table}[htb]
\label{wcmudep}
\begin{center}
\begin{tabular}{| c | c | r | r | r | }
\hline
 $C_i(\mu)$ & $\mu=m_W$ & $\mu=10.0$ GeV
                              & $\mu=5.0$ GeV
                              & $\mu=2.5$ GeV\\
\hline \hline
$C_1$ & $0.0$ & $-0.158$ & $-0.235$ & $-0.338$ \\
$C_2$ & $1.0$ & $1.063$ & $1.100$ & $1.156$ \\
$C_3$ & $0.0$ & $0.007$ & $0.011$ & $0.016$ \\
$C_4$ & $0.0$ & $-0.017$ & $-0.024$ & $-0.034$ \\
$C_5$ & $0.0$ & $0.005$ & $0.007$ & $0.009$ \\
$C_6$ & $0.0$ & $-0.019$ & $-0.029$ & $-0.044$ \\
$C_7$ & $-0.193$ & $-0.290$ & $-0.333$ & $-0.388$ \\
$C_8$ & $-0.096$ & $-0.138$ & $-0.153$ & $-0.171$ \\
$C_7^{\mathit{eff}}$ & $-0.193$ & $-0.273$ & $-0.306$ & $-0.347$ \\
$C_8^{\mathit{eff}}$ & $-0.096$ & $-0.132$ & $-0.146$ & $-0.162$ \\
\hline
\end{tabular} 
\end{center}
\caption{Wilson coefficients $C_i(\mu )$
                                   at the scale $\mu=m_W=80.33$ GeV
(``matching conditions") and at three other scales, $\mu = 10.0$ GeV,
$\mu =5.0$ GeV and $\mu = 2.5$ GeV,
 evaluated with two-loop $\beta$-function and the
 leading-order anomalous-dimension matrix. The entries correspond to the top
quark mass
 $\overline{m_t}(m_t^{pole})=170$ GeV (equivalently,
$m_t^{pole}= 180 $ GeV) and the QCD coupling constant
 $\alpha_s(m_Z^2)=0.117$, both in the $\overline{MS}$ scheme.}   
\end{table}
\subsection {Real and virtual $O(\a_s)$ corrections for the matrix element}
\par
 Now, we discuss the real and virtual $O(\as)$ corrections to the matrix
element for $b \to s + \gamma$ at the scale $\mu \approx m_b$, which by 
themselves form a 
well-defined gauge invariant albeit scheme-dependent set of corrections.
This scheme dependence will be cancelled against the one in the anomalous
dimension $\gamma^{(1)}$, as discussed in the context of the dominant
contributions to the non-leptonic decays of the $B$ hadron  
earlier. The results presented here correspond to the NDR scheme.

\par
\begin{figure}[htb]
\vspace{0.10in}
\centerline{\epsfig{file=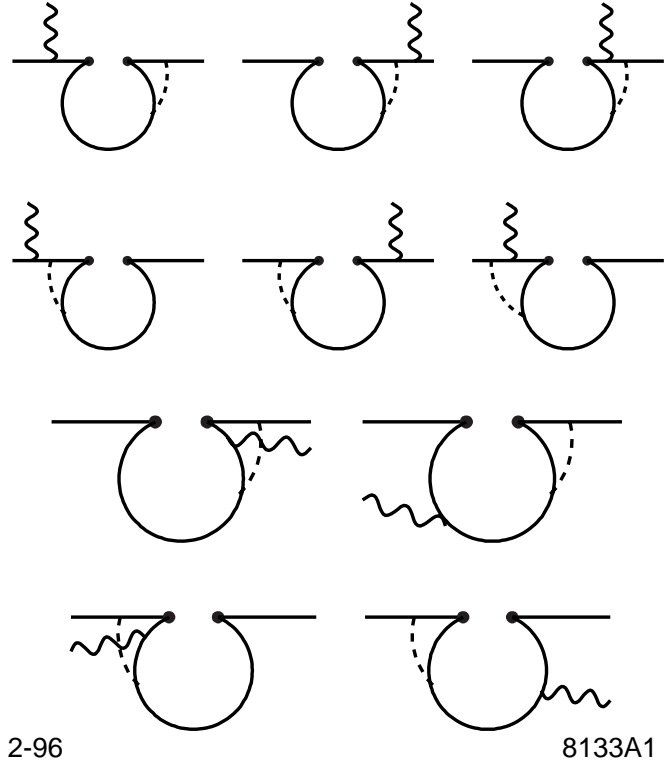,height=4in,angle=0}}
\vspace{0.08in}
\caption[]{Non-vanishing two-loop diagrams
associated with the operator $O_2$ calculated in \protect\cite{GHW96}.
The fermions ($b$, $s$ and $c$ quark) are represented by solid lines.
The wavy (dashed) line represents the photon (gluon).
\label{fig1ghw96}}
\end{figure}

Recapitulating the essential steps, we recall that
the Bremsstrahlung corrections  in $b \to s \gamma + g$, calculated 
in \cite{ag1} - \cite{ag3} and \cite{ag95}, were aimed at
 getting a non-trivial photon energy spectrum at the partonic level.
In these papers, the virtual corrections to $b \to s \g$ in $O(\as)$
were included only by taking into account those virtual diagrams which 
are needed to cancel the infrared
singularities (and also the collinear ones in the limit $m_s \to 0$)
 generated by the 
Bremsstrahlung diagram. The emphasis was on deriving the photon energy 
spectrum
in $B \to X_s + \gamma$ away from the end-point $x_\gamma \to 1$ and the
Sudakov-improved photon energy spectrum in the region $x_\gamma \to 1$.
Clearly, the left-out virtual diagrams
shown in Fig.~\ref{fig1ghw96} do not contribute either to the 
Sudakov 
spectrum or to the region $x_\gamma \neq 1$ at the parton level. They, 
however, do contribute to the overall decay rate in $B \to X_s + \gamma$.
Recently, these additional virtual correction have been evaluated in
\cite{GHW96}, neglecting the contributions of the small operators 
$O_3$--$O_6$. This additional contribution reduces substantially
the scale dependence of the leading order (or partial next-to-leading order)
 decay 
width $\Gamma(B \to X_s + \gamma)$, which previously was found to be 
substantial 
and constituted a good fraction of the theoretical uncertainty in the
inclusive decay rate
\cite{AGM92,Buras94,Ciuchini94,ag95}.
%
%

 We shall follow closely the derivation of the virtual corrections given in 
\cite{GHW96}. Concentrating on the dominant operators
$O_2, ~O_7$ and $O_8$,  the contribution of the next-to-leading order  
correction to the matrix element part in $b \to s+\gamma$ can be
expressed as follows: 
 \begin{equation}
{\cal M} = {\cal M}_2 + {\cal M}_7 + {\cal M}_8
\end{equation}
and the various terms (including appropriate counterterm 
contributions) can be summarized as \cite{GHW96}:
\be
\label{m2ren}
{\cal M}_2 = \bra s \g |O_7| b \ket _{tree} \, \frac{\a_s}{4 \p} \,
\left( \ell_2 \log \frac{m_b}{\mu}  + r_2 \right) \quad ,
\ee
with
\be
\label{l2}
\ell_2 = \frac{416}{81}.
\ee
\begin{eqnarray}
\label{rer2ndr}
\Re r_2 &=& \frac{2}{243} \, \left\{- 833 + 144 \pi^2 z^{3/2}
\right. \nonumber \\
&& \hspace{0.3cm}
+ \left[ 1728 -180 \pi^2 -1296 \zeta (3) + (1296-324 \pi^2) L +
108 L^2 + 36 L^3 \right] \, z \nonumber \\
&& \hspace{0.3cm}
+ \left[ 648 + 72 \pi^2 + (432 - 216 \pi^2) L + 36 L^3 \right] \, z^2
\nonumber \\ 
&& \hspace{0.3cm}        \left.                 +
\left[ -54 - 84 \pi^2 + 1092 L - 756 L^2 \right] \, z^3 \, \right\}
\end{eqnarray}
\begin{eqnarray}
\label{imr2ndr}
\Im r_2 &=& \frac{16 \p}{81} \, \left\{- 5
+ \left[ 45-3 \pi^2 + 9 L +
9 L^2 \right] \, z
+ \left[ -3 \pi^2 + 9 L^2 \right] \, z^2 +
\left[ 28 - 12 L  \right] \, z^3 \, \right\}~.
\end{eqnarray}
Here, $\Re r_2$ and $\Im r_2$ denote the real and the imaginary part
of $r_2$, respectively, $z=(m_c/m_b)^2$ and $L=\log (z)$. 

The real and virtual corrections associated with the
operator $O_7$, calculated in \cite{ag1,ag2,ag95} can be combined
into a \underline{modified matrix element} for
$b \to s \g$,
in such a way that its square reproduces
the result derived in these papers. This modified 
matrix element ${\cal M}_7^{mod}$ reads \cite{GHW96}:
\be
\label{m7lr}
{\cal M}_7^{mod} = \bra s \g|O_7| b \ket _{tree}
\, \left( 1+ \frac{\a_s}{4\p} \left( \ell_7 \, \log \frac{m_b}{\mu}
+r_7
\right) \right)
\ee
with
\be
\label{l7r7}   
\ell_7 = \frac{8}{3}  \quad , \quad
r_7 = \frac{8}{9} \, (4 - \pi^2) \quad .
\ee

Finally, the result for ${\cal M}_8$ is \cite{GHW96}:
\be
\label{m8lr}
{\cal M}_8 = \bra s \g |O_7| b \ket _{tree} \, \frac{\a_s}{4 \p} \,
\left( \ell_8 \log \frac{m_b}{\mu}  + r_8 \right) \quad ,
\ee
with
\be
\label{l8r8}
\ell_8 = - \frac{32}{9} \quad , \quad
r_8 = - \frac{4}{27} \,
\left( -33 +2\p^2  -6 i \p   
\right)
\quad .
\ee
\par
With the results given above,
one can write down the amplitude ${\cal M}(b \to s \g )$ 
by summing the various contributions already mentioned.
Since the relevant scale for a $b$ quark decay is expected to
be $\mu \sim m_b$,  the matrix elements of the
operators may be expanded around
$\mu=m_b$ up to order $O(\a_s)$  
and the next-to-leading order result can be written as:
\be
\label{amplitudevirtuell}
{\cal M}(b \to s \g ) = -\frac{4 G_F \l_t}{\sqrt{2}} \, D \,
\bra s \g|O_7(m_b)|b \ket _{tree}
\ee
with
\be
\label{d}
D = C_7^{\mathit{eff}}(\mu) + \frac{\a_s(m_b)}{4\p} \left(
C_i^{(0)eff}(\mu) \gamma_{i7}^{(0)eff} \log \frac{m_b}{\mu} +
C_i^{(0)eff} r_i
\right)         \quad ,
\ee
where the quantities $\gamma_{i7}^{(0)eff}=\ell_i + 8 \, \delta_{i7}$
are  the entries of the (effective) leading order anomalous
dimension matrix 
and  the quantities $\ell_i$ and $r_i$ are given
for $i=2,7,8$ in eqs. (\ref{l2},\ref{rer2ndr}),
(\ref{l7r7}) and (\ref{l8r8}), respectively.
The first term, $C_7^{\mathit{eff}}(\mu)$,
on the r.h.s. of Eq.~(\ref{d})  has to be
taken up to  next-to-leading logarithmic precision in order
to get the full next-to-leading logarithmic result, whereas
it is sufficient to use the leading logarithmic values of
the other Wilson coefficients in Eq.~(\ref{d}).

As pointed out by Buras et al. \cite{Buras94},
the explicit logarithms of the form $\a_s(m_b) \log(m_b/\mu)$ in Eq. 
(\ref{d}) are cancelled by the $\mu$-dependence of 
$C_7^{(0)\mathit{eff}}(\mu)$. Therefore, the scale dependence is
significantly reduced by including the virtual corrections
completely to this order.

The decay width $\G^{virt}$ which follows
from ${\cal M}(b \to s \g)$ in Eq.~(\ref{amplitudevirtuell}) reads 
\be
\label{widthvirt}
\G^{virt} = \frac{m_{b,pole}^5 \, G_F^2 \l_t^2 \a_{em}}{32 \p^4}
\, F \, |D|^2 \quad ,
\ee
where the terms of $O(\a_s^2)$
in $|D|^2$ have been discarded.
The factor $F$ in Eq.~(\ref{widthvirt}) is
\be
F = \left( \frac{m_b(\mu=m_b)}{m_{b,pole}} \right)^2 =
1- \frac{8}{3} \,  \frac{\a_s(m_b)}{\p} \quad ,  
\ee
and its origin lies in the explicit presence of $\mb$ in the operator
$O_7$.
To get the inclusive decay width for $b \to s \g (g)$, also
the Bremsstrahlung corrections (except the part
already absorbed above) must be added. The contribution of the operators
${\cal O}_2$ and ${\cal O}_7$ was  calculated already in \cite{ag1}.

\subsection{Estimating long-distance effects in $\BGAMAXS$}

\par
 In order to get the complete amplitude for $\BGAMAXS$ one has to include
also the effects of the long-distance contributions, which arise from 
the   
matrix elements of the four-quark operators  in ${\cal H}_{eff}$,
$\langle X_s \gamma \vert {\cal O}_i \vert B \rangle$. We shall
discuss such contributions in the exclusive radiative   
decays $B \to (\rho,\omega) + \gamma$ subsequently, which have been
calculated in a more robust theoretical framework using QCD sum rules.
This framework (likewise lattice-QCD) is not applicable in inclusive 
decays in a straightforward manner. Lacking anything better, we
discuss phenomenological models used in the literature in
estimating the long-distance amplitude in radiative decays $\BGAMAXS$.

 In calculating the matrix elements of four-Fermion operators, it is 
usual to
invoke the hypothesis of factorization. In the present context,
factorization is combined with the additional assumption of  vector
meson dominance, involving the decays
 $ B \to \sum_{i} V_{i} + X_s \to \gamma +X_s$, where $V_{i}= J/\psi,
\psi^\prime,...$ \cite{bsgamld} - \cite{Ricciardi}.
 It should be remarked that non-leptonic decays, such as $B \to
(J/\psi,\psi^\prime + X_s)$, by themselves are not under the quantitative
control of the factorization-based
 framework due to the presence of significant non-factorizing pieces in the
non-leptonic amplitudes - perhaps a sign of the breakdown of
local quark-hadron duality (?).
Phenomenologically, however, data on two-body and quasi two-body decays
are well accounted for in terms of the Bauer-Stech-Wirbel effective
parameters $a_1\equiv c_1/N_c + c_2$ and $a_2 \equiv c_1+c_2/N_c$ \cite{BSW}.
The details of this analysis
 and a host of other related decay modes can be seen in \cite{BH95}.
Concerning the use of vector meson dominance, one has to ensure that the
resulting amplitude  ${\cal M}(\BGAMAXS)$  remains
manifestly gauge invariant. In the present model
this amounts to discarding the longitudinal polarization contribution in
the non-leptonic decays $B \to (J/\psi, \psi^\prime,...) +X_s$, which
in fact dominates the decay widths \cite{BH95}, and keeping only the smaller
contribution from the transverse polarization of $J/\psi, \psi^\prime,...$.
 Following
\cite{DHT95,GP95}, one can write the decay amplitude as:
\begin{eqnarray}
{\cal M}(b\rightarrow s J/\psi )_T
=  {G_F\over \sqrt{2}}a_2V_{cb}V_{cs}^* {g_{J/\psi}(m_{J/\psi}^2)\over m_b}
\bar s \sigma^{\mu\nu}(1+\gamma_5)b q_\nu \epsilon^\dagger_\mu(q)\;,
\end{eqnarray}
where $g_\psi$ is defined as $\langle \psi(q) \vert \bar{c}\gamma_\mu c\vert
0 \rangle =-i g_\psi(q^2) \epsilon_\mu^\dagger (q)$.
 For the
decays under consideration one needs the value of $a_2$, which has been
found to be $\vert a_2 \vert = 0.24 \pm 0.04$ \cite{BH95}.
 One also needs to evaluate the coupling constant
$g_V(q^2)$ at the point $q^2=0$. From leptonic decays of vector mesons, one
gets, however,  $g_V(q^2=M_V^2)$.
It has been remarked in literature, in particular in \cite{DHT95,GP95},
 that using $g_V(q^2=0)=g_V(q^2=M_V^2)$ would
substantially overestimate the long-distance contribution due to
the expected dynamical suppression of the effective coupling 
$g_V(q^2)$,   
as one extrapolates to the point $q^2=0$. Taking all this into account, an
estimate of the long-distance amplitude from the intermediate $J/\psi$ state
 is \cite{DHT95}:
\begin{eqnarray}
{\cal M}_{LD}(b\rightarrow s J/\psi\rightarrow s \gamma )
=  {G_F\over 2\sqrt{2}}a_2V_{tb}V_{ts}^* ({2\over 3}e{g_{J/\psi}^2(0)\over
m_{J/\psi}^2
m_b})
\bar s \sigma^{\mu\nu}(1+\gamma_5)b F_{\mu\nu}\;,
\end{eqnarray}
where use has been made of the
 $\psi$ to $\gamma$ conversion vertex,
\begin{equation}
 <0|eJ^\mu_{em}|\psi>
=(2e/3)<0|\bar c\gamma^\mu c|\psi> = -(2e/3)ig_\psi(0)\epsilon_\mu~,
\end{equation}
and we must use the value $g_{J/\psi}(0)$ for the conversion coupling
constant.
Including all the ($c\bar c$) resonances and the short distance contribution
${\cal M}_{SD}$, the two-body  decay amplitude ($b \to s \gamma)$ can be
written as \begin{eqnarray}
{\cal M}(b\rightarrow s\gamma)
=  -{eG_F\over 2\sqrt{2}}V_{tb}V_{ts}^*[{1\over 4\pi^2}m_b D(\mu)
-a_2{2\over 3} \sum_i{g_{\psi_i}^2(0)\over m_{\psi_i}^2m_b} ]
\bar s \sigma^{\mu\nu}(1+\gamma_5)b F_{\mu\nu}\;,
\end{eqnarray}
where $\psi_i$ represents the following  vector $c\bar c$ resonant
states:
$\psi(1S)$, $\psi(2S)$, $\psi(3770)$, $\psi(4040)$, $\psi(4160)$, and
$\psi(4415)$, and $D$ is the function given earlier. Since in the
LD-amplitude, gluon Bremsstrahlung is neglected, it is consistent
to neglect this also in the SD-piece.
 Taking this estimate
 as giving the right order of magnitude for the long-distance
contribution, Deshpande et al. \cite{DHT95} conclude that such
long-distance effects can be as large as 10\%. Other estimates, in
particular by Golowich and Pakvasa \cite{GP95}, lead to an even smaller
long-distance contribution. Clearly, one can not argue very conclusively
if such estimates are completely quantitative due to the assumptions
involved. In future, one
could improve these estimates by  using  data from HERA on elastic $J/\psi
-$, and $\psi^\prime$-photoproduction
to get $g_{J/\psi}$ and $g_{\psi^\prime}(0)$ directly, reducing at least the
extrapolation uncertainties involved in the presently adopted procedure
of extracting these coupling constants from the leptonic decay widths of
each state and extrapolating to the point $q^2=0$ using an Ansatz.
%
\subsection{Estimates of $\BBGAMAXS$ in the Standard Model}
 In the quantitative
estimates of the SM branching ratio $\BBGAMAXS$ given below we have 
neglected the  
LD-contributions. This is an assumption, which as we discussed above,
we do not expect to work much better than $O(10 \%)$.
 It is 
theoretically preferable to calculate this quantity in terms of the
semileptonic decay branching ratio
\begin{equation}
\label{brdef}
{\cal B} ( B \ra  X_{s} \g) = [\frac{\Gamma(B \ra  
\gamma + X_{s})}{\Gamma_{SL}}]^{th}
\, {\cal B} (B \to X\ell \nu_\ell)  \qquad ,
\end{equation}  
where, the leading-order QCD corrected
 $\G_{SL}$ has been given earlier. The leading order power 
corrections in the heavy quark expansion, discussed in the context of the 
semileptonic decay rate, are identical 
in the inclusive decay rates for  $\BGAMAXS$ and $B \to X \ell \nu_\ell$, 
entering in the
numerator and denominator in the square bracket, respectively, and hence 
drop out \cite{Chayetal,Bigietal}.

\begin{figure}[htb]
\vspace{0.10in}
\centerline{\epsfig{file=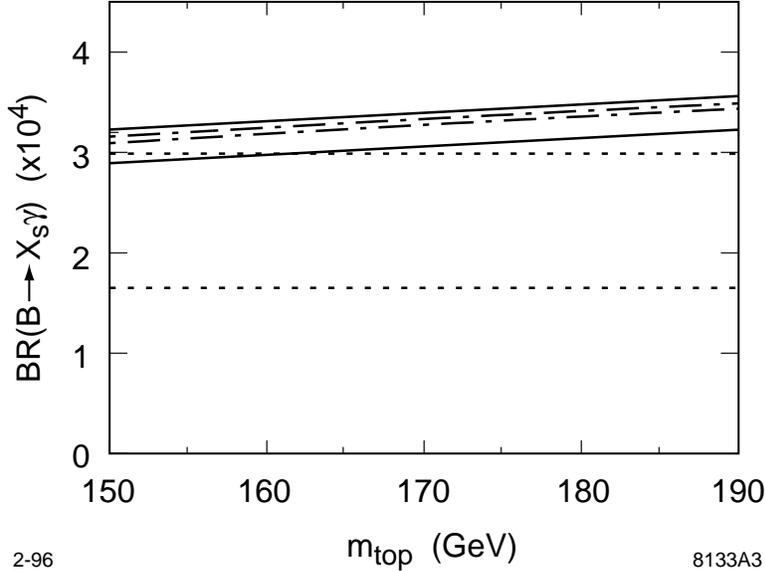,height=3in}}
\vspace{0.08in}
\caption[]{Branching ratio for $b \to s \g (g)$
calculated in \protect\cite{GHW96} with the parameters
$\Vtsabs/\Vcbabs =1,~\Vtbabs =1, ~m_b^{pole}=4.8$ GeV and $m_c/m_b=0.29$. 
The different curves are explained in the text.
\label{GHRfig3}}
\end{figure}

   The error on the branching ratio ${\cal B} ( B \ra  X_{s} \g)$ comes
from four different sources. We list them below:
\begin{enumerate}
\item $\Delta m_t$:  The present value of $m_t$ is $m_t=175 \pm 9$ GeV
\cite{DPG96}, which is usually interpreted as the pole mass
although this is not unambiguous. With this
the running top quark mass in the $\overline{MS}$ scheme is
$\overline{m_t} = 166 \pm 9$ GeV. However, in Eq.~(\ref{d}), there is no
distinction made between $m_t^{pole}$ and $\overline{m_t}$. To take
both the theoretical and experimental errors into account, we take
$m_t=170 \pm 15$ GeV. This leads to an error of $\pm 5\%$
in ${\cal B} ( B \ra  X_{s} \g)$.
\item $\Delta \mu$: The scale dependence is now reduced thanks to the work 
done in \cite{GHW96}. It is usual
to estimate the residual $\mu$-dependence by taking a range $m_b/2 \leq \mu 
\leq 2 m_b$. This is plotted in
Fig. ~\ref{GHRfig3} (solid curves), which shows  that
it yields an error of $\pm 6 \%$ on ${\cal B}(B \to X_s + \gamma)$.
\item Errors from the extrinsic parameters, \\
 (i) $\Delta (m_b)$ (equivalently from $\Delta(m_c/m_b)$),\\
(ii) $\Delta(\as(m_Z)$ (equivalently
$\Delta\Lambda_5$, the uncertainty on the QCD-scale parameter), 
(iii) $\Delta(BR_{SL})$, the experimental uncertainty on the semileptonic
 branching ratio.\\
Taking $m_c/m_b=0.29 \pm 0.02$, ${\cal B}(B \to X \ell \nu_\ell) 
=(10.4\pm 0.4)\%$, $ \as(m_Z)=0.117 \pm 0.006$ 
\cite{PDG94}, an uncertainty of $\pm 12\%$ was estimated 
on ${\cal B}(B \to X_s + \gamma)$ in \cite{ag95}.
\item Last, but by no means least, there is at present an incalculable error
due to the fact that $C_7^{\mathit{eff}}$ is not yet available in the 
next-to-leading order.
\end{enumerate}
In view of this missing piece, it is not possible to give a completely
corrected NLL prediction for ${\cal B}(\BGAMAXS)$ at present.
In what follows, we shall replace $C_7^{\mathit{eff}}$ by its leading log 
value, which yields the branching ratio \cite{ag96}:
\begin{equation}\label{smbsgbr}
{\cal B} (\BGAMAXS )= (3.20 \pm 0.30 \pm 0.38) \times 10^{-4}
\end{equation}
where the first error comes from the combined error on $\Delta m_t$ and
$\Delta \mu$, as can be seen in Fig. ~\ref{GHRfig3},
and the second from the extrinsic source. This estimate 
is subject to an additive renormalization due to the missing NLL anomalous
dimension piece and the somewhat less tractable contribution of the 
LD-effect, discussed in the previous subsection but not included in
the numerical estimates of the branching ratio (\ref{smbsgbr}).  
However,   
if the NLL anomalous dimensions of the four-quark operators ${\cal O}_1,...
{\cal O}_6$, which are known \cite{BW90}, are any useful guide for
estimating $C_7^{\mathit{eff}}$, we do not
expect a substantial change of this branching ratio in the complete NLL. Of
course, the big challenge is to show that this indeed is the case.
In view of this, we propose to add  an additional theoretical uncertainty of 
order $\pm 10\%$  on ${\cal B} (\BGAMAXS )$, yielding
${\cal B} (\BGAMAXS )= (3.20 \pm 0.30 \pm 0.38 \pm 0.32) \times 10^{-4}$.
Combining the theoretical errors in quadrature gives
\begin{equation}\label{smbsgbrf}
{\cal B} (\BGAMAXS )= (3.20 \pm 0.58) \times 10^{-4},
\end{equation}
which we consider is a fairly robust estimate of the SM branching ratio at 
present. This can be improved only after the two estimated pieces are 
replaced by the actual calculations. The SM branching ratio
${\cal B} (\BGAMAXS )$ is compatible with the present measurement
${\cal B} (\BGAMAXS )= (2.32 \pm 0.67) \times 10^{-4}$ \cite{CLEOrare2}.
Expressed in terms of the CKM matrix element ratio, one gets
\begin{equation}\label{vtscb}
\frac{\Vtsabs}{\Vcbabs} = 0.85 \pm 0.12 (\mbox{expt}) \pm 0.10 (\mbox{th}),
\end{equation}
which is within errors consistent with unity, as expected from the
unitarity of the CKM matrix.
 
%
%
%
%
%

\subsection{Photon energy spectrum in $\BGAMAXS$}
The two-body partonic process $b \to s \gamma$
yields a photon energy spectrum $1/(\Gamma) d \Gamma (b \to s 
\gamma) = \delta(1-x)$, where 
the scaled photon energy $x$ is defined as
$ E_\g = (m_b^2-m_s^2)/(2 \, m_b )\, x $.
Perturbative QCD corrections, such as
$b \to s \gamma + g$, give a
characteristic Bremsstrahlung spectrum in $x$ in the interval $[0,1]$
peaking near the end-points, $E_\gamma \to
E_\gamma ^{max}$ (or $x \to 1$) and $E_\gamma \to 0$ (or $x \to 0$), arising 
from the soft-gluon and soft-photon configurations, respectively.
Near the end-points, one 
has to improve the spectrum obtained in fixed order perturbation 
theory. This is usually done in the region $x \to 1$ by isolating and 
exponentiating the leading behaviour in $\alpha_{em}\alpha_s(\mu)^m 
\log^{2n} (1-x)$  with $m\leq n$,
where $\mu$ is a typical momentum in the decay $\BGAMAXS$. The running of 
$\alpha_s$ is a non-leading effect, but  as it is characteristic of QCD it
modifies the Sudakov-improved end-point photon energy spectrum
 \cite{KS94,Shifmangamma} 
compared to its  analogue in QED \cite{Sudakov}.
 As long as the $s$-quark mass
is non-zero, there is no collinear singularity in the spectrum.
However, parts of the spectrum have large logarithms of the form
$\as \log (m_b^2/m_s^2)$, which are important near the end-point
$x \to 0$ but their influence persists also in the
intermediate photon energy region and they have to be resummed
\cite{ag95,klp95}.

\par
 It has been observed in a number of papers
\cite{KS94,neubertbsg,Bigietal2}, that the $x$-moments
(scaled photon energy)  in $\BGAMAXS$ and those involving
lepton energy in the decay $B \to X_u \ell \nu_\ell$ are related.
Defining the moments as:
\bea
 \M^{(n)}(\BGAMAXS) &\equiv & \frac{1}{\Gamma} \int_{0}^{M_B/m_b} dx x^{n-1}
                   \frac{d \Gamma}{dx} \\ \nonumber
\M^{(n)} (B \to X_u \ell \nu_\ell) &\equiv&  - \int_{0}^{M_B/m_b} dx x^n
\frac{d}{dx}\big(\frac{1}{\Gamma_\ell}\frac{d \Gamma_\ell}{dx}\big) \\  
\nonumber
&=& \frac{n}{\Gamma_\ell} \int_{0}^{M_B/m_b} dx x^{n-1} \frac{d
\Gamma_\ell}{dx}~, \label{moments}
\eea
The moments $\M^{(n)}$ have been worked out in
the leading non-trivial order in perturbation theory and
the results can be expressed as:
\be
\M^{(n)} \sim 1 + \frac{\alpha_s}{2 \pi} C_F(A\log^2n + B \log n + 
\mbox{const.}) \ee
where $C_F=4/3$,
the leading coefficient is universal with $A=-1$ \cite{Sudakov}, and the
non-leading coefficients are process dependent;
$B=7/2$ \cite{ag1} and $B
=31/6$ \cite{JK89},
for $\BGAMAXS$ and $B \to X_u \ell \nu_\ell$, respectively.
 Measurements of
the moments could eventually be used to relate the CKM matrix element
$V_{ts}$ and
$V_{ub}$.
   
\par
  How about calculating the photon energy spectrum in $\BGAMAXS$
and the lepton energy spectrum in $B \to X \ell \nu_\ell$ completely
in QCD? Concentrating on $\BGAMAXS$, it is known that there
is a gap between the end-points of the physical spectrum $(E_\gamma^{max}
= (m_B^2-m_X^2)/(2m_B))$, where $X=(m_K+m_\pi)$, and the partonic
spectrum $(E_\gamma^{max}=(m_b^2-m_s^2)/(2m_b))$, and hence in the
 parton-model description there is a window in $E_\gamma$ which remains 
empty.
This gap can only be filled up by non-perturbative effects, which in
their simple form can be attributed to the $b$-quark motion
in the $B$ hadron. Phenomenological models were already proposed in the 
infancy of $B$ physics which took into account non-perturbative
smearing of the partonic spectra to fill this gap. The 
question is if this effect can be computed in QCD proper.

\par
 Attempts to calculate the photon and lepton energy spectra
in the heavy quark expansion method lead to formal expressions which near 
the end-point are divergent \cite{MW94}, \cite{Bigietal2} - \cite{JR94}.
The point is that near the end-point, the energy release for the light 
quark system for the final state is not of $O(\mb)$ but of the order of
the parameter
 $\bar{\Lambda}=M_B-m_b \sim O(\Lambda_{\mbox{\small QCD}})$. Thus, the
expansion parameter is no longer $1/\mb^2$ and the operator product
expansion - the backbone of the heavy quark expansion - breaks down.
This divergent series in the effective theory has to be  
 cleverly resummed and the distributions averaged over momentum bins 
\cite{neubertbsg} to smooth the
increasing number of derivatives of the delta function, $\delta^n(1-x)$.
 This resummation allows to define
 an effective non-perturbative shape function 
\cite{neubertbsg,Shifmangamma}, which can not be calculated in the
effective theory, but one could use this concept advantageously to relate
the energy spectra in the semileptonic decays $B \to X_u \ell \nu_\ell$
and $B \to X_s + \gamma$. It has been further argued in \cite{Shifmangamma} 
that
a Gaussian distribution, such as the one used in the Fermi motion model in
\cite{Alipiet,ACCMM}, with an appropriate definition of the
effective momentum-dependent $b$-quark mass, as prescribed by the
heavy quark effective theory, is a good approximation to
the shape function in $B \to X_u \ell \nu_\ell$ and $B \to X + \gamma$. 
 
\par
  We shall confine ourselves 
to the discussion of the photon energy spectrum, which combines   
the perturbatively computed spectrum, discussed in the previous section, 
with a model of the quark motion which fills the mentioned energy 
gap in the end-point region. We will then show a
comparison of this spectrum with
 the measured photon energy spectrum in $\BGAMAXS$ \cite{CLEOrare2}.
 In this model 
\cite{Alipiet}, which admittedly is simplistic but has received some
theoretical support in the HQET approach subsequently \cite{MW94, Bigietal2},
  the $b$ quark in $B$ hadron is assumed to have a Gaussian
distributed Fermi motion determined by a non-perturbative parameter, $p_F$,
\begin{equation}
\label{lett13}
 \phi(p)= \frac {4}{\sqrt{\pi}{p_F}^3} \exp (\frac {-p^2}{{p_F}^2})
\quad , \quad p = |\vec{p}|
\end{equation}
with the wave function normalization 
$ \int_0^\infty \, dp \, p^2 \, \phi(p) = 1.$
The photon energy spectrum from
the decay of the $B$-meson at rest is then given by
\begin{equation}
\label{lett15}
 \frac{d\Gamma}{dE_\gamma}= \int_0^{p_{max}} \, dp \, p^2 \, \phi(p)
  \frac {d\Gamma_b}{dE_\gamma}(W,p,E_\g) \quad ,
\end{equation}
where $p_{max}$ is the maximally allowed value of $p$ and
$ \frac{d\Gamma_b}{dE_\g}$
 is the photon energy spectrum from the decay of the $b$-quark in
flight, having a momentum-dependent mass $W(p)$. This is calculated
in perturbation theory taking into account the appropriate Sudakov
behaviour in the $E_\gamma$ end-point region at the partonic level.

\begin{figure}[htb]
\vspace{0.10in}
\centerline{
\epsfysize=3in
\rotate[r]{
\epsffile{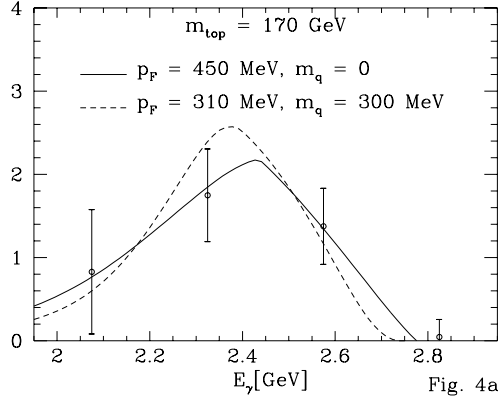}
}
}
\vspace{0.08in}
\caption[]{Comparison of the normalized photon energy distribution
using the
 CLEO data \protect\cite{CLEOrare2} corrected for detector effects and 
theoretical
distributions from \protect\cite{ag95} , both  normalized to unit area in
the photon energy interval between 1.95 GeV and 2.95 GeV. The solid 
 curve corresponds to the values with the minimum $\chi^2$,
 $(m_q,p_F)$=(0, ~450 MeV), and the dashed curve to the values
 $(m_q,p_F)$=(300 MeV, 310 MeV).
\label{agfig4}}
\end{figure}

An analysis of the CLEO photon 
energy spectrum has been undertaken in \cite{ag95}  to determine 
the non-perturbative parameters of this model, namely  
$m_b(pole)$ and  $p_F$. The latter is related to the
kinetic energy parameter $\lambda_1$ defined earlier in the HQET
approach. 
The experimental
errors are still large and the fits result in relatively small $\chi^2$ 
values; the minimum, $\chi^2_{min}=0.038$, is obtained
for $p_F=450$ MeV and 
$m_b(pole)=4.77$ GeV, in good agreement with theoretical estimates of
the same, namely
$m_b(pole)= 4.8\pm 0.15$ GeV \cite{Neubert94,bqmass} and
$p_F^2=-\lambda_1/2= 0.25 \pm 0.05$ GeV$^2$ 
obtained from the QCD sum rules \cite{BB94}. The central value,
 $\lambda_1=-0.4$ GeV$^2$, obtained
from the photon energy spectrum in $\BGAMAXS$
 is also in good agreement with a recent
determination of the same from the lepton energy spectrum in
$B \to X \ell \nu_\ell$ in which partially integrated spectrum and its 
first moment are analyzed in terms of the HQET parameters $\bar{\Lambda}$
and $\lambda_1$ using CLEO data, getting $\lambda_1=-0.35\pm 0.05$ GeV$^2$
and $\bar{\Lambda} =0.55 \pm 0.05$ GeV \cite{GKLW96}.
In Fig. \ref{agfig4} we have plotted the photon energy spectrum normalized
to unit area in the interval between 1.95 GeV and 2.95 GeV for
the parameters which correspond to the minimum $\chi^2$ (solid curve)
and for another set of parameters that lies near the
$\chi^2$-boundary defined by $\chi^2=\chi^2_{min} +1$.
(dashed curve). Data from CLEO \cite{CLEOrare2} are also 
shown. Further details of this analysis can be seen in \cite{ag95}.

%
%

\subsection{Inclusive radiative decays \bgamaxd }

\par
The theoretical interest in studying the 
(CKM-suppressed) inclusive radiative decays
\bgamaxd\ lies in the first place in the 
 possibility of determining the CKM-Wolfenstein 
parameters $\rho$ and $\eta$.  
 The relevant region in the
decays $B \to X_d + \gamma$ is the end-point photon energy spectrum,
which has to be measured requiring that
 the hadronic system $X_d$ recoiling against the
photon does not contain strange hadrons to suppress the large-$E_\g$
photons from the decay $\BGAMAXS$. Assuming that this is feasible,
one can determine 
 from the ratio of the decay rates
$\BBGAMAXD/\BBGAMAXS$ the CKM-Wolfenstein parameters.
 This measurement was proposed in \cite{ag2}, where
the photon energy spectra were also worked out. In spirit, such an 
analysis is very similar to the already undertaken for the inclusive 
semileptonic
decays in which the end-point lepton energy spectrum is solved for the
ratio $\Vubabs/\Vcbabs$. Now, the end-point spectra in $B \to X + \gamma$
will be analyzed in terms of the ratio $\Vtdabs/\Vtsabs$. Of course, the 
experimental issues involved in the two cases are quite different and
measuring $\BGAMAXD$ is lot more challenging than measuring $B \to X_u 
\ell \nu_\ell$.

\indent
 In close analogy
with the \bgamaxs\ case discussed earlier,
the complete set of dimension-6 operators relevant for
the processes $b \to d \gamma$ and $b \to d \gamma g$ 
can be written as:
\begin{equation}
\label{heffd}
{\cal H}_{eff}(b \to d)=
 - \frac{4 G_{F}}{\sqrt{2}} \, \xi_{t} \, \sum_{j=1}^{8}
C_{j}(\mu) \, \hat{O}_{j}(\mu),\quad
\end{equation}
where $\xi_{j} = V_{jb} \, V_{jd}^{*}$ for $j=t,c,u$. The operators
 $\hat{O}_j, ~j=1,2$, have implicit in them CKM factors. In the
Wolfenstein parametrization \cite{Wolfenstein}, one can express these
factors as :
\begin{equation} 
\xi_u = A \, \lambda^3 \, (\rho - i \eta),
~~~\xi_c = - A \, \lambda^3 ,
~~~\xi_t=-\xi_u - \xi_c.
\end{equation}
We note that all three CKM-angle-dependent quantities
$\xi_j$ are of the
same order of magnitude, $O(\lambda^3)$, where $\lambda =\sin \theta_C 
\simeq 0.22$.
This is an important difference as compared to the effective
Hamiltonian ${\cal H}_{eff}(b \to s)$ written earlier,
 in which case the
 effective Hamiltonian
 factorizes into an overall CKM factor $\lambda_t = V_{tb} \, 
V_{ts}^{*}$. For calculational ease,  this difference
can be implemented by defining the operators $\hat{O}_1$ and $\hat{O}_2$
entering in ${\cal H}_{eff}(b \to d)$ as follows \cite{ag2}:
\begin{eqnarray}
\label{basis}
&&\hat{O}_{1} =
 -\frac{\xi_c}{\xi_t}(\bar{c}_{L \beta} \go{\mu} b_{L \alpha})
(\bar{d}_{L \alpha} \gu{\mu} c_{L \beta})
 -\frac{\xi_u}{\xi_t}(\bar{u}_{L \beta} \go{\mu} b_{L \alpha})
(\bar{d}_{L \alpha} \gu{\mu} u_{L \beta}) ,\nonumber \\
&& \hat{O}_{2} =
-\frac{\xi_c}{\xi_t}(\bar{c}_{L \alpha} \go{\mu} b_{L \alpha})
(\bar{d}_{L \beta} \gu{\mu} c_{L \beta}) 
 -\frac{\xi_u}{\xi_t}(\bar{u}_{L \alpha} \go{\mu}
b_{L \alpha}) (\bar{d}_{L \beta} \gu{\mu} u_{L \beta}) ,
\end{eqnarray}
and the rest of the operators $(\hat{O}_j;~j=3...8)$ are
defined like their
counterparts ${O}_j$ in ${\cal H}_{eff}(b \to s)$, with the obvious 
replacement
$s \to d$. With this definition, the matching conditions $C_j(m_W)$
 and the solutions
of the RG equations yielding $C_j(\mu)$ become
identical for the two operator bases $O_j$ and $\hat{O}_j$.

\par
 It has been explicitly checked in the $O(\alpha_s)$ calculations
of the decay rate and photon energy spectrum  involving $b \to d \gamma$ and
$b \to d g \gamma$ transitions that the limit $m_u \to 0$
for the decay rate $\GGAMAXD$ exists \cite{ag2}.
This implies that the rates and $E_\gamma$-spectrum are free of
mass singularities. In a recent paper
by Greub, Hurth and Wyler \cite{GHW96}, the finiteness proof has been 
extended to the NLL order, in which the virtual corrections to the matrix
elements are calculated completely. 
  From this it follows that there are no 
logarithms of the type
 $\alpha_{em}\alpha_s \log (m_u^2/m_c^2)$ \cite{Ricciardi,GHW96}.
Some papers, estimating LD-contributions in radiative $B$ decays, seem
to contradict this by assuming light-quark contributions which have such
spurious log-dependence. There is no calculational basis for this 
assumption.  
 On the other hand, as far as the dependence of the decay rate and spectra 
on the external light quark masses is concerned,
one encounters logarithms of the
type $\alpha_{em}\alpha_s [(1+(1-x)^2)/x] \log (m_b^2/m_s^2)$
 (for $b \to s g \gamma)$ and
$\alpha_{em}\alpha_s [(1+(1-x)^2)/x] \log (m_b^2/m_d^2)$ (for $b \to d g 
\gamma$), which are important
 near the soft-photon
($x \to 0)$ region \cite{ag95} and must also be exponentiated 
\cite{klp95}.

The essential difference between  $\GGAMAXS$ and $\GGAMAXD$ 
lies in the matrix elements of the first two operators $O_1$ and $O_2$
(in $H_{eff}(b \to s)$) and $\hat{O}_1$ and $\hat{O}_2$ (in $H_{eff}(b 
\to d)$).
The derivation of the inclusive decay rate and the final-state distributions
in \bgamaxd\ otherwise 
goes along very similar lines as for the decays \bgamaxs\ .
The branching ratio  $\BBGAMAXD$
 in the SM  can be written as:
\begin{eqnarray}
\label{branstruc}
&& \BBGAMAXD = D_1 |\xi_t|^2 \nonumber \\
&&                     \{
1 - \frac{1-\rho}{(1-\rho)^2 + \eta^2} \, D_2
  - \frac{\eta}{(1-\rho)^2 + \eta^2} \, D_3 
 + \frac{D_4}{(1-\rho)^2 + \eta^2} \} , \quad
\end{eqnarray}
where the functions $D_i$ depend on the parameters $\mt,\mb,\mc,\mu$,
as well as the others we discussed in the context of ${\cal B}(\BGAMAXS)$
in the previous section.
 For the central values of these parameters , one gets	
$D_1=0.21, ~D_2=0.17, ~D_3=0.03, ~D_4=0.10$ \cite{ag2}. This analysis has 
to be
updated taking into account the complete NLL virtual calculations in 
\cite{GHW96}, and so the numbers being quoted for $D_i$ are subject to some
change. To get the 
inclusive branching 
ratio the CKM parameters $\rho$ and $\eta$ have to be constrained from the
unitarity fits. Taking the parameters from a recent fit, one gets
$ 5.0 \times 10^{-3} \leq \vert \xi_t  \vert \leq 1.4 \times
10^{-2}$
(at 95\% C.L.) \cite{al95},
yielding an order of magnitude uncertainty in $\BBGAMAXD$ -
hence the interest in measuring this branching ratio. Taking the central 
values of the fitted CKM parameters discussed later in these lectures
$A=0.8, \lambda=0.2205, \eta =0.34$ and $\rho=-0.07$ \cite{al95},
 one gets
\begin{equation}
 \BBGAMAXD = (1.7\pm 0.85) \times 10^{-5},
\end{equation}
 which is
approximately a factor $10 - 20$ 
smaller than the CKM-allowed branching ratio $\BBGAMAXS$, measured by CLEO
\cite{CLEOrare2}.

%

\vspace*{3.0ex}
\subsection{ Estimates of ${\cal B}(B \to V + \gamma )$
 and constraints on the CKM parameters}

\par
Exclusive radiative
 $B$ decays $B \to V + \gamma$, with $V=K^*,\rho,\omega$, are also 
potentially
very interesting from the point of view of determining the CKM parameters
\cite{abs93}. The extraction of these parameters would, however,  involve a 
trustworthy 
estimate of the SD- and LD-contributions in the decay amplitudes.
\par
  The SD-contribution in the 
 exclusive decays $(B_u, B_d) \to (K^*,\rho) + \gamma$,
$B_d \to \omega + \gamma$  and the
corresponding $B_s$ decays, $B_s \to (\phi,K^*) + \gamma $,
involve the magnetic moment operator ${\cal O}_7$ and the related one 
obtained by the obvious change $s \to d$, $\hat{O}_7$.
The transition form factors governing the radiative $B$ decays
 $B \to V + \gamma$ can be generically  defined as:
\be
 \langle V,\lambda |\frac{1}{2} \bar \psi \sigma_{\mu\nu} q^\nu b
 |B\rangle  =
     i \epsilon_{\mu\nu\rho\sigma} e^{(\lambda)}_\nu p^\rho_B p^\sigma_V
F_S^{B\rightarrow V}(0).
\label{defF}
\ee
Here $V$ is a vector meson
with the polarization vector $e^{(\lambda)}$,
$V=\rho, \omega, K^*$ or $\phi$;
$B$ is a generic
$B$-meson $B_u, B_d$ or $B_s$, and $\psi$ stands for the
field of a light $u,d$ or $s$ quark. The vectors $p_B$, $p_V$ and
$q=p_B-p_V$
correspond to the 4-momenta of the initial $B$-meson and the
outgoing vector
meson and photon, respectively. In (\ref{defF}) the QCD
renormalization of the $\bar \psi \sigma_{\mu\nu} q^\nu b$ operator
is implied.
 Keeping only the SD-contribution 
 leads to obvious relations among the exclusive 
decay rates, exemplified here by the decay
rates for $(B_u,B_d) \to \rho + \gamma$ and $(B_u,B_d) \to K^* + \gamma$:
\be
\frac{\Gamma ((B_u,B_d) \to \rho + \gamma)}
     {\Gamma ((B_u,B_d) \to K^* + \gamma)} 
  = \frac{\vert \xi_t \vert^2}{\vert\lambda_t \vert ^2}
      \frac{\vert F_S^{B \to \rho }(0)\vert^2}
          {\vert F_S^{B \to K^* }(0)\vert^2} \Phi_{u,d}
  \simeq \kappa_{u,d}\left[\frac{\Vtdabs}{\Vtsabs}\right]^2 \,,
\label{SMKR}
\ee
where $\Phi_{u,d}$ is a phase-space factor which in all cases is close to 1
and
\begin{equation}
 \kappa_{i} \equiv [\frac{F_S^{B_i \to \rho \gamma}}{F_S^{B_i \to K^* 
\gamma}}]^2
\end{equation}
 is the ratio of the (SD) form factors squared.
The transition form factors $F_S$
 are model dependent, and since the exclusive-to-inclusive ratio $R_{K^*}$ 
has been measured,
one could use data to distinguish among models.
This aside,  the ratios of the form factors, i.e. $\kappa_i$,
 should be more reliably calculable as they depend
essentially only on the SU(3)-breaking effects.
 If the SD-amplitudes were the only contributions, the measurements of the
 CKM-suppressed radiative decays $(B_u,B_d) \to \rho + \gamma ,
~B_d \to \omega + \gamma$ and $B_s \to K^* + \gamma$ could be
used in conjunction with the decays $(B_u,B_d) \to K^* + \gamma$ to 
determine 
the CKM parameters. The present experimental upper limits on the CKM ratio
$\Vtdabs/\Vtsabs$ from radiative $B$ decays 
are indeed based on this assumption, yielding at 90\% C.L.\cite{Tomasz95}:
\be
\left\vert {V_{td} \over V_{ts}} \right\vert \leq 0.64 - 0.76~,
\ee
depending on the models used for the $SU(3)$ breaking effects
in the form factors \cite{abs93,bksnsr}.

  The possibility of significant
LD-contributions in
radiative $B$ decays from the light quark intermediate states
has been raised in a number of papers
\cite{bsgamld} -- \cite{Ricciardi}.
Their amplitudes necessarily involve other CKM matrix elements and hence the
simple factorization of the decay rates in terms of the CKM factors
involving $\Vtdabs$ and $\Vtsabs$ no longer holds thereby
 invalidating the relationships
given above. The discussion about the LD-contribution that we presented
in the context of the inclusive decays $\BGAMAXS$ applies in its essence
also for the exclusive decays, such as $B \to K^* + \gamma$, with
appropriate modifications. In line with this discussion, we shall
assume that the LD-contributions are small also in exclusive
decays $B \to K^* + \gamma$.
 In what follows, we discuss some
 charged and neutral exclusive $B$-decays,
$B^\pm \to \rho^\pm \gamma$ and
 $B$-decays $B \to (\rho^0,\omega) \gamma$,
involving the CKM-suppressed transitions and estimate the
SD- and LD-contributions.

\begin{figure}[htb]
\vspace{0.10in}
\vspace*{-3cm}
\centerline{
\epsfysize=6in
{
\epsffile{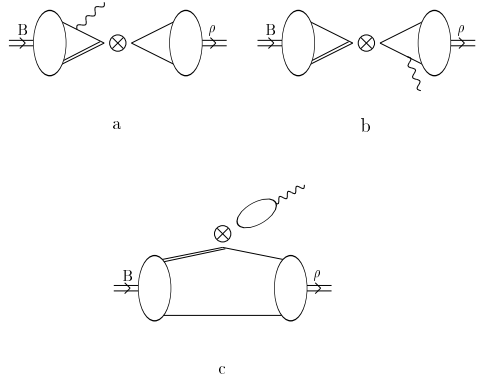}
}
}
\vspace{0.08in}
\vspace*{-8cm}
\caption[]{
 Weak annihilation contributions in $B_u \to \rho  \gamma$
involving the operators $O^\prime_1$ and $O^\prime_2$ denoted by 
$\bigotimes$ with the
photon emission from a) the loop containing the $b$ quark, b) the loop
containing the light quark, and c) the tadpole which contributes only
with additional gluonic corrections.
\label{abfig1}}
\end{figure}

The LD-contributions in $B \to V + \gamma$  are
induced by the matrix elements of the
four-Fermion operators $\hat{O}_1$ and $\hat{O}_2$ (likewise $O_1$ and 
$O_2$). Estimates of these contributions
 require non-perturbative methods.
 This problem has been investigated recently
in \cite{wyler95,ab95} using 
 a technique \cite{BBK89}
which treats the photon emission from the light quarks in a theoretically
consistent and model-independent way. This has been combined
with the light-cone QCD sum rule approach to calculate both the SD and LD
--- parity conserving and parity violating --- amplitudes
in the decays $B_{u,d} \to \rho(\omega) + \gamma$.
To illustrate this, we concentrate on the $B_u^\pm$ decays,
$B_u^\pm \to \rho^\pm + \gamma$ and take up the neutral $B$ decays
$B_d \to \rho (\omega) + \gamma$ at the end.
The LD-amplitude of the four-Fermion operators $\hat{O}_1$, $\hat{O}_2$
is dominated by  the
 contribution of the weak annihilation
of valence quarks in the $B$ meson. It is color-allowed for the
decays of charged $B^\pm$ mesons, as shown in Fig. ~\ref{abfig1}, where 
also the tadpole diagram is shown, which, however, contributes only in the
presence of gluonic corrections, and hence neglected.
In the factorization approximation, one may write the dominant contribution
in the operator $\hat{O_2}$ (here $O^\prime_2$ is the part of $\hat{O_2}$ 
with the CKM factor $\xi_u/\xi_t$ in Eq.~(\ref{basis})
\begin{equation}\label{factor}
\langle \rho\gamma | O^\prime_2|B\rangle =
\langle \rho | \bar d \Gamma_\mu u|0\rangle
\langle \gamma | \bar u \Gamma^\mu b|B\rangle
+
\langle \rho\gamma | \bar d \Gamma_\mu u|0\rangle
\langle 0 | \bar u \Gamma^\mu b|B\rangle ~,
\end{equation}
and make use of the definitions of the decay constants
\begin{eqnarray}
\langle 0 | \bar u \Gamma_\mu b|B\rangle  & =& i p_\mu f_B,
\nonumber\\
\langle \rho | \bar d \Gamma_\mu u|0\rangle &=&
\varepsilon^{(\rho)}_\mu m_\rho f_\rho,
\end{eqnarray}
to reduce the problem at hand to the calculation of simpler form factors
induced by vector and axial-vector currents.
%
%

The factorization approximation assumed in \cite{wyler95,ab95} has
not been tested experimentally in
radiative $B$ decays. From a theoretical point of view,
non-factorizable
contributions belong to either the $O(\alpha_s)$ (and higher order) 
radiative corrections or to contributions of higher-twist operators to
the sum rules. Their inclusion should not change the conclusions
substantially.

The LD-amplitude in the decay $B_u \to \rho^\pm + \gamma$ can be written 
in terms of the form factors $F_1^L$ and $F_2^L$,
\begin{eqnarray}\label{Along}
{\cal A}_{long} &=&
-\frac{e\,G_F}{\sqrt{2}} V_{ub}V_{ud}^\ast
\left( C_2+\frac{1}{N_c}C_1\right) m_\rho
\varepsilon^{(\gamma)}_\mu \varepsilon^{(\rho)}_\nu
\nonumber\\&&{}\times
 \Big\{-i\Big[g^{\mu\nu}(q\cdot p)- p^\mu q^\nu\Big] \cdot 2 F_1^{L}(q^2)
  +\epsilon^{\mu\nu\alpha\beta} p_\alpha q_\beta
 \cdot 2 F_2^{L}(q^2)\Big\}\,.
\end{eqnarray}
Again, one has to invoke a model to calculate the form factors. Estimates
 from the light-cone QCD sum rules give
\cite{ab95}:
\begin{equation}\label{result}
 F^L_1/F_S = 0.0125\pm 0.0010\,,\quad F^L_2/F_S = 0.0155\pm 0.0010 ~,
\end{equation}
where the errors correspond to the variation of the
Borel parameter in the QCD sum rules. Including other possible 
uncertainties, 
 one expects an accuracy of the ratios in (\ref{result}) of order 20\%.
Estimates of $F_S$ in the normalization of Eq. (\ref{defF}) range
between $F_S(B \to K^* \gamma) = 0.31$ (Narison in \cite{bksnsr}) to
$F_S(B \to K^* \gamma) = 0.37$ (Ball in \cite{bksnsr}), with a typical
error of $\pm 15\%$, and hence are all consistent with each other.
This, for example, gives $R_{K^*}=0.16 \pm 0.05$,
using the result from \cite{abs93}, which is in good agreement with data.

Returning to the discussion of the LD-contribution, we note that 
 the parity-conserving and parity-violating amplitudes turn out
to be numerically close to each other, $F_1^L\simeq F^L_2 \equiv F_L$,
hence the ratio of the LD- and the SD- contributions reduces to a number 
\cite{ab95}
 \begin{equation}\label{ratio2p}
{\cal A}_{long}/{\cal A}_{short}=
R_{L/S}^{B_u\to\rho\gamma}
\cdot\frac{V_{ub}V_{ud}^\ast}{V_{tb}V_{td}^\ast} ~.
\end{equation}
Using $C_2=1.10$, $C_1=-0.235$, $C_7^{\mathit{eff}}=-0.306$
from Table \ref{wcmudep} (corresponding to the scale $\mu=5$ GeV)
 \cite{ag95} gives:
\begin{equation}\label{result2}
R_{L/S}^{B_u\to\rho\gamma} \equiv
 \frac{4 \pi^2 m_\rho(C_2+C_1/N_c)}{m_b C_7^{\mathit{eff}}}
\cdot\frac{F_L^{B_u \to \rho \gamma}}{F_S^{B_u \to \rho \gamma}}=-0.30\pm 
0.07 ~, \end{equation}
which is not negligible.
 To get a ball-park estimate of the ratio
${\cal A}_{long}/{\cal A}_{short}$, we take the central values of
the CKM matrix elements,
 $V_{ud}=0.9744\pm 0.0010$ \cite{PDG94},
$|V_{td}|=(1.0\pm 0.2)\times 10^{-2}$,
$|V_{cb}|=0.039\pm 0.004$ and $|V_{ub}/V_{cb}|=0.08\pm 0.02$ \cite{alpisa95},
 yielding,
\begin{equation}
|{\cal A}_{long}/{\cal A}_{short}|^{B_u\to\rho\gamma}
= |R_{L/S}^{B_u\to\rho\gamma}|
\frac{|V_{ub}V_{ud}|}{|V_{td}V_{bt}|} \simeq 10\% ~.
\end{equation}
Thus, the CKM factors suppress the LD-contributions.

The analogous LD-contributions to the neutral $B$ decays
$B_d\to\rho\gamma $ and $B_d\to\omega\gamma $ are
expected to be much smaller, a point
that has also been noted in the context of the VMD and quark model
based estimates \cite{bsgamld}. In the present approach,
 the corresponding form factors for the decays
$B_d \to \rho^0(\omega)  \gamma$ are obtained from
the ones for the decay $B_u\to\rho^\pm \gamma$ discussed above by the
replacement of the light quark charges
 $e_u\to e_d$, which gives the factor $-1/2$; in addition,
and more importantly, the
LD-contribution to the neutral $B$ decays
is colour-suppressed, which reflects itself
through the replacement of the factor
$a_1$  by $a_2$. This yields for the ratio
\begin{equation}
\frac{R_{L/S}^{B_d\to\rho\gamma}}{R_{L/S}^{B_u\to\rho\gamma}}=
\frac{e_d a_2}{e_u a_1} \simeq -0.13 \pm 0.05 ,
\end{equation}
where the numbers are based on using
$a_2/a_1 = 0.27 \pm 0.10$ \cite{BH95}. This would then yield at most
$R_{L/S}^{B_d\to\rho\gamma} \simeq R_{L/S}^{B_d\to\omega\gamma}=0.05$,
which in turn gives
\begin{equation}
 \frac{{\cal A}_{long}^{B_d\to\rho\gamma}}{{\cal 
A}_{short}^{B_d\to\rho\gamma}}\leq 0.02.
\end{equation}
 Even if this underestimates the LD-contribution by a factor 
2, due to the approximations made in \cite{wyler95,ab95},
we conclude that it is quite safe to neglect
the LD-contribution in the neutral $B$-meson radiative decays.

\begin{figure}[htb]
\vskip -2.4truein
\centerline{\epsfysize=7in
{\epsffile{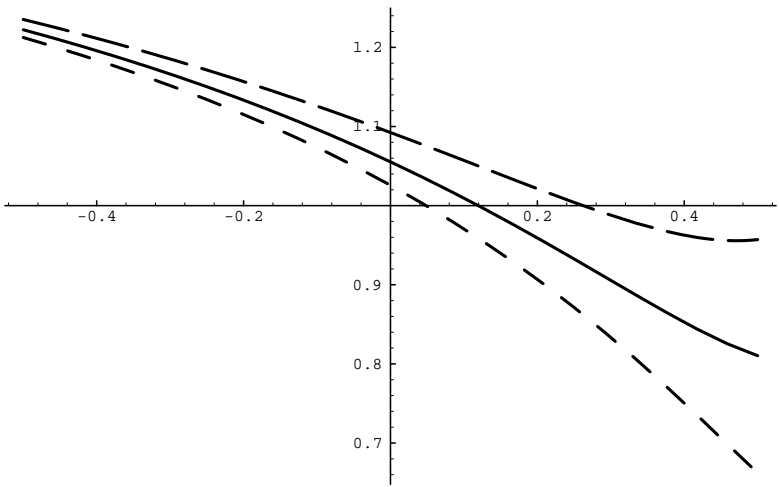}}}
\vskip -1.0truein
\caption[]{
 Ratio of the neutral and charged $B$-decay rates
 $\Gamma (B_u \to \rho \gamma)/2\Gamma (B_d \to \rho \gamma)$ as a function
of the Wolfenstein parameter $\rho$, with $\eta =0.2$ (short-dashed curve),
$\eta =0.3$ (solid curve), and $\eta =0.4$ (long-dashed curve). (Figure taken
from \protect\cite{ab95}.)
\label{abfig2}}
\end{figure}

Restricting to the colour-allowed LD-contributions, the relations, which
obtains ignoring such contributions (and isospin invariance),
\beq\label{ratio2}
\Gamma(B_{u} \to \rho^+ \gamma)=2 ~\Gamma(B_{d} \to \rho^0  \gamma)
    = 2 ~\Gamma (B_{d} \to \omega  \gamma)~,
\eeq
get modified to
\begin{eqnarray}\label{ratio5}
\lefteqn{\frac{\Gamma(B_u\to \rho\gamma)}{2\Gamma(B_d\to \rho\gamma)}
=\frac{\Gamma(B_u\to \rho\gamma)}{2\Gamma(B_d\to \omega\gamma)}
 =\left|1+R_{L/S}^{B_u\to\rho\gamma}
\frac{V_{ub}V_{ud}^\ast}{V_{tb}V_{td}^\ast}\right|^2 =
}
\nonumber\\&&{}
=1+2\cdot R_{L/S} V_{ud}\frac{\rho(1-\rho)-\eta^2}{(1-\rho)^2+\eta^2}
+(R_{L/S})^2 V_{ud}^2\frac{\rho^2+\eta^2}{(1-\rho)^2+\eta^2}\,.
\end{eqnarray}
where $R_{L/S}\equiv R_{L/S}^{B_u\to\rho\gamma}$.  
The ratio
$\Gamma(B_u\to \rho\gamma)/2\Gamma(B_d\to \rho\gamma)
(=\Gamma(B_u\to \rho\gamma)/2\Gamma(B_d\to \omega\gamma))$
is shown in Fig. ~\ref{abfig2}
 as a function of the parameter $\rho$, with
 $\eta= 0.2, ~0.3$  and $0.4$.
This suggests that
a measurement of this ratio would constrain the Wolfenstein parameters
$(\rho, \eta)$, with the dependence on $\rho$ more marked
than on $\eta$. In particular,
a negative value of $\rho$ leads to a
 constructive interference in
$B_u\to\rho\gamma$ decays, while large positive values of $\rho$ give 
a destructive interference. This behaviour is in qualitative agreement
with what has been also pointed out in \cite{Cheng}.

\par
\begin{figure}[htb]
\vskip -2.4truein
\centerline{\epsfysize=7in
{\epsffile{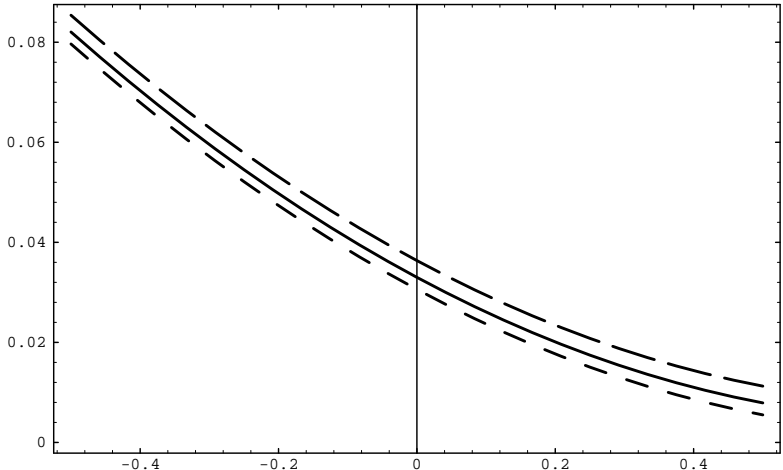}}}
\vskip -1.0truein
\caption[]{
 Ratio of the CKM-suppressed and CKM-allowed radiative $B$-decay
rates\\
$\Gamma (B_u \to \rho \gamma)/\Gamma (B \to K^* \gamma)$ (with $B=B_u$ or
$B_d$) as a function of the Wolfenstein parameter $\rho$,
a) with $\eta =0.2$ (short-dashed curve), $\eta =0.3$ (solid curve), and
$\eta =0.4$ (long-dashed curve). (Figure taken from \protect\cite{ab95}.)
\label{abfig3}}
\end{figure}

The ratio of the CKM-suppressed and CKM-allowed
 decay rates  for charged $B$ mesons
 gets modified due to the LD contributions. Following \cite{GP95},
we ignore the LD-contributions in $\Gamma(B \to K^*\gamma)$. The ratio of
the decay rates in question can therefore be written as:
\begin{eqnarray}\label{ratio3}
\lefteqn{\frac{\Gamma(B_u\to \rho\gamma)}{\Gamma(B_u\to K^*\gamma)}
= \kappa_u \lambda^2[(1-\rho)^2+\eta^2]
}
\nonumber\\&&{}
\times\Bigg\{
1+2\cdot R_{L/S} V_{ud}\frac{\rho(1-\rho)-\eta^2}{(1-\rho)^2+\eta^2}
+(R_{L/S})^2 V_{ud}^2\frac{\rho^2+\eta^2}{(1-\rho)^2+\eta^2}\Bigg\}\,,
\end{eqnarray}
 Using the central value from the estimates of the ratio of the
  form factors squared 
$\kappa_u=0.59 \pm 0.08$
  \cite{abs93}, we show the ratio (\ref{ratio3}) in Fig. 
~\ref{abfig3} as a function of $\rho$ for $\eta=0.2,0.3$, and $0.4$.
It is seen that the dependence of this ratio is rather weak on $\eta$
but it depends on $\rho$ rather sensitively.
The effect of the LD-contributions is modest but not negligible, introducing
an uncertainty  
comparable to the $\sim 15\%$ uncertainty in the overall normalization
due to the $SU(3)$-breaking effects in the quantity $\kappa_u$.
%
%
%
%

\indent
Neutral $B$-meson radiative decays are less-prone to the LD-effects,
 as argued above, and hence one expects that to a good approximation
the ratio of the decay rates for neutral $B$ meson obtained in the
approximation of SD-dominance remains valid \cite{abs93}:
\begin{equation}
\frac{\Gamma(B_d\to \rho\gamma,\omega\gamma)}{\Gamma(B\to K^*\gamma)}
 = \kappa_d\lambda^2 [(1-\rho)^2+\eta^2]~,
\end{equation}
where this relation holds for each of the two decay modes separately.
 It is a realistic hope that this relation is 
theoretically (almost) on the same footing in the standard model 
as the one for the ratio of the $B^0$-$\overline{B^0}$ mixing-induced
mass differences, which satisfies the relation \cite{al95}:
\be
\frac{\delms}{\delmd} = \kappa_{sd}
\left\vert \frac{V_{ts}}{V_{td}} \right\vert^2
= \kappa_{sd}\frac{1}{\lambda^2 [(1-\rho)^2+\eta^2]} ~.
\ee
The hadronic uncertainty in this ratio 
 is in the SU(3)-breaking factor $\kappa_{sd}\equiv (f_{B_s}^2 \hat{B}_{B_s}/
f_{B_d}^2 \hat{B}_{B_d})$, which involves the pseudoscalar coupling
constants and the so-called bag constants. This quantity is
 estimated as $\kappa_{sd}=1.35 \pm 0.25$ in the QCD sum 
rules and lattice QCD approaches. (For details and references, see 
\cite{al95}).
The present upper limit for the mass-difference ratio $\delms/\delmd > 12.3
$ at 95 \% C.L. from the ALEPH data \cite{ALEPHxs} provides a better
constraint on the CKM parameters, yielding $\Vtdabs/\Vtsabs < 0.35$,
 than the corresponding constraints from the
rare radiative decays $B \to (\rho,\omega) + \gamma$, which give an upper 
limit of 0.75 for the same CKM-ratio. We expect experimental sensitivity to
increase in both measurements,
reaching the level predicted for this ratio 
in the standard model, $\Vtdabs/\Vtsabs =0.24 \pm 0.05$ \cite{al95},
in the next several years in the ongoing experiments at CLEO, SLC and 
Tevatron, and the forthcoming ones at the $B$ factories, HERA-B and LHC.  

\indent

 Finally, combining the estimates for the LD- and SD-form factors in
\cite{ab95} and
\cite{abs93}, respectively, and restricting the Wolfenstein
parameters in the range $-0.4 \leq \rho \leq 0.4$ and $ 0.2 \leq \eta
\leq 0.4$, as suggested by the CKM-fits \cite{al95}, we give the
following ranges for the absolute branching ratios:
\begin{eqnarray}\label{ratio4}
{\cal B}(B_u\to \rho\gamma)
&=& (1.9 \pm 1.6) \times 10^{-6} ~,
\nonumber\\
{\cal B}(B_d\to \rho\gamma) &\simeq& {\cal B}(B_d \to \omega \gamma)
= (0.85 \pm 0.65) \times 10^{-6} ~,
\end{eqnarray}
where we have used the experimental value for the branching ratio
${\cal B} (B \to K^* + \gamma)$
\cite{CLEOrare1},
adding the errors in quadrature. The large error reflects the poor
knowledge of the CKM matrix elements and hence experimental determination
of these branching ratios will put rather stringent constraints on the
Wolfenstein parameter $\rho$.

Summarizing the effect of the LD-contributions in
radiative $B$ decays, we note that they are
dominantly given by the annihilation diagrams. QCD sum-rule-based 
estimates are encouraging in that they strengthen the hope
that such contributions are modest in exclusive radiative $B$ decays,
in particular in the neutral $B$-decays $B^0 \to (\rho^0,\omega) + \gamma$,
with $B^\pm \to
\rho^\pm \gamma$ modified by $O(20)\%$ from its SD-rate.
This should be checked in other theoretically sound frameworks.
Of course, forthcoming data on charged and neutral $B$-meson decays will be 
able to determine the LD-contribution directly. 
Presently available data  in general suggest that
the contribution  of annihilation diagrams in $B$ decays is not significant,
 as seen through the near equality of the lifetimes
for the $B^\pm, ~B_d^0$ and $B_s^0$ mesons and the near equality of 
the observed $B^\pm$ and $B^0$ radiative decay rates. In terms of the
operator product expansion, they all involve the four-Fermi operators,
sandwiched between various inclusive and exclusive states. The matrix 
elements are process-dependent and have to be estimated as well as possible.
Some illustrative examples were given here. 
%
%
%

\subsection{Inclusive rare decays $B \to X_s \ell^+ \ell^-$ in the SM}

\par
The decays \bxsll, with $\ell=e,\mu,\tau$, provide a more sensitive search
strategy for finding new physics in rare $B$ decays
than for example the decay \bxsg , which constrains
only the magnitude of $C_7^{\mathit{eff}}$. This experimental constraint has
triggered a lot of theoretical investigations on beyond-the-SM frameworks,
as can be judged from the incomplete list of papers in 
\cite{hewett,giudice}. The sign of $C_7^{\mathit{eff}}$, which
 depends on the underlying physics, is not
determined by the measurement of ${\cal B}(\BGAMAXS)$. This sign in our
convention is negative in the SM (see Table \ref{wcmudep}).
 However, it is known (see for example \cite{AGM94}) that
in supersymmetric models, both the negative and positive signs are allowed
as one scans over the allowed parameter space of this model.
The determination of the sign of $C_7^{\mathit{eff}}$ is an important matter
as this will impose further constraints on the parameters of many models.
It will also test the prediction of the SM, by the same token.

\par
  Continuing this discussion, we recall that
a part of the amplitude for \bxsll  ~involving the coupling of the virtual 
photon to the charged lepton pair depends on the
effective Wilson coefficient $C_7^{\mathit{eff}}$ encountered in the 
electromagnetic
penguin decays $\BGAMAXS$. For low dilepton masses, the differential decay 
rate
for \bxsll ~is dominated by this contribution.
However,  as we shall see below, the \bxsll ~amplitude in
the standard model
has two additional terms, arising from the two FCNC four-Fermi operators,
 \footnote{This also
holds for a large class of models such as MSSM and two-Higgs doublet
models but not for all. In LR symmetric models, for example, there
are additional FCNC four-Fermi operators involved \cite{LRsymmetry}.}
which are not constrained by the $\BGAMAXS$ data.  
Calling their coefficients $C_{9}$ and $C_{10}$, it has been argued in
\cite{AGM94} that the signs and
magnitudes of all three coefficients $C_7^{\mathit{eff}}$, $C_{9}$ and 
$C_{10}$
can, in principle,  be determined from the decays $\BGAMAXS$ and \bxsll .
The coefficient $C_8$, which governs the strength of the chromomagnetic
moment transition, can, in principle, also be determined by measuring
the photon energy spectrum in $\BGAMAXS$ in low-to-intermediate photon
energy region.  

\par
 The SM-based rates for the decay \bsll , calculated in the free quark decay
approximation, have been known in the LO approximation for some time
\cite{BSGAM}. The LO calculations have the unpleasant
feature that the decay distributions and rates are scheme-dependent.
 The required NLO calculation is in the meanwhile
available, which reduces the scheme-dependence of the LO effects in these
decays \cite{Misiak1,BM94}. In addition,
long-distance (LD) effects, which are expected to be very important in the
decay \bxsll  \cite{long}, have also been estimated from data
 on the assumption that they arise dominantly due to
the charmonium resonances $ J/\psi$ and $\psi'$ through the decay chains
$B \rightarrow X_s J/\psi (\psi',...) \rightarrow X_s \ell^+ \ell^-$.
The effect of these resonances persists even far away from the resonant
masses deforming the short-distance based distributions appreciably
\cite{long}.
Likewise, the  leading $(1/{m_b}^2)$ power corrections
to the partonic decay rate and the dilepton invariant mass distribution
have been calculated with the help of the operator product expansion in the 
effective heavy quark theory
\cite{falketalbsll}.
A theoretically complete description including the
improved perturbative treatment of the decay \bxsll , the LD-effects
and power corrections mentioned above, to the
best of our knowledge, is still lacking.
 In what follows, we shall 
not take into account the power corrections  but include the 
rest to get estimates for the decay rates and distributions in \bxsll.
 
The amplitude for \bxsll is calculated in the effective theory
approach, which we have discussed earlier,  by
extending the operator basis of the effective Hamiltonian
introduced in Eq.~(\ref{heffbsg}):
\begin{equation}\label{heffbsll}
{\cal H}_{eff}(b \to s + \gamma ; b \to s + \ell^+\ell^- )
  = {\cal H}_{eff} (b \to s + \gamma) -\frac{4 G_F}{\sqrt{2}} V_{ts}^* V_{tb}
\left[ C_9 {\cal O}_9 +C_{10}{\cal O}_{10} \right],
\end{equation}
where the two additional operators are:
\begin{eqnarray}
{\cal O}_9 &=& \frac{\alpha}{4 \pi} \bar{s}_\alpha \gamma^{\mu} P_L b_\alpha 
\bar{\ell} \gamma_{\mu} \ell , \nonumber\\
{\cal O}_{10} &=& \frac{\alpha}{4 \pi} \bar{s}_\alpha \gamma^{\mu} P_L 
b_\alpha \bar{\ell} \gamma_{\mu}\gamma_5 \ell ~.
\end{eqnarray}

The analytic expressions for $C_{9}(m_W)$ and $C_{10}(m_W)$ can be seen
in \cite{BBL95,AGM94} and will not be given here.
 We recall that the
coefficient $C_9$ in LO is scheme-dependent. However, this is compensated
by an additional scheme-dependent part in the
(one loop) matrix element of ${\cal O}_9$. We call the
sum  $C_9^{\mathit{eff}}$, which is scheme-independent and enters in the 
physical decay amplitude given below,
                
\begin{eqnarray}
\lefteqn{{\cal M}(b \to s +\ell^+\ell^-) =
 \frac{4 G_F}{\sqrt{2}} V_{ts}^* V_{tb}\frac{\alpha}{\pi}} \nonumber\\
& &\left[ C_9^{\mathit{eff}}\bar{s} \gamma^{\mu} P_L b \bar{\ell} 
\gamma_{\mu} \ell
 +C_{10}\bar{s} \gamma^{\mu} P_L b \bar{\ell} \gamma_{\mu}\gamma_5 \ell
- 2C_7^{\mathit{eff}} \bar{s} i\sigma_{\mu \nu} 
\frac{q^\nu}{q^2}(m_bP_R+m_sP_L)b
\bar{\ell} \gamma^{\mu} \ell \right],\nonumber\\
&& {}
\end{eqnarray}
with
\begin{equation}
C_9^{\mathit{eff}} (\hat{s}) \equiv C_9\eta({\hat{s}}) + Y(\hat{s}).
\end{equation}
The function $Y(\hat{s})$ is the one-loop matrix element of ${\cal O}_9$
 and is defined as:
 \begin{eqnarray}   
        Y(\s) & = & g(\mc,\s)
                \left(3 \, C_1 + C_2 + 3 \, C_3
                + C_4 + 3 \, C_5 + C_6 \right) 
\nonumber \\
        & & - \frac{1}{2} g(1,\s)
                \left( 4 \, C_3 + 4 \, C_4 + 3 \,
                C_5 + C_6 \right) \nonumber \\
        & & - \frac{1}{2} g(0,\s) \left( C_3 +
                3 \, C_4 \right) \nonumber \\
        & &     + \frac{2}{9} \left( 3 \, C_3 + C_4 +
                3 \, C_5 + C_6 \right) \nonumber \\
        & &     - \xi \, \frac{4}{9} \left( 3 \, C_1 +
                C_2 - C_3 - 3 \, C_4 \right),
                \label{eqn:y} \\
        \eta(\s) & = & 1 + \frac{\alpha_s(\mu)}{\pi}
                \omega(\s) ~. 
\end{eqnarray}
Here $\s \equiv (\hat{p}_+ + \hat{p}_-)^2/m_b^2$ is the scaled dilepton 
mass,
 with
$p_\pm = (E_\pm, \mbox{\boldmath $p_\pm$})$ denoting four-momenta of 
$\ell^\pm$, $\hat{m_i}$ are the scaled quark masses, $\hat{m_i}=m_i/m_b$,
 and the function $\omega(\hat{s})$ represents the
$O(\alpha_s)$ correction from one-gluon exchange in the matrix element of 
${\cal O}_9$, the analogue of
which we have discussed in the context of the semileptonic $B$ decays
earlier, derived in \cite{JK89}:
\begin{eqnarray}
\omega(\hat{s}) &=& -\frac{2}{9}\pi^2 -\frac{4}{3}{\mbox Li}_2(s)-\frac{2}{3}
\ln \hat{s} \ln(1-\hat{s}) - 
\frac{5+4\hat{s}}{3(1+2\hat{s})}\ln(1-\hat{s})\nonumber\\
&-& \frac{2\hat{s}(1+\hat{s})(1-2\hat{s})}{3(1-\hat{s})^2(1+2\hat{s})}
\ln \hat{s} + \frac{5 + 9\hat{s} -6\hat{s}^2}{6(1-\hat{s})(1+2 \hat{s})}~.
\label{omegahats}
\end{eqnarray}

 The function $g(z,\hat{s})$ includes the
charm quark-antiquark pair contribution 
\cite{Misiak1,BM94}:
 \begin{eqnarray}
g(z,\hat{s}) &=& -\frac{8}{9}\ln (\frac{\mb}{\mu})
 -\frac{8}{9} \ln z + \frac{8}{27} +\frac{4}{9}y
-\frac{2}{9}(2 + y) \sqrt{\vert 1-y \vert}\nonumber\\
&\times & \left[\Theta(1-y)(\ln\frac{1+\sqrt{1-y}}{1-\sqrt{1-y}} -i\pi )
+\Theta(y-1) 2 \arctan \frac{1}{\sqrt{y-1}} \right] ~,
\end{eqnarray}
\begin{equation}
g(0,\hat{s}) = \frac{8}{27}-\frac{8}{9}\ln (\frac{\mb}{\mu})
              -\frac{4}{9}\ln \hat{s} + \frac{4}{9}i\pi ~,
\end{equation}
where $y=4z^2/\hat{s}$.
We recall from the discussion in \cite{Misiak1,BM94} that
$\xi$ is dependent on the
dimensional regularization scheme, with, \begin{equation}
        \xi = \left\{
                \begin{array}{ll}
                        0       & \mbox{(NDR)} \\   
                        -1      & \mbox{(HV)}.
                \end{array}
                \right.
\end{equation}
For numerical estimates, the Wilson coefficients $C_1,...C_6$ and
 $C_7^{\mathit{eff}}$
are given in Table \ref{wcmudep} and we 
 give here the value for $C_9$ (in the NDR-scheme) and $C_{10}$
for the central values of the parameters used in Table \ref{wcmudep} 
($\overline{m_t} =170$ GeV and $\Lambda_{QCD}=0.195$ GeV):
\begin{eqnarray}\label{c910}
 C_9^{NDR}(\mu=5.0^{+5.0}_{-2.5} ~\mbox{GeV}) &=& 
4.09^{+0.27}_{-0.41},  \nonumber\\
C_{10} (m_W) &=& -4.32 ~.
\end{eqnarray}
Note that the coefficient $C_{10}(m_W)$ does not get renormalized by
QCD corrections. 

\par
 Following refs. \cite{AGM94} and \cite{amm91}, 
the differential decay rates in \bxsll (ignoring lepton masses) are,
\begin{eqnarray}
	{{\rm d}{\cal B}(\hat{s}) \over {\rm d}\hat{s}} & = &
		{\cal B}_{sl} \frac{\alpha^2}{4 \pi^2} \frac{ 
		\lambda_t^2}{\Vcbabs^2} \frac{1}{f(\hat{m}_c) \kappa(\hat{m}_c)}
		u (\hat{s}) \left[ \vphantom{\frac{1}{1}}
		\left( |C_9^{\mathit{eff}}(\hat{s})|^2
 		+ C_{10}^2 \right) \alpha_1 (\hat{s},\hat{m}_s)
		\right. \nonumber \\
& & \left. + \frac{4}{\hat{s}} (C^{eff}_7)^2 \alpha_2 (\hat{s},\hat{m}_s)
+ 12 \alpha_3 (\hat{s},\hat{m}_s) C^{eff}_7 {\cal 
\Re}(C_9^{\mathit{eff}}(\hat{s}))
		\right] ,
	\label{eqn:dbrs}
\end{eqnarray}
with $u(\hat{s})=\sqrt{\left[\hat{s}-(1+\hat{m_s})^2\right]
\left[\hat{s}-(1-\hat{m_s})^2 \right]} $,
$f(z)$ has been defined earlier as we discussed $\Gamma_{\small SL}$,
$\kappa(z)=1-2\as(\mu)/3\pi \left[(\pi^2-31/4)(1-z)^2 + 3/2 \right]$, and 
 \begin{eqnarray}
	\alpha_1 (\hat{s},\hat{m}_s) & = &
		- 2 \hat{s}^2 + \hat{s} (1+ \hat{m}_s^2)
       		+(1-\hat{m}_s^2)^2 ,
		\label{eqn:alpha1} \\
	\alpha_2 (\hat{s},\hat{m}_s) & = &
		-(1+ \hat{m}_s^2) \hat{s}^2
           	- (1+14 \hat{m}_s^2+\hat{m}_s^4) \hat{s}
            	+ 2 (1+ \hat{m}_s^2)(1-\hat{m}_s^2)^2 ,
		\label{eqn:alpha2} \\
	\alpha_3 (\hat{s},\hat{m}_s) & = &
		(1-\hat{m}_s^2)^2 - (1+ \hat{m}_s^2) \hat{s}. 
		\label{eqn:alpha3}
\end{eqnarray}
Here ${\cal \Re}(C_7^{\mathit{eff}})$ represents the real part of 
$C_7^{\mathit{eff}}$.
A useful quantity is the  differential FB asymmetry in the c.m.s. of the
dilepton
defined in refs. \cite{amm91}:
\begin{equation}\label{FBasym}
\frac{d {\cal A}(\hat{s})}{d\hat{s}} = \int_0^1 \frac{d{\cal B}}{dz}
                                      -\int_0^{-1} \frac{d{\cal B}}{dz},
\end{equation}
where $z=\cos \theta$, which can be expressed as:
\begin{eqnarray}
	{{\rm d}{\cal A}(\hat{s}) \over {\rm d}\hat{s}} & = &
		- {\cal B}_{sl} \frac{3 \alpha^2}{4 \pi^2}
                \frac{1}{f(\hat{m}_c)} u^2 (\hat{s})
		C_{10} \left[ \hat{s}{\cal \Re} ( C_9^{\mathit{eff}}(\hat{s})) +
		2 C^{eff}_7 (1 + \hat{m}_s^2) \right] .
	\label{eqn:dasym}
\end{eqnarray}
 The Wilson coefficients
$C^{eff}_7$, $C^{eff}_9$ and $C_{10}$ appearing in the above equations
can be determined from data by solving the partial branching ratio
${\cal B}(\Delta \hat{s})$ and partial FB asymmetry
${\cal A}(\Delta \hat{s})$, where $\Delta \hat{s}$ defines an
interval in the dilepton invariant mass \cite{AGM94}.

 There are
other quantities which one can measure in the decays $B \to X_s \ell^+ 
\ell^-$ to disentangle the underlying dynamics.
 We mention here the longitudinal polarization
of the lepton in \bxsll, in particular in $B \to X_s \tau^+ \tau^-$,
proposed by Hewett \cite{Hewettpol}. In a recent paper, Kr\"uger and Sehgal
\cite{KS96} have stressed that complementary information is contained in
the two orthogonal components of polarization ($P_T$, the component in the
decay plane, and $P_N$, the component normal to the decay plane), both of
which are proportional to $m_\ell/m_b$, and therefore significant
for the $\tau^+ \tau^-$ channel. A third quantity, called energy asymmetry,
proposed by Cho, Misiak and Wyler \cite{CMW96}, defined as
\begin{equation}
{\cal A}=\frac{N(E_{\ell^-} > E_{\ell^+}) - N(E_{\ell^+} > E_{\ell^-})}
              {N(E_{\ell^-} > E_{\ell^+}) + N(E_{\ell^+} > E_{\ell^-})}~,
\end{equation}
where $N(E_{\ell^-} > E_{\ell^+})$ denotes the number of lepton pairs
where $\ell^+$ is more energetic than $\ell^-$ in the $B$-rest frame,
is, however, not an independent measure, as
it is directly proportional to the FB asymmetry discussed above. The relation
is \cite{AHHM96}:
\begin{equation}
\int {\cal A}(\hat{s})= {\cal B} \times A~.
\end{equation}
This is easy to notice if one writes the Mandelstam variable
$u(\hat{s})$ in the
dilepton c.m. and the $B$-hadron rest systems. 

\begin{figure}[htb]
\vskip -1.9truein
\centerline{\epsfysize=7in
{\epsffile{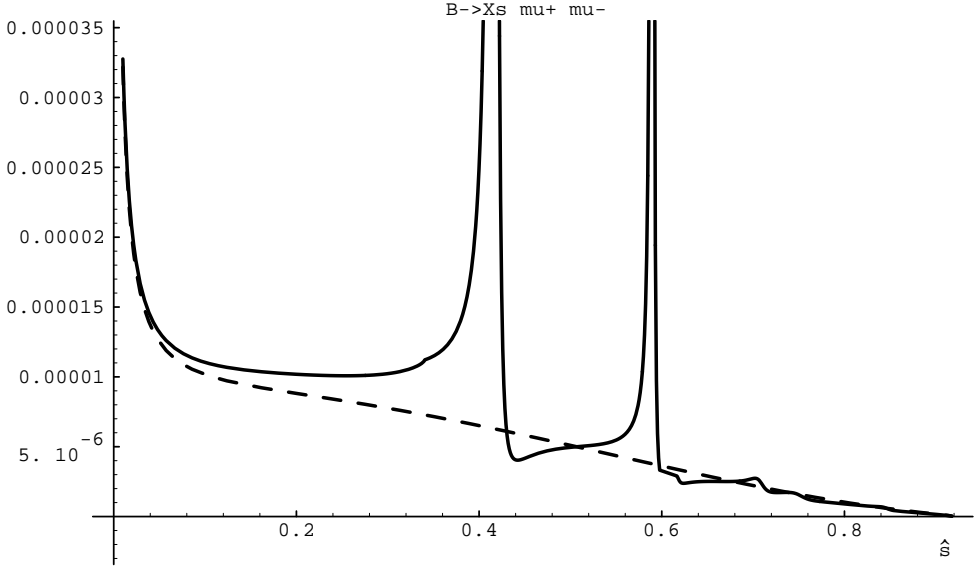}}}
\vskip -2.0truein
\caption[]{
Dimuon invariant mass distribution in $B \to X_s \mu^+ \mu^-$ in
 the SM including
next-to-leading order QCD correction. The dashed curve corresponds to the
short-distance contribution only and the solid curve is
the sum of the long-distance and short-distance contributions. (Figure taken
from \protect\cite{AHHM96}.)}
\label{fig:dbrnsm}
\end{figure}

\begin{figure}[htb]
\vskip -2.0truein
\centerline{\epsfysize=7in
{\epsffile{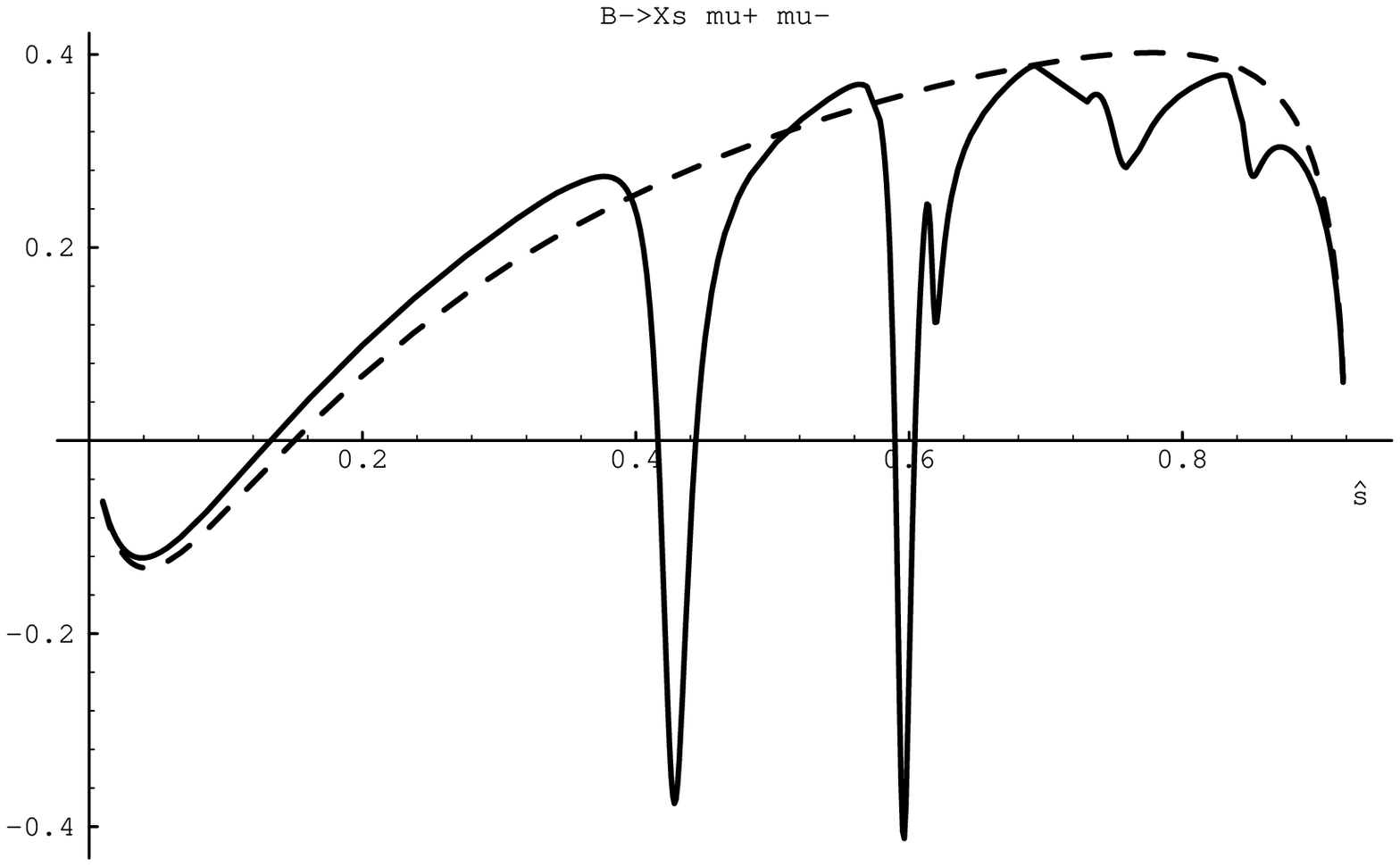}}}
\vskip -2.0truein
\caption[]{FB asymmetry for $B \to X_s \mu^+ \mu^-$ in the SM 
as a function of the dimuon invariant mass  including the
                next-to-leading order QCD correction.
The dashed curve corresponds to the
short-distance contribution only and the solid curve is
the sum of the long-distance and short-distance contributions. (Figure taken
from \protect\cite{AHHM96}.)}
\label{fig:asymmnsm}
\end{figure}

Next, we discuss the effects of LD contributions in the
processes $B \to X_s \ell^+ \ell^-$. Note that the
 LD contributions due to the vector mesons such as $J/\psi$ and 
$\psi^\prime$, as well as the continuum $c\bar{c}$ contribution already
discussed, 
appear as an effective $(\bar{s}_L \gamma_\mu b_L)(\bar{\ell} \gamma^\mu 
\ell)$ interaction term only, i.e. in the operator ${\cal O}_9$.
 This implies that the LD-contributions should change
$C_9$ effectively,  $C_7$ as discussed earlier is dominated by the
SD-contribution, and 
$C_{10}$ has no LD-contribution. In accordance with this, 
the function $Y(\hat{s})$ is replaced by,
\begin{equation}
	Y(\hat{s}) \rightarrow Y^\prime(\hat{s}) \equiv Y(\hat{s}) + 
		Y_{\mbox{res}}(\hat{s}),
\end{equation}
where $Y_{\mbox{res}}(\hat{s})$ is given as \cite{amm91},
\begin{equation}
	Y_{\mbox{res}}(\hat{s}) = \frac{3}{\alpha^2} \kappa 
		\left(3 C_1 + C_2 + 3 C_3 + C_4 + 3 C_5 + C_6 \right)
		\sum_{V_i = J/\psi, \psi^\prime,...}
		\frac{\pi \Gamma(V_i \rightarrow l^+ l^-) M_{V_i}}{
		M_{V_i}^2 - \hat{s} m_b^2 - i M_{V_i} \Gamma_{V_i}} ,
\end{equation}
where $\kappa$ is a fudge factor, which appears due to the inadequacy
of the factorization framework in describing data on $B \to J/\psi X_s$.
Here we use 
$\kappa \left( 3 C_1 + C_2 + 3 C_3 + C_4 + 3 C_5 + C_6 \right) = +0.88 $ 
for the numerical calculation, which reproduces (in average) the measured 
branching ratios for  $B \to J/\psi 
X_s$ and $B \to \psi' X_s$, after the contributions from the $\chi_c$ states
have been subtracted. This is consistent with the treatment of LD-effects
in \cite{KS96} and \cite{LW95}, where further 
discussions and theoretical uncertainties on the LD-contributions can be 
seen. The long-distance effects lead to significant interference effects
in the  dilepton invariant mass
distribution and the FB asymmetry in \bxsll shown in Figs. \ref{fig:dbrnsm}
and \ref{fig:asymmnsm}, respectively. This can be used to
test the SM, as the signs of the Wilson coefficients in
general are model dependent.

 The leading $(1/m_Q)$ power correction to the 
dilepton mass spectrum in $B \to X_s \ell^+ \ell^-$  
has been worked out in \cite{FLS94} using heavy quark expansion. The
correction  was found to be positive and  
around $10\%$ over a good part of the dilepton mass spectrum with the
particular choice of the parameters $\lambda_1$ and 
$\lambda_2$ adopted in this paper. In the meanwhile, the estimate of the
parameter $\lambda_1$ has changed (it has a negative value now) compared
to what has been used in \cite{FLS94}. Apart from this, the analytic
form of the power corrected result in \cite{FLS94}
is also somewhat puzzling and remains to be verified as, on general
grounds, one expects
the heavy quark expansion to break down near the end-point of the dilepton 
mass spectrum, i.e., as $\hat{s} \to \hat{s}^{max}$, which is not 
suggested by the power-corrected  spectrum reported in 
\cite{FLS94}. In the opinion of this author, the end-point spectrum
requires integration over a range of dilepton masses (or smearing) to be
calculable in the heavy quark expansion method. A detailed discussion of
this point and a new derivation of power corrections
to the dilepton mass spectrum and FB asymmetry will be presented in
\cite{AHHM96}. 

%
%
%
%

Taking into account the spread in the values of the input parameters,
$\mu, ~\Lambda, ~\mt$, and ${\cal B}_{SL}$
discussed in the previous section in the context of ${\cal B}(B \to X_s + 
\gamma)$, we estimate the following branching ratios for the SD-piece
only (i.e., from the intermediate top quark contribution only) 
\cite{AHHM96}:
 \begin{eqnarray}\label{brbsll}
{\cal B}(\bxsee) &=& (8.4 \pm 2.2) \times 10^{-6}, \nonumber\\
{\cal B}(\bxsmm) &=& (5.7 \pm 1.2) \times 10^{-6}, \nonumber\\
{\cal B}(\bxstt) &=& (2.6 \pm 0.5) \times 10^{-7}, 
\end{eqnarray}
where theoretical errors and the error on ${\cal B}_{SL}$ have been added in 
quadrature. These estimates are consistent with the results presented in 
\cite{BM94,KS96}.
 The present experimental limit for the inclusive branching ratio in
\bxsll is actually still the one set by the UA1 collaboration some time
ago \cite{UA1R}, namely ${\cal B}(\bxsmm) > 5.0 \times 10^{-5}$. As far
as we know, there are no interesting limits on the other two modes,
involving $X_s e^+e^-$ and $X_s \tau^+ \tau^-$.

  It is obvious from Fig.~\ref{fig:dbrnsm} that   
only in the dilepton mass region far away from
the resonances is there a hope of extracting the Wilson coefficients
governing the short-distance physics. The region below the $J/\psi$ resonance
is well suited for that purpose as the dilepton invariant 
mass distribution there is dominated by the SD-piece.
Including the LD-contributions, following branching ratio has been
estimated for the dilepton mass range $0.2 \leq \hat{s} \leq 0.36$ in 
\cite{AHHM96}:
\be \label{brbslld}
{\cal B}(\bxsmm) = (1.3 \pm 0.3) \times 10^{-6}, 
\ee
with ${\cal B}(\bxsee) \simeq {\cal B}(\bxsmm)$. The FB-asymmetry is 
estimated to be 
in the range $10\%$ - $27\%$, as can be seen in Fig.~\ref{fig:asymmnsm}. 
These branching ratios and the FB asymmetry
are expected to be measured within the next several years at
BABAR, BELLE, CLEO, CDF, D0, and HERA-B.
 In the high invariant mass region, the short-distance contribution 
dominates. However, the rates are down by roughly an order of magnitude
compared to the region below the $J/\psi$-mass. Estimates of the
branching ratios are of $O(10^{-7})$, which should be accessible at
the LHC.

\par
 The experimental limits on the decay rates of the exclusive decays $B \to 
(K,K^*)
\ell^+ \ell^-$, being searched for by the CLEO and CDF collaborations
\cite{Tomasz95,BH95,sphicas}, while arguably closer to the SM-based 
estimates,
can only be interpreted in specific models of form factors, which hinders
somewhat their transcription in terms of the information on
the underlying Wilson coefficients. However, many of these form factors
can be related to the known ones using ideas from heavy quark effective
theory augmented with QCD sum rule estimates in
studying the heavy to light form factors in $B$ decays \cite{am91}. 
We will not take up such exclusive decays, important as they are, in
these lectures and content ourselves with
presenting the expected branching ratios for some of the experimentally
interesting decay modes. Using the exclusive-to-inclusive ratios
$R_{K\ell\ell} \equiv \Gamma(B \to K \ell^+ \ell^-)/\Gamma (B \to X_s 
\ell^+ \ell^-) =0.07 \pm 0.02$ and
$R_{K^*\ell\ell} \equiv \Gamma(B \to K^* \ell^+ \ell^-)/\Gamma (B \to X_s
\ell^+ \ell^-) =0.27 \pm 0.0.07$, which were estimated in\cite{AGM92}, 
the results are presented in Table \ref{tab9}.
 
 In conclusion, the semileptonic FCNC decays $B \to X_s \ell^+
\ell^-$ (and also the exclusive decays)
 will provide very precise tests of the SM, as they will determine
the signs and magnitudes of the three Wilson coefficients, $C_7,
~C_9^{\mathit{eff}}$ and $C_{10}$.
This, perhaps, may also reveal physics beyond-the-SM
if it is associated with not too high a scale. The MSSM model is a
good case study where measurable deviations from the SM
are anticipated and worked out \cite{AGM94,CMW96}.

\subsection{Summary and overview of rare $B$ decays in the SM}

\par
 The rare $B$ decay mode $B \to X_s \nu \bar{\nu}$, and some of the
exclusive channels associated with it,
 have comparatively larger branching ratios. The estimated inclusive 
branching ratio in the SM is \cite{AGM92,BBL95,Grossman}:
\begin{equation}\label{bxsnunu}
 {\cal B}(B \to X_s \nu \bar{\nu}) = (4.0 \pm 1.0) \times 10^{-5}~,
\end{equation}
where the main uncertainty in the rates is due
to the top quark mass, as the top quark contribution completely dominates
the decay rate,  and a residual one from 
the semileptonic branching ratio ${\cal B}_{SL}$. The scale-dependence,
which enters indirectly through the top quark mass, has 
been brought under control through the NLL corrections, calculated in
\cite{BuBu93}. The corresponding CKM-suppressed decay $B \to X_d \nu 
\bar{\nu}$ is related by the ratio of the CKM matrix element
squared \cite{AGM92}:
\begin{equation}\label{bxsdnunu}
 \frac{{\cal B}(B \to X_d \nu \bar{\nu})}
  {{\cal B}(B \to X_s \nu \bar{\nu})} = \left[ 
\frac{\Vtdabs}{\Vtsabs}\right]^2 ~.
 \end{equation}
Similar relations hold for the ratios of the exclusive decay rates, in which
the r.h.s.\\ depends additionally on the ratios of the form factors squared,
which deviate  from unity through $SU(3)$-breaking terms, in close
analogy with the exclusive radiative decays discussed earlier.
 These decays are particularly attractive  probes of
the short-distance physics, as  the long-distance
contributions are practically absent in such decays. Hence, relations
such as the one in (\ref{bxsdnunu}) provide, in principle, one of 
the best methods for the
 determination of the CKM matrix element ratio $\Vtdabs/\Vtsabs$ 
\cite{AGM92}. From the practical point of view, however, these decay 
modes are rather difficult to measure, in particular
at the hadron colliders and probably also at the $B$ factories. The 
best chances are at the $Z^0$-decays at LEP, and indeed the present
best upper limit is based on the analysis of LEP data \cite{Grossman}.
This  derived limit \cite{Grossman}
\begin{equation}\label{bsnunulim}
 {\cal B}(B \to X \nu \bar{\nu}) < 3.9 \times 10^{-4},
\end{equation}
will be hard to improve significantly in the foreseeable future. 
The estimated branching ratios in a number of inclusive and
exclusive decay modes are given  in Table \ref{tab9}, updating the 
estimates in \cite{AGM92}.

  Further down the entries in Table \ref{tab9} are listed some two-body 
rare decays, such as $(B_s^0, B_d^0) \to \gamma \gamma$, studied in
 in \cite{LLY90} - \cite{DJL96}, where only the lowest order contributions
are calculated, i.e., without any QCD corrections, 
 and the decays $(B_s^0,B_d^0) \to
\ell^+\ell^-$, studied in the 
next-to-leading order QCD in \cite{BuBu93}. Some of them,
in particular, the decays $B_s^0 \to \mu^+ \mu^-$ and perhaps also
the radiative decay $B_s^0 \to \gamma \gamma$, have a fighting chance to be
measured at LHC. The estimated decay rates, which depend on the 
pseudoscalar coupling constant $f_{B_s}$ (for $B_s$-decays) and 
$f_{B_d}$ (for $B_d$-decays), together with the present experimental
bounds are listed in Table \ref{tab9}. Since no QCD corrections have been
included in the rate estimates of $(B_{s}, B_{d}) \to \gamma \gamma$,
the branching ratios are somewhat uncertain.
 The constraints on beyond-the-SM physics that will
eventually follow from these decays are qualitatively similar to the
ones that (would) follow from the decays $\BGAMAXS$ and $B \to X_s \ell^+ 
\ell^-$, which we have discussed at length earlier.

\begin{table}[htb]
\begin{center}
\begin{tabular}{|c|c|c|}  
\hline
Decay Modes & ${\cal B}$(SM) & Measurements and Upper Limits (90\% C.L.)\\
  \hline
$ (B_{d},B_{u}) \to X_{s} \gamma $
& $(3.2 \pm 0.58)  \times 10^{-4}$ & $(2.32 \pm 0.67) \times 10^{-4}$
~CLEO \cite{CLEOrare2}\\
  \hline
$ (B_{d},B_{u}) \to K^* \gamma $
& $(4.0 \pm 2.0)  \times 10^{-5}$ & $(4.5 \pm 1.5 \pm 0.9)\times 10^{-5}$
                                                ~CLEO \cite{CLEOrare1}\\
  \hline
$ (B_{d},B_{u}) \to X_{d} \gamma $
& $(1.0 \pm 0.8) \times 10^{-5}$ & --\\
  \hline
$ B_u  \to \rho^\pm + \gamma $
& $(1.9 \pm 1.6)  \times 10^{-6}$ & $ < 2.0 \times
10^{-5}$ ~CLEO \cite{BH95}\\ \hline
$ B_d  \to \rho^0 + \gamma $
& $(0.85 \pm 0.65)  \times 10^{-6}$ & $ < 2.4 \times 10^{-5}$ ~CLEO
\cite{BH95} \\ \hline
$ B_d  \to \omega + \gamma $
& $(0.85 \pm 0.65)  \times 10^{-6}$ & $ <1.05 \times 10^{-5}$
~CLEO \cite{BH95} \\
\hline
$ (B_{d},B_{u}) \to X_{s} e^+ e^- $
                    & $(8.4 \pm 2.2)  \times 10^{-6}$ & --\\
\hline
$ (B_{d},B_{u}) \to X_{d} e^+ e^- $
                    & $(4.9 \pm 4.3) \times 10^{-7}$ & --\\
\hline
$ (B_{d},B_{u}) \to X_{s} \mu^+ \mu^- $
& $(5.7 \pm 1.3)  \times 10^{-6}$ & $  < 5.0 \times 10^{-5}$
~UA1 \cite{UA1R}\\ \hline
$ (B_{d},B_{u}) \to X_{d} \mu^+ \mu^- $
& $(3.3 \pm 2.8)  \times 10^{-7}$ &  --\\
\hline
$ (B_{d},B_{u}) \to X_{s} \tau^+ \tau^- $
& $(2.6 \pm 0.5)  \times 10^{-7}$ & --\\
 \hline
$ (B_{d},B_{u}) \to X_{d} \tau^+ \tau^- $
& $(1.5 \pm 1.3)  \times 10^{-8}$ &  --\\
\hline  
$ (B_{d},B_{u}) \to K e^+ e^- $   
& $(5.9 \pm 2.3)  \times 10^{-7}$ & $ < 1.2 \times 10^{-5}$
~CLEO \cite{Tomasz95}\\ \hline
$ (B_{d},B_{u}) \to K \mu^+ \mu^- $
 & $(4.0 \pm 1.5)  \times 10^{-7}$ & $ < 0.9 \times 10^{-5}$
~CLEO \cite{Tomasz95}\\ \hline
$ (B_{d},B_{u}) \to K^* e^+ e^- $
                    & $(2.3 \pm 0.9)  \times 10^{-6}$ & $ < 1.6 \times
10^{-5}$ ~CLEO \cite{Tomasz95}\\  
\hline
$ (B_{d},B_{u}) \to K^* \mu^+ \mu^- $
& $(1.5 \pm 0.6)\times 10^{-6}$ & $ <2.5 \times 10^{-5}$ ~CDF \cite{CDF}\\
\hline
$ (B_{d},B_{u}) \to X_{s} ~\nu \bar{\nu} $
& $(4.0 \pm 1.0)  \times 10^{-5}$ & $< 3.9 \times 10^{-4}$ ~\cite{Grossman}\\
 \hline
$ (B_{d},B_{u}) \to X_{d} ~\nu \bar{\nu} $
                    & $(2.3 \pm 2.0)  \times 10^{-6}$ & -- \\
\hline
$ (B_{d},B_{u}) \to K ~\nu \bar{\nu} $
                    & $(3.2 \pm 1.6)  \times 10^{-6}$ & -- \\
\hline
$ (B_{d},B_{u}) \to K^* ~\nu \bar{\nu} $
                    & $(1.1 \pm 0.55)  \times 10^{-5}$ & -- \\
\hline
$ B_{s} \to \gamma \gamma $
    & $(3.0 \pm 1.0)  \times 10^{-7}$ & $ < 1.1 \times 10^{-4}$
~L3 \cite{L3}\\ \hline
$ B_{d} \to \gamma \gamma $
    & $(1.2 \pm 1.1) \times 10^{-8}$ & $< 3.8 \times 10^{-5}$
~L3 \cite{L3}\\ \hline
$ B_{s} \to \tau^+ \tau^- $
                    & $(7.4 \pm 2.1)  \times 10^{-7}$ & --\\
\hline
$ B_{d} \to \tau^+ \tau^- $
                    & $(3.1 \pm 2.9)  \times 10^{-8}$ & --\\
\hline
$ B_{s} \to \mu^+ \mu^- $
                    & $(3.5 \pm 1.0)  \times 10^{-9}$ & $<8.4 \times 
10^{-6}$ ~CDF \cite{CDF}\\ \hline
$ B_{d} \to \mu^+ \mu^- $
                    & $(1.5 \pm 1.4)  \times 10^{-10}$ & $< 1.6 \times
10^{-6}$ ~CDF \cite{CDF}\\
\hline
$ B_{s} \to e^+ e^- $
                    & $(8.0 \pm 3.5)  \times 10^{-14}$ & --\\
\hline
$ B_{d} \to e^+ e^- $
                    & $(3.4 \pm 3.1)  \times 10^{-15}$ & --\\
\hline
\end{tabular}
\end{center}
\caption{Estimates of the branching fractions for FCNC $B$-decays
in the standard model taking into account the uncertainties in the
input parameters as discussed in the text. The entries in the second column 
correspond 
to the short-distance contributions only except for the radiative decays
$ B_u  \to \rho^\pm + \gamma $ and $B_d \to (\rho^0, \omega) + \gamma$,
where long-distance effects have also been included.
For the two-body branching 
ratios, we have used $f_{B_d}= 200$ MeV and $f_{B_s}/f_{B_d}=1.16$. The
CKM matrix element ratio is taken as $\Vtdabs/\Vtsabs=0.24 \pm 0.11$,
as determined from the present fits, and this error is folded in quadrature
with the other errors in estimating the relevant branching fractions.
 Experimental measurements and upper
limits are also listed. Note that the limit quoted from 
\protect\cite{Grossman} is an indirect one.}
\label{tab9}
\end{table}
%

\section{An Update of the CKM Matrix}

In updating the CKM matrix elements, the Wolfenstein
parametrization \cite{Wolfenstein} has been used, which has been given
earlier. The emphasis here is on
those quantities which constrain the CKM
parameters, $\lambda$, $A$, $\rho$ and $\eta$. However, for the sake
of completeness we also quote the values for the other elements of the
CKM matrix.

We recall that $\vert V_{us}\vert$ has been extracted with good accuracy
from $K\to\pi e\nu$ and hyperon decays \cite{PDG94} to be
\be
\vert V_{us}\vert=\lambda=0.2205\pm 0.0018~.
\ee
This agrees quite well with the determination of $V_{ud}\simeq 1-{1\over
2}\lambda^2$ from $\beta$-decay,
\be
\vert V_{ud}\vert=0.9744\pm 0.0010~.
\ee

The values of the matrix elements involving the charm quark row are: 
\cite{PDG94}:
 \be
\vert V_{cs} \vert =1.01\pm 0.18~; ~~~~\vert V_{cd} \vert = 0.204 \pm 0.017
\ee
The value of the third matrix element involving the charm quark $V_{cb}$
is rather important for the CKM fits, as it determines the
parameter $A$. This has been discussed earlier, yielding:
\be
  \vert V_{cb} \vert = 0.0388 \pm 0.0036   \Longrightarrow  ~A = 0.80 \pm 
0.075~.
 \label{Avalue}
\ee

The other two CKM parameters $\rho$ and $\eta$ are constrained by the
measurements of $\vert V_{ub}/V_{cb}\vert$, $\abseps$ (the CP-violating
parameter in the kaon system), $\xd$ (\bdbdbar\ mixing) and (in principle)
$\epsilon^\prime/\epsilon$ ($\Delta S=1$ CP-violation in the kaon system).
The constraints from $\epsilon^\prime/\epsilon$ are not included, due
to the various experimental and theoretical uncertainties surrounding it at
present. For an up to date review of this topic, we refer to \cite{BBL95},
which contains an exhaustive list of references to the original literature
on this and other related topics.
 
The ratio $\vert V_{ub}/V_{cb}\vert$ is determined to be:
\be
\left\vert \frac{V_{ub}}{V_{cb}} \right\vert = 0.08\pm 0.02 \Longrightarrow
~\sqrt{\rho^2 + \eta^2} = 0.36 \pm 0.08~.
\ee
The experimental value of $\abseps$ is \cite{PDG94}
\be
\abseps = (2.26\pm 0.02)\times 10^{-3}~.
\ee
Theoretically, $\abseps$ is essentially proportional to the imaginary part
of the box diagram for \kkbar\ mixing and is given by \cite{Burasetal}
\begin{eqnarray}
\abseps &=& \frac{G_F^2f_K^2M_KM_W^2}{6\sqrt{2}\pi^2\Delta M_K}
\hat{B}_K\left(A^2\lambda^6\eta\right)
\bigl(y_c\left\{\hat{\eta}_{ct}f_3(y_c,y_t)-\hat{\eta}_{cc}\right\}
 \nonumber \\
&~& ~~~~~~~~~~~~~~+ ~\hat{\eta}_{tt}y_tf_2(y_t)A^2\lambda^4(1-\rho)\bigr),
\label{eps}
\end{eqnarray}
where $y_i\equiv m_i^2/M_W^2$, and the functions $f_2$ and $f_3$ can be
found in Ref.~\cite{al95}. Here, the $\hat{\eta}_i$ are QCD correction
factors, of which $\hat{\eta}_{cc}$ \cite{HN94} and $\hat{\eta}_{tt}$
\cite{etaB} were calculated some time ago to next-to-leading order, and
$\hat{\eta}_{ct}$ was known only to leading order \cite{Burastop,Flynn}.
Recently, this last renormalization constant was also calculated to
next-to-leading order \cite{HN95}. In the CKM fits published in
\cite{alpisa95}, the 
following values for the renormalization-scale-invariant coefficients 
have been used:
 $\hat{\eta}_{cc}\simeq 1.32
$, $\hat{\eta}_{tt}\simeq 0.57$, $\hat{\eta}_{ct}\simeq 0.47 $, calculated
for $\hat{m}_c= 1.3$ GeV and  the NLO QCD parameter 
$\Lambda_{\overline{MS}}=310$ MeV in Ref.~\cite{HN95}. 

The final parameter in the expression for $\abseps$ is the
renormalization-scale independent parameter $\hat{B}_K$, which represents
our ignorance of the hadronic matrix element\\ $\langle K^0 \vert
{({\overline{d}}\gamma^\mu (1-\gamma_5)s)}^2 \vert
{\overline{K^0}}\rangle$. The evaluation of this matrix element has been
the subject of much work. The earlier results are summarized in
Ref.~\cite{AL92}. Some recent calculations of $\hat{B}_K$ using the
lattice QCD methods \cite{Soni95} and $1/N_c$ approach \cite{BP95} are:
$\hat{B}_K=0.83 \pm 0.03$ [Sharpe \cite{Sharpe94}],
$\hat{B}_K=0.86 \pm 0.15$ [APE Collaboration \cite{Crisafulli95}],
$\hat{B}_K=0.67 \pm 0.07$ [JLQCD Collaboration \cite{JLQCD}],
$\hat{B}_K=0.78 \pm 0.11$ [Bernard and Soni \cite{JLQCD}], and
$\hat{B}_K=0.70 \pm 0.10$ [Bijnens and Prades \cite{BP95}].

We now turn to \bdbdbar\ mixing. The present world average of $\xd\equiv
\Delta M_d/\Gamma_d$, which is a measure of this mixing, is \cite{Saulanwu95}
\be
\xd = 0.71 \pm 0.04~,
\label{xdvalue}
\ee
which is based on time-integrated measurements which directly measure
$\xd$, and on time-dependent measurements which measure the mass difference
$\Delta M_d$ directly. This is then converted to $\xd$ using the $B_d^0$
lifetime, which is known very precisely
 $(\tau(B_d)=1.56 \pm 0.05~\mathrm{ps}$) \cite{Kroll95}.

From a theoretical point of view it is better to use the mass difference
$\Delta M_d$, as it liberates one from the errors on the lifetime
measurement. In fact, the present precision on $\Delta M_d$, pioneered by
time-dependent techniques at LEP, is quite competitive with the precision
on $\xd$. The LEP average for $\Delta M_d$ has been combined with
the one from CDF and that
derived from time-integrated measurements yielding the present world
average \cite{Saulanwu95}
\be
\Delta M_d = 0.457 \pm 0.019~\mathrm{(ps)}^{-1} ~.
\label{deltamd}
\ee

The mass difference $\Delta M_d$ is calculated from the \bdbdbar\ box
diagram. Unlike the kaon system, where the contributions of both the $c$-
and the $t$-quarks in the loop were important, this diagram is dominated by
$t$-quark exchange:
\be
\label{bdmixing}
\Delta M_d = \frac{G_F^2}{6\pi^2}M_W^2M_B\left(\fbb\right)\hat{\eta}_B y_t
f_2(y_t) \vert V_{td}^*V_{tb}\vert^2~, \label{xd}
\ee
where, using Eq.~(\ref{Vwolf}), $\vert V_{td}^*V_{tb}\vert^2=
A^2\lambda^{6}\left[\left(1-\rho\right)^2+\eta^2\right]$. Here,
$\hat{\eta}_B$ is the QCD correction. In Ref.~\cite{etaB}, this correction
is analyzed including the effects of a heavy $t$-quark. It is found that
$\hat{\eta}_B$ depends sensitively on the definition of the $t$-quark mass,
and that, strictly speaking, only the product $\hat{\eta}_B(y_t)f_2(y_t)$
is free of this dependence. In the fits presented here we use the value
$\hat{\eta}_B=0.55$, calculated in the $\overline{MS}$ scheme, following
Ref.~\cite{etaB}. Consistency requires that the top quark mass be rescaled
from its pole (mass) value  to the value
$\overline{\mt}(\mt(\mathrm{pole}))$ in the $\overline{MS}$ scheme, which is
typically about 10 GeV smaller \cite{mtmsbar}.

For the $B$ system, the hadronic uncertainty is given by $\fbb$, analogous
to $\hat{B}_K$ in the kaon system, except that in this case, also $\fbd$ is
not measured. In the CKM fits \cite{alpisa95}, the following ranges for 
$\fbb$ and
$\hat{B}_{B_d}$, which are compatible with results from both lattice-QCD
\cite{CMichel,Soni95,Shigemitsu} and QCD sum rules \cite{Narison}, were
used:
\begin{eqnarray}
\fbd &=& 180 \pm 50 ~\mbox{MeV}~, \nonumber \\
\hat{B}_{B_d} &=& 1.0 \pm 0.2 ~.
\label{FBrange}
\end{eqnarray}
With this theoretical input, one can get an estimate of the CKM matrix
element $\Vtdabs$. With $\Delta M_d$ known very precisely,
and $\Delta m_t/m_t= \pm 6\%$ \cite{DPG96}, the error on
$\Vtdabs$ is dominated by theoretical error:
\begin{equation}\label{vtddeltamd}
\Vtdabs =(0.92 \pm 0.02 \pm 0.10 \pm 0.28) \times 10^{-2}~.
\end{equation}
The three errors are from $\Delta M_d$ (expt), $\Delta m_t$ and theory, 
respectively. Adding the errors in quadrature, this gives
\begin{equation}\label{vtdquad}
\Vtdabs =(0.92 \pm 0.30) \times 10^{-2}~,
\end{equation}
which is better than the range $ 0.004 \leq \Vtdabs \leq 0.015$
following from unitarity \cite{PDG94}.

To complete the estimates of all the nine CKM matrix elements,
we list here the determination of $\Vtsabs/\Vcbabs$ from the inclusive
branching ratio ${\cal B}(\BGAMAXS)$, given in Eq.~(\ref{vtscb}): 
\begin{equation}
\frac{\Vtsabs}{\Vcbabs} = 0.85 \pm 0.22 (\mbox{expt} + \mbox{th}) ,
 \nonumber
\end{equation}
where, like in $\Vcbabs$, we have added the experimental and theoretical
errors linearly. Combining it with $\Vcbabs=0.0388 \pm 0.0036$, gives
\begin{equation}
\Vtsabs = 0.033 \pm 0.009 (\mbox{expt} + \mbox{th}) .
 \nonumber
\end{equation}
A determination of $\Vtbabs$ can be derived from
the CDF measurements Eq.~(\ref{rtbcdf}). 
While compatible with the unitarity bound \cite{PDG94}
\begin{equation}
0.9988 \leq \Vtbabs \leq 0.9995~,
\end{equation}  
the direct determination of $\Vtbabs$ from CDF is far less 
accurate.
 The present knowledge of the nine CKM matrix elements is summarized in
Table \ref{tabckm}.

\begin{table}
\hfil
\vbox{\offinterlineskip
\halign{&\vrule#&
   \strut\quad#\hfil\quad\cr
\noalign{\hrule}
height2pt&\omit&&\omit&\cr
& $\vert V_{ij} \vert$ && Present Value & & \cr
height2pt&\omit&&\omit&\cr
\noalign{\hrule}
height2pt&\omit&&\omit&\cr
& $\Vudabs$ && $ 0.9744 \pm 0.0010 ~\cite{PDG94}$ & \cr
& $\Vusabs$ && $ 0.2205 \pm 0.0011 ~\cite{PDG94}$ & \cr
& $\Vubabs$ && $ (3.1 \pm 0.8) \times 10^{-3} ~\cite{PDG94}$ & \cr
& $\Vcdabs$ && $ 0.204 \pm 0.017 ~\cite{PDG94}$ & \cr
& $\Vcsabs$ && $ 1.01\pm 0.18 ~\cite{PDG94}$ & \cr
& $\Vcbabs$ && $ 0.0388 \pm 0.0036 ~\cite{alpisa95}$ & \cr
& $\Vtdabs$ && $ (9.2 \pm 3.0) \times 10^{-3} $ & \cr
& $\Vtsabs$ && $ 0.033 \pm 0.009 $ & \cr
& $\Vtbabs$ && $ 0.9991 \pm 0.0004 ~\cite{PDG94}$ & \cr
height2pt&\omit&&\omit&\cr
\noalign{\hrule}}}
\caption{Present values of the CKM matrix elements $\vert V_{ij} \vert$.
Note that the value of $\Vtbabs$ follows from unitarity. All others are
measured in decays discussed in the text and in the PDG review 
\protect\cite{PDG94}, from  which references to the original literatue 
can be traced.} \label{tabckm} 
\end{table}


\subsection{The present profile of the Unitarity Triangle}

 The entries in Table \ref{tabckm} provide a test of unitarity. Since,
the value quoted for $\Vtbabs$ is from unitarity, the other two 
constraints are:
\begin{eqnarray}
\label{ckmunit}
\Vudabs^2 + \Vusabs^2 + \Vubabs^2 &=& 0.998 \pm 0.002 ~, \nonumber\\
\Vcdabs^2 + \Vcsabs^2 + \Vcbabs^2 &=& 1.03 \pm 0.18~.
\end{eqnarray}
This shows that unitarity is well satisfied. One could now ask the
question, how well are the parameters $\rho$ and $\eta$  determined
at present, which in turn determine the profile of the unitarity triangles.
This is done by fitting the CKM parameters using the quantities we have
discussed.
 
In order to find the allowed unitarity triangles, the computer program
MINUIT was used to fit the CKM parameters $A$, $\rho$ and $\eta$ to the
experimental values of $\Vcbabs$, $\vert V_{ub}/V_{cb}\vert$, $\abseps$ and
$\xd$. Since $\lambda$ is very well measured, it was fixed to its
central value given above. 
Two types of fits can be attempted \cite{alpisa95}:
\begin{itemize}
\item
Fit 1: the ``experimental fit.'' Here, only the experimentally measured
numbers are used as inputs to the fit with Gaussian errors; the coupling
constants $f_{B_d} \sqrt{\hat{B}_{B_d}}$ and $\hat{B}_K$ are given fixed
values.
\item
Fit 2: the ``combined fit.'' Here, both the experimental and theoretical
numbers are used as inputs assuming Gaussian errors for the theoretical
quantities.
\end{itemize}

We first discuss the ``experimental fit" (Fit 1). The goal here is to
restrict the allowed range of the parameters ($\rho,\eta)$ for given values
of the coupling constants $f_{B_d} \sqrt{\hat{B}_{B_d}}$ and $\hat{B}_K$.
For each value of $\hat{B}_K$ and $f_{B_d}\sqrt{\hat{B}_{B_d}}$, the CKM
parameters $A$, $\rho$ and $\eta$ are fit to the experimental numbers given
earlier and the $\chi^2$ is calculated.
In the CKM fits performed in \cite{alpisa95}, specific values
in the range 0.4 to 1.0 for $\hat{B}_K$ were considered and it was shown that
for $\hat{B}_K = 0.4$ a very poor fit
to the data is obtained, so that such small values are quite disfavoured.
The fits performed in the range $0.6 \leq \hat{B}_K \leq 1.0$, which
adequately cover the more recent predictions given above, had
a good quality.

\begin{figure}
\vskip -2.4truein
\centerline{\epsfxsize 7.0 truein \epsfbox {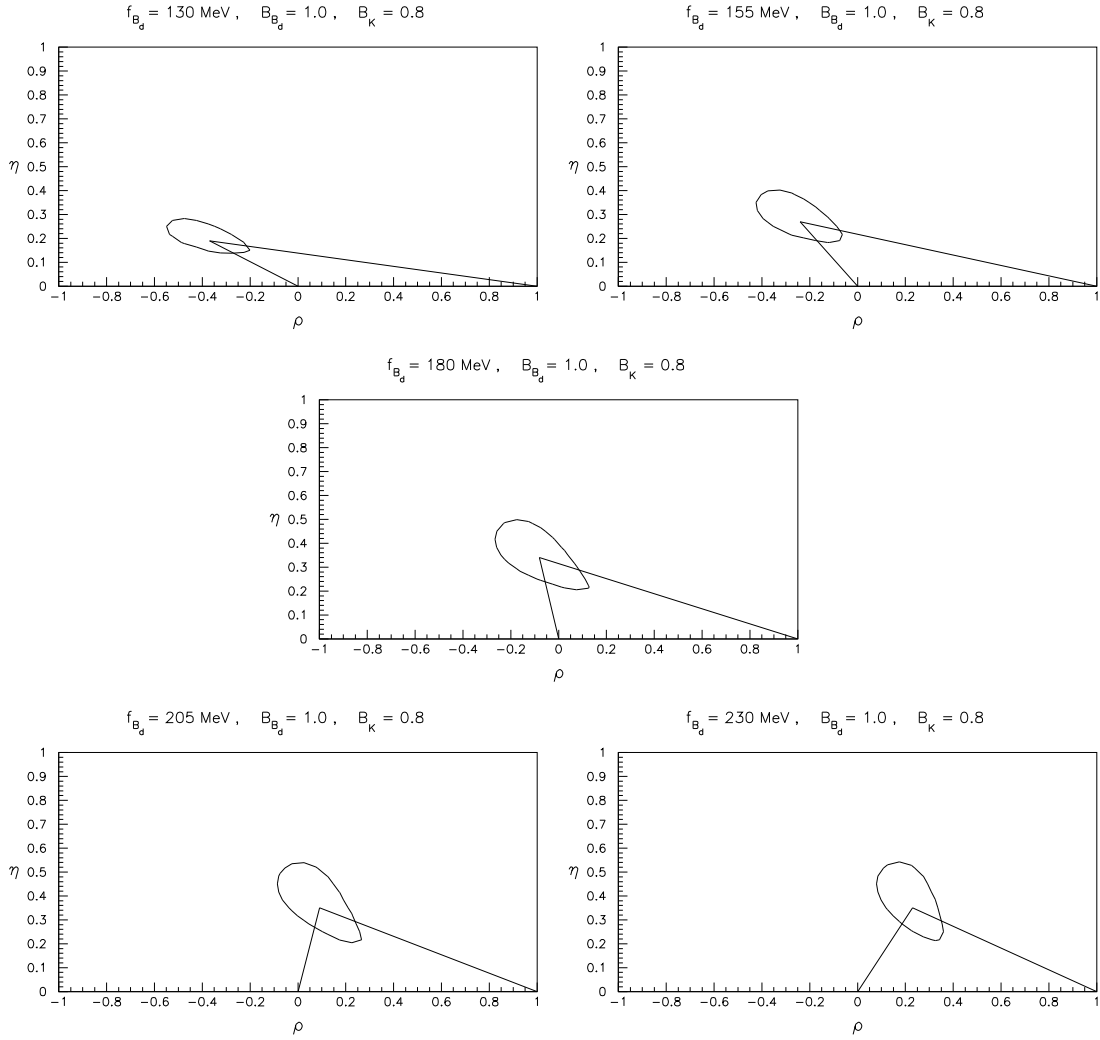}}
\vskip -1.0truein
\caption{Allowed region in $\rho$-$\eta$ space, from a fit to the 
experimental values given in the text. We have fixed
$\hat{B}_K=0.8$ and vary the coupling constant product
$\fbd\protect\sqrt{\hat{B}_{B_d}}$ as indicated on the figures. The solid
line represents the region with $\chi^2=\chi_{min}^2+6$ corresponding to
the 95\% C.L.\ region. The triangles show the best fit. (Figure taken from
\protect\cite{alpisa95}.) }
\label{rhoeta1}
\end{figure}

In Fit 2 in \cite{alpisa95},
 a central value plus an error to $\hat{B}_K$ was assigned and
 two ranges for $\hat{B}_K$ were considered:
\be
\hat{B}_K = 0.8 \pm 0.2 ~,
\label{BKrange1}
\ee
which reflects the estimates of this quantity in lattice QCD, or 
\be
\hat{B}_K = 0.6 \pm 0.2 ~,
\label{BKrange2}
\ee
which overlaps with the values suggested by the earlier chiral perturbation
theory estimates
\cite{Pich94}. It was shown that there is not an enormous difference
in the results for the two ranges. However, as now there seems to be a
theoretical consensus emerging, with $\hat{B}_K \simeq 0.8$,
 we show the case where
$f_{B_d}\sqrt{\hat{B}_{B_d}}$ is varied in 
the range 130 MeV to 230 MeV. The fits are presented as an allowed region
in $\rho$-$\eta$ space at 95\% C.L. ($\chi^2 = \chi^2_{min} + 6.0$). The
results are shown in Fig.~\ref{rhoeta1}. As we pass from
Fig.~\ref{rhoeta1}(a) to Fig.~\ref{rhoeta1}(e), the unitarity triangles
represented by these graphs become more and more obtuse. Even more striking
than this, however, is the fact that the range of possibilities for these
triangles is quite large. There are two things to be learned from this. 
First, our knowledge of the unitarity triangle is at present rather poor. 
Second, unless our knowledge of hadronic matrix elements improves
considerably, measurements of $\abseps$ and $x_d$, no matter how precise,
will not help much in further constraining the unitarity triangle. This is
why measurements of CP-violating rate asymmetries in the $B$ system are so
important \cite{BCPasym,AKL94}. Being largely independent of theoretical
uncertainties, they will allow us to accurately pin down the unitarity
triangle. With this knowledge, we could deduce the correct values of
$\hat{B}_K$ and $f_{B_d}\sqrt{\hat{B}_{B_d}}$, and thus rule out or confirm
different theoretical approaches to calculating these hadronic quantities.

%
%
Despite the large allowed region in the $\rho$-$\eta$ plane, certain values
of $\hat{B}_K$ and $f_{B_d}\sqrt{\hat{B}_{B_d}}$ are disfavoured since they
do not provide a good fit to the data. For example, fixing $\hat{B}_K=0.8$,
which is presently theoretically favoured,
we can use the fitting program to provide the minimum $\chi^2$ for various
values of $f_{B_d}\sqrt{\hat{B}_{B_d}}$. The results are shown in Table
\ref{tabbk8}, along with the best fit values of $(\rho,\eta)$. Since we
have two variables ($\rho$ and $\eta$), we use $\chi^2_{min}<2.0$ as our
``good fit" criterion, and we see that $f_{B_d} \sqrt{\hat{B}_{B_d}} < 130$
MeV and $f_{B_d} \sqrt{\hat{B}_{B_d}} > 240$ MeV give poor fits to the
existing data.

\begin{table}
\hfil
\vbox{\offinterlineskip
\halign{&\vrule#&
 \strut\quad#\hfil\quad\cr
\noalign{\hrule}
height2pt&\omit&&\omit&&\omit&\cr
& $\fbd\sqrt{\hat{B}_{B_d}}$ (MeV) && $(\rho,\eta)$ && $\chi^2_{min}$ & \cr
height2pt&\omit&&\omit&&\omit&\cr
\noalign{\hrule}
height2pt&\omit&&\omit&&\omit&\cr
& $120$ && $(-0.42,~0.16)$ && $3.04$ & \cr
& $130$ && $(-0.37,~0.19)$ && $1.32$ & \cr
& $140$ && $(-0.32,~0.23)$ && $0.43$ & \cr
& $150$ && $(-0.27,~0.26)$ && $0.07$ & \cr
& $160$ && $(-0.22,~0.29)$ && $1.4 \times 10^{-3}$ & \cr
& $170$ && $(-0.15,~0.32)$ && $0.05$ & \cr
& $180$ && $(-0.08,~0.34)$ && $0.09$ & \cr
& $190$ && $(-0.01,~0.35)$ && $0.06$ & \cr
& $200$ && $(0.06,~0.35)$ && $0.01$ & \cr
& $210$ && $(0.13,~0.35)$ && $0.02$ & \cr
& $220$ && $(0.18,~0.35)$ && $0.2$ & \cr
& $230$ && $(0.23,~0.35)$ && $0.61$ & \cr
& $240$ && $(0.28,~0.35)$ && $1.29$ & \cr
& $250$ && $(0.32,~0.35)$ && $2.22$ & \cr
height2pt&\omit&&\omit&&\omit&\cr
\noalign{\hrule}}}
\caption{The ``best values'' of the CKM parameters $(\rho,\eta)$ as a
function of the coupling constant $\fbd\protect\sqrt{\hat{B}_{B_d}}$,
obtained by a minimum $\chi^2$ fit to the experimental data, including the
renormalized value of $m_t=170 \pm 11$ GeV. We fix $\hat{B}_K=0.8$. The
resulting minimum $\chi^2$ values from the MINUIT fits are also given.
(Table taken from \protect\cite{alpisa95}.)} 
\label{tabbk8}
\end{table}

\begin{figure}
\vskip -1.0truein
\centerline{\epsfxsize 3.5 truein \epsfbox {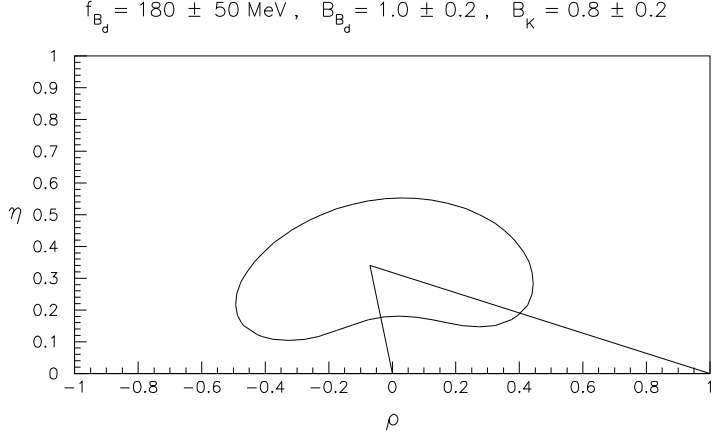}}
\vskip -1.4truein
\caption{Allowed region in $\rho$-$\eta$ space, from a simultaneous fit to
both the experimental and theoretical quantities discussed in the text.
 The theoretical errors are treated as Gaussian for
this fit. The solid line represents the region with $\chi^2=\chi_{min}^2+6$
corresponding to the 95\% C.L.\ region. The triangle shows the best fit.
(Figure taken from \protect\cite{alpisa95}.)}
\label{rhoeta2}
\end{figure}

We now discuss the ``combined fit" (Fit 2). Since the coupling constants
are not known and the best we have are estimates given in the range in
Eq.~(\ref{BKrange1}), a reasonable profile of the
unitarity triangle at present can be obtained by letting the coupling
constants vary in this range. The resulting CKM triangle region is shown
in Fig.~\ref{rhoeta2}. As is clear from this figure, the allowed region is
rather large at present. The
preferred values obtained from the ``combined fit" are
\be
(\rho,\eta) = (-0.07,0.34) ~~~(\mbox{with}~\chi^2 = 6.6\times 10^{-2})~.
\ee
%
%


\subsection{$\xs$ and the Unitarity Triangle}

Mixing in the \bsbsbar\ system is quite similar to that in the \bdbdbar\
system. The \bsbsbar\ box diagram is again dominated by $t$-quark exchange,
and the mass difference between the mass eigenstates $\delms$ is given by a
formula analogous to that of Eq.~(\ref{xd}):
\be
\delms = \frac{G_F^2}{6\pi^2}M_W^2M_{B_s}\left(\fbbs\right)
\hat{\eta}_{B_s} y_t f_2(y_t) \vert V_{ts}^*V_{tb}\vert^2~.
\label{xs}
\ee
Using the fact that $\vert V_{cb}\vert=\vert V_{ts}\vert$,
it is clear that one of the sides of the unitarity triangle, $\vert
V_{td}/\lambda V_{cb}\vert$, can be obtained from the ratio of $\delmd$ and
$\delms$,
\be
\frac{\delms}{\delmd} =
 \frac{\hat{\eta}_{B_s}M_{B_s}\left(\fbbs\right)}
{\hat{\eta}_{B_d}M_{B_d}\left(\fbb\right)}
\left\vert \frac{V_{ts}}{V_{td}} \right\vert^2.
\label{xratio}
\ee
All dependence on the $t$-quark mass drops out, leaving the square of the
ratio of CKM matrix elements, multiplied by a factor which reflects
$SU(3)_{\rm flavour}$ breaking effects. The only real uncertainty in this
factor is the ratio of hadronic matrix elements. Whether or not $\xs$ can
be used to help constrain the unitarity triangle will depend crucially on
the theoretical status of the ratio $\fbbs/\fbb$. In what follows, we will
take $\xi_s \equiv (f_{B_s} \sqrt{\hat{B}_{B_s}}) / (f_{B_d}
\sqrt{\hat{B}_{B_d}}) = (1.16 \pm 0.1)$, consistent with both lattice-QCD
\cite{Shigemitsu,CMichel}
 and QCD sum rules \cite{Narison}. (The SU(3)-breaking
factor in $\delms/\delmd$ is $\xi_s^2$.)

The mass and lifetime of the $B_s$ meson have now been measured at LEP and
Tevatron and their present values are $M_{B_s}=5370.0 \pm 2.0$ MeV and
$\tau(B_s)= 1.55 \pm 0.10 ~ps$ \cite{Kroll95}. 
The QCD correction factor $\hat{\eta}_{B_s}$ is equal to its $B_d$
counterpart, i.e.\ $\hat{\eta}_{B_s} =0.55$. The main uncertainty in $\xs$
(or, equivalently, $\delms$) is now $\fbbs$. Using the determination of $A$
given previously, $\tau_{B_s}= 1.55 \pm 0.10~(ps)$ and $\overline{\mt}=170
\pm 11$ GeV, we obtain
\begin{eqnarray}
\delms &=& \left(13.1 \pm 2.8\right)\frac{\fbbs}{(230~\mbox{MeV})^2} 
~(ps)^{-1}~, \nonumber \\
\xs &=& \left(20.3 \pm 4.5\right)\frac{\fbbs}{(230~\mbox{MeV})^2}~.
\end{eqnarray}
The choice $f_{B_s}\sqrt{\hat{B}_{B_s}}= 230$ MeV corresponds to the
central value given by the lattice-QCD estimates, and with this the CKM-fits
discussed earlier
give $\xs \simeq 20$ as the preferred value in the SM. Allowing the
coefficient to vary by $\pm 2\sigma$, and 
taking the central value for
$f_{B_s}\sqrt{\hat{B}_{B_s}}$, this gives 
\begin{eqnarray} 
11.4 &\leq & \xs \leq 29.4~, \nonumber\\
7.5 ~(ps)^{-1} &\leq & \delms \leq 18.7 ~(ps)^{-1}~.
 \label{bestxs}
\end{eqnarray}
It is difficult to ascribe a confidence level to this range due to the
dependence on the unknown coupling constant factor. In particular, the 
upper limit is scaling as
 $[f_{B_s}\sqrt{\hat{B}_{B_s}}/(230 \mathrm{MeV})]^2$.
 All one can say is that
the standard model predicts large values for $\xs$, most of which are above
the present experimental limit $\xs > 8.8$ (equivalently $\delms > 6.1
~(ps)^{-1}$) \cite{ALEPHxs,Saulanwu95}.

The ALEPH lower bound $\delms > 6.1~(ps)^{-1}$ (95\% C.L.) \cite{ALEPHxs}
and the present world average $\delmd = (0.457 \pm 0.019)~(ps)^{-1}$ can
be used to put a bound on the ratio $\delms/\delmd$. The lower limit on
$\delms$ is correlated with the value of $f_s$, the fraction of $b$ quark
fragmenting into $B_s$ meson, as shown in the ALEPH analysis
\cite{ALEPHxs}. The value obtained from the measurement  of the quantity
$f_s BR(B_s \to D_s \ell \nu_\ell)$ is $f_s=(11.0 \pm 2.8) \%$.  The
time-integrated mixing ratios $\bar{\chi}$ and $\chi_d$, assuming maximal
mixing in the $B_s$-$\overline{B_s}$ system $\chi_s =0.5$, give $f_s=(9.9\pm
1.9)\%$. The weighted average of these numbers is $f_s=(10.2 \pm 1.6)\%$
\cite{Saulanwu95}. With $f_s=10 \%$, one gets $\delms > 5.6~(ps)^{-1}$ at 
95\% C.L., yielding $\delms/\delmd > 11.8$ at 95\% C.L. Assuming, however,
$f_s=12\%$ gives $\delms > 6.1~(ps)^{-1}$, yielding $\delms/\delmd >
12.8$ at 95\% C.L. We will use this latter number.

\begin{figure}
\vskip -1.0truein
\centerline{\epsfxsize 3.5 truein \epsfbox {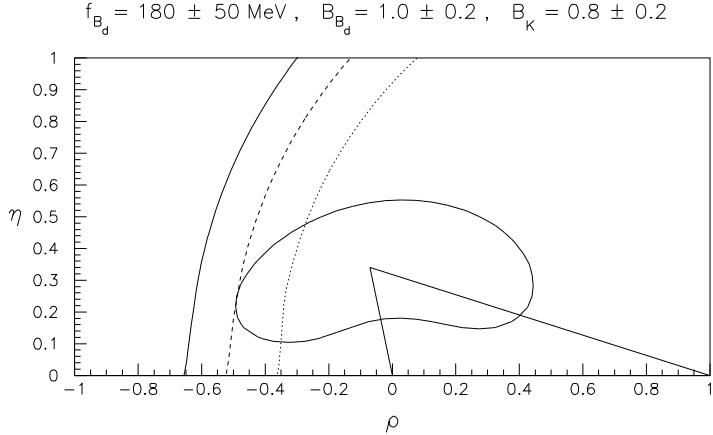}}
\vskip -1.4truein
\caption{Further constraints in $\rho$-$\eta$ space from the ALEPH bound
on $\delms$. The bounds are presented for 3 choices of the SU(3)-breaking
parameter: $\xi_s^2 = 1.1$ (dotted line), $1.35$ (dashed line) and $1.6$
(solid line). In all cases, the region to the left of the curve is ruled
out. (Figure taken from \protect\cite{alpisa95}.)}
\label{xslimit}
\end{figure}

The 95\% confidence limit on $\delms/\delmd$ can be turned into a bound on
the CKM parameter space $(\rho,\eta)$ by choosing a value for the
SU(3)-breaking parameter $\xi_s^2$. We assume three representative values:
$\xi_s^2 = 1.1$, $1.35$ and $1.6$, and display the resulting constraints in
Fig.~\ref{xslimit}. From this graph we see that the ALEPH bound marginally
restricts the allowed $\rho$-$\eta$ region for small values of $\xi_s^2$,
but does not provide any useful bounds for larger values.

%
%

Summarizing the discussion on $\xs$, we note that the lattice-QCD-inspired
estimate $f_{B_s} \sqrt{\hat{B}_{B_s}} \simeq 230$ MeV and the CKM fit
predict that $\xs$ lies between 12 and 30, with a central value around 20.
The upper and lower bounds and the central value scale as
$(f_{B_s}\sqrt{\hat{B}_{B_s}}/230 ~\mbox{MeV})^2$. The present constraints
from the lower bound on $\xs$ on the CKM parameters are marginal but this
would change with improved data. In particular, one expects to reach a
sensitivity $\xs \simeq 15$ (or $\delms \simeq 10~ps^{-1})$ at LEP 
combining all data and
 tagging techniques \cite{Saulanwu95}. One expects comparable sensitivity
at the SLC, where the beam polarization can be used advantageously, and
at the HERA-B, CDF and D0 where the $B_s$ mesons will be produced copiously. 
The entire range for $x_s$ 
for the SM given in Eq.~(\ref{bestxs}) will be accessible at the LHC.  
 A measurement of $\xs$ (equivalently $\Delta M_s$)
would be very helpful in further constraining the CKM parameter space.


\section{CP Violation in the $B$ System}

This topic has been reviewed in \cite{alpisa95}, and since only marginal
changes have taken place in the experimental situation and our present
understanding of CP violation in the $B$ system, we reproduce essentially 
this discussion here.

It is expected that the $B$ system will exhibit large CP-violating effects,
characterized by nonzero values of the angles $\alpha$, $\beta$ and
$\gamma$ in the unitarity triangle (Fig.~\ref{triangle}) \cite{BCPasym}.
The most promising method to measure CP violation is to look for an
asymmetry between $\Gamma(B^0\to f)$ and $\Gamma({\overline{B^0}}\to f)$,
where $f$ is a CP eigenstate. If only one weak amplitude contributes to the
decay, the CKM phases can be extracted cleanly (i.e.\ with no hadronic
uncertainties). Thus, $\sin 2\alpha$, $\sin 2\beta$ and $\sin 2\gamma$ can
in principle be measured in $\bdbarp \to \pi^+ \pi^-$, $\bdbarp\to J/\psi
K_S$ and $\bsbarp\to\rho K_S$, respectively.

Unfortunately, the situation is not that simple. In all of the above cases,
in addition to the tree contribution, there is an additional amplitude due
to penguin diagrams \cite{penguins}. In general, this will introduce some
hadronic uncertainty into an otherwise clean measurement of the CKM phases.
In the case of $\bdbarp\to J/\psi K_S$, the penguins do not cause any
problems, since the weak phase of the penguin is the same as that of the
tree contribution. Thus, the CP asymmetry in this decay still measures
$\sin 2\beta$. 

For $\bdbarp \to \pi^+ \pi^-$, however, although the penguin is expected to
be small with respect to the tree diagram, it will still introduce a
theoretical uncertainty into the extraction of $\alpha$. Fortunately, this
uncertainty can be removed by the use of isospin \cite{isospin}. The key
observation is that the $I=2$ component of the $B\to\pi\pi$ amplitude is
pure tree (i.e., it has no penguin contribution) and therefore has a
well-defined CKM phase. By measuring the rates for $B^+\to\pi^+\pi^0$,
$B^0\to\pi^+\pi^-$ and $B^0\to\pi^0\pi^0$,  as well as their CP-conjugate
counterparts, it is possible to isolate the $I=2$ component and obtain
$\alpha$ with no theoretical uncertainty. Thus, even in the presence of
penguin diagrams, $\sin 2\alpha$ can in principle be extracted from the
decays $B\to\pi\pi$. It must be admitted, however, that this isospin
program is ambitious experimentally. If it cannot be carried out, the error
induced on $\sin 2\alpha$ is of order $|P/T|$, where $P$ ($T$) represents
the penguin (tree) diagram. The ratio $|P/T|$ is difficult to estimate --
it is dominated by hadronic physics. However, one ingredient is the ratio
of the CKM elements of the two contributions: $|V_{tb}^* V_{td} / V_{ub}^*
V_{ud} | \simeq |V_{td}/V_{ub}|$. From the fits in \cite{alpisa95},
 the allowed range for the ratio of these CKM matrix elements is
\be
1.2 \leq \left\vert {V_{td}\over V_{ub}} \right\vert \leq 5.8 ~,
\ee
with the central value close to 3.

It is $\bsbarp\to\rho K_S$ which is most affected by penguins. In fact,
the penguin contribution is probably larger in this process than the tree
contribution. This decay is clearly not dominated by one weak (tree)
amplitude, and thus cannot be used as a clean probe of the angle $\gamma$.
Instead, two other methods have been devised, not involving CP-eigenstate
final states. The CP asymmetry in the decay $\bsbarp\to D_s^\pm K^\mp$ can
be used to extract $\sin^2 \gamma$ \cite{ADK}. Similarly, the CP asymmetry
in  $B^\pm\to\dcp K^\pm$ also measures $\sin^2 \gamma$ \cite{growyler}.
Here, $\dcp$ is a $D^0$ or $\overline{D^0}$ which is identified in a 
CP-eigenstate mode (e.g.\ $\pi^+\pi^-$, $K^+K^-$, ...). 

These CP-violating asymmetries can be expressed straightforwardly in terms
of the CKM parameters $\rho$ and $\eta$. The 95\% C.L.\ constraints on
$\rho$ and $\eta$ found previously can be used to predict the ranges of
$\sin 2\alpha$, $\sin 2\beta$ and $\sin^2 \gamma$ allowed in the standard
model. The allowed ranges which correspond to each of the figures in
Fig.~\ref{rhoeta1}, obtained from Fit 1, are found in Table \ref{cpasym1}.
In this table we have assumed that the angle $\beta$ is measured in
$\bdbarp\to J/\Psi K_S$, and have therefore included the extra minus sign
due to the CP of the final state.

\begin{table}
\hfil
\vbox{\offinterlineskip
\halign{&\vrule#&
 \strut\quad#\hfil\quad\cr
\noalign{\hrule}
height2pt&\omit&&\omit&&\omit&&\omit&\cr
& $\fbd\sqrt{\hat{B}_{B_d}}$ (MeV) && $\sin 2\alpha$ &&
$\sin 2\beta$ && $\sin^2 \gamma$ & \cr
height2pt&\omit&&\omit&&\omit&&\omit&\cr
\noalign{\hrule}
height2pt&\omit&&\omit&&\omit&&\omit&\cr
& $130$ && 0.46 -- 0.88 && 0.21 -- 0.37 && 0.12 -- 0.39  & \cr
& $155$ && 0.75 -- 1.0 && 0.31 -- 0.56 && 0.34 -- 0.92 & \cr
& $180$ && $-$0.59 -- 1.0 && 0.42 -- 0.73 && 0.68 -- 1.0 & \cr
& $205$ && $-$0.96 -- 0.92 && 0.49 -- 0.86 && 0.37 -- 1.0 & \cr
& $230$ && $-$0.98 -- 0.6 && 0.57 -- 0.93 && 0.28 -- 0.97 & \cr
height2pt&\omit&&\omit&&\omit&&\omit&\cr
\noalign{\hrule}}}
\caption{The allowed ranges for the CP asymmetries $\sin 2\alpha$, $\sin
2\beta$ and $\sin^2 \gamma$, corresponding to the constraints on $\rho$ and
$\eta$ shown in Fig.~\protect\ref{rhoeta1}. Values of the coupling constant
$\fbd\protect\sqrt{\hat{B}_{B_d}}$ are stated. We fix $\hat{B}_K=0.8$. The
range for $\sin 2\beta$ includes an additional minus sign due to the CP of
the final state $J/\Psi K_S$. (Table taken from \protect\cite{alpisa95}.)}
\label{cpasym1}
\end{table}

\begin{figure}
\vskip -2.4truein
\centerline{\epsfxsize 7.0 truein \epsfbox {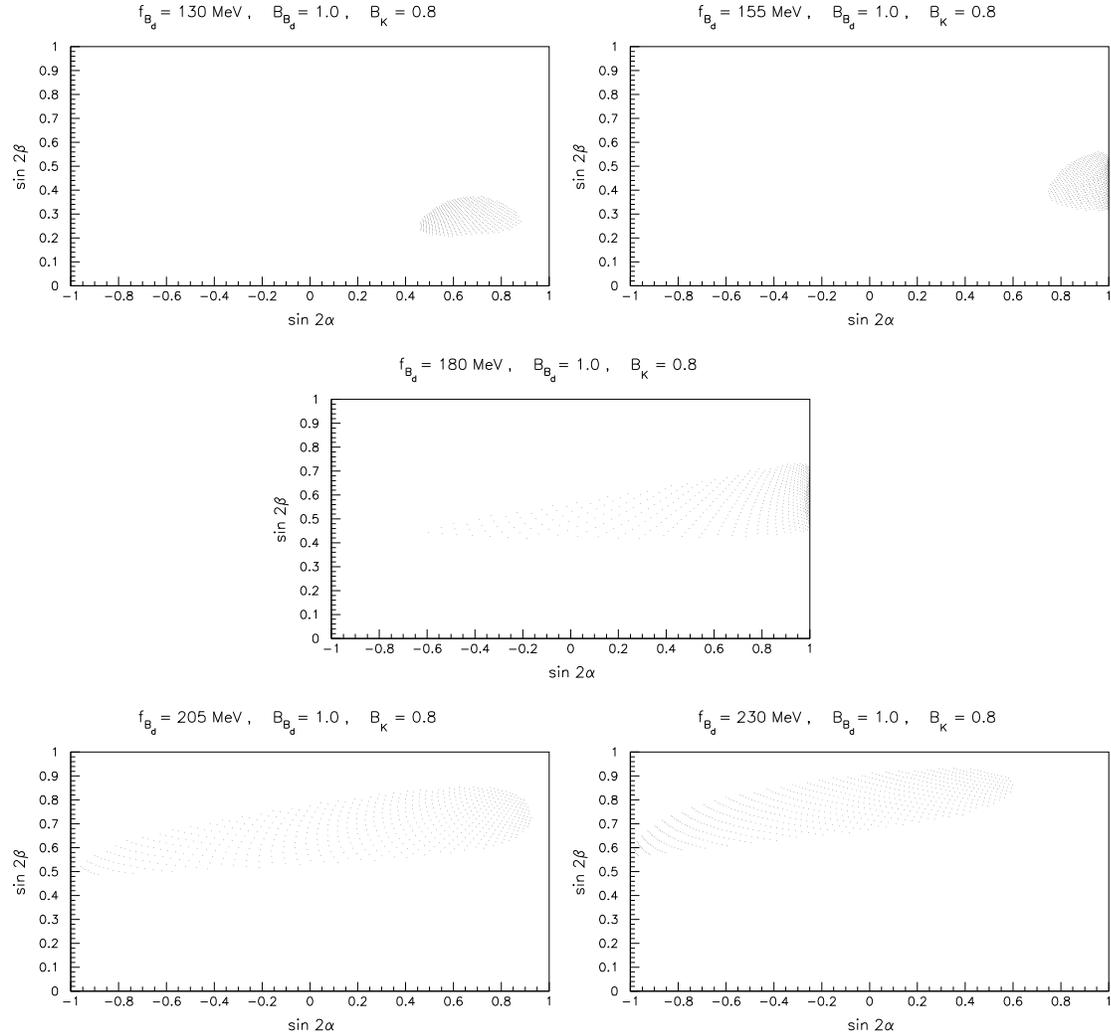}}
\vskip -1.0truein
\caption{Allowed region of the CP asymmetries $\sin 2\alpha$ and $\sin
2\beta$ resulting from the ``experimental fit" of the data for different
values of the coupling constant $\fbd\protect\sqrt{\hat{B}_{B_d}}$
indicated on the figures a) -- e). We fix $\hat{B}_K=0.8$.
(Figure taken from \protect\cite{alpisa95}.)}
\label{alphabeta1}
\end{figure}

\begin{figure}
\vskip -1.0truein
\centerline{\epsfxsize 3.5 truein \epsfbox {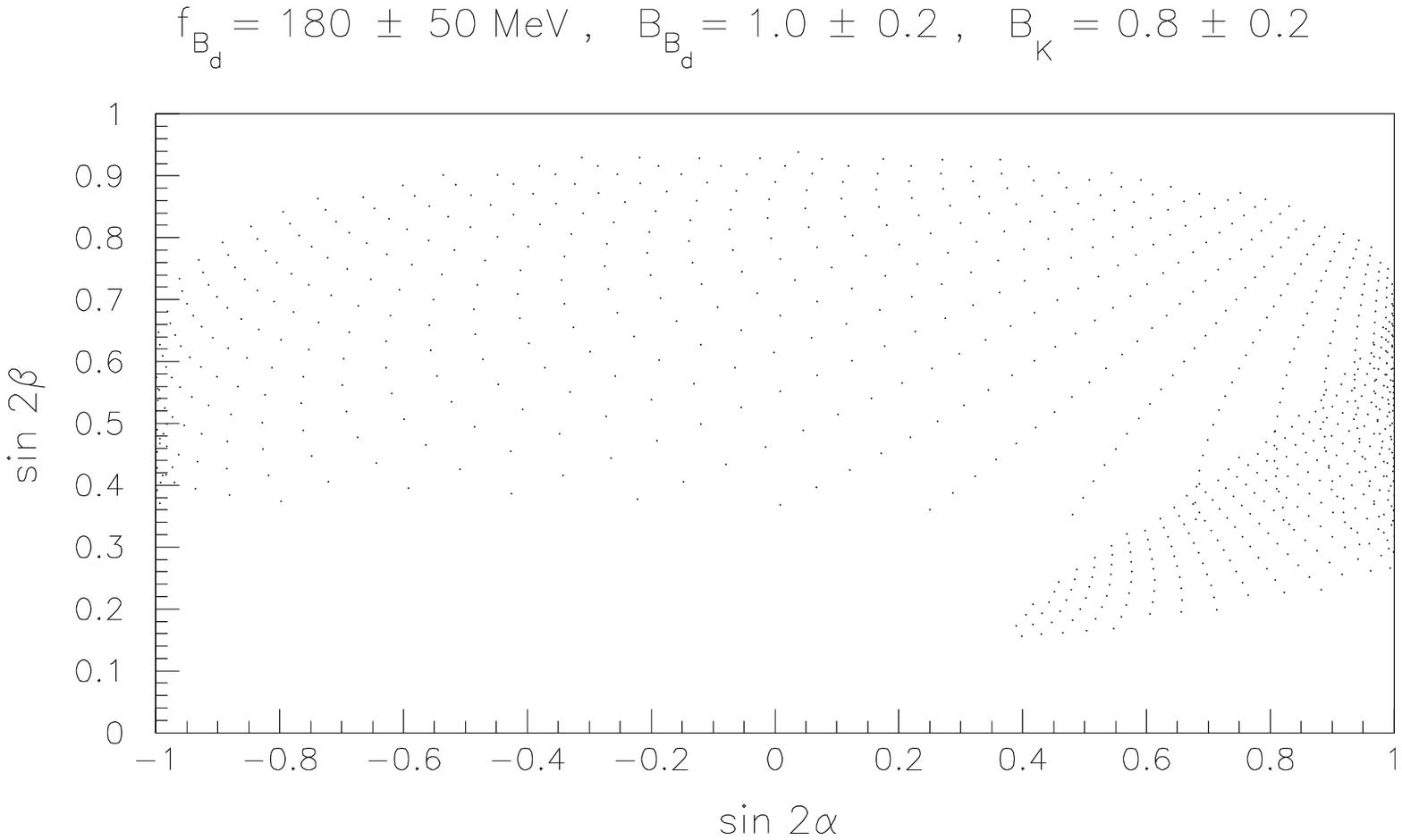}}
\vskip -1.4truein
\caption{Allowed region of the CP asymmetries $\sin 2\alpha$ and $\sin
2\beta$ resulting from the ``combined fit" of the data for the ranges for
$\fbd\protect\sqrt{\hat{B}_{B_d}} $ and $\hat{B}_K$ given in the text.
(Figure taken from \protect\cite{alpisa95}.)}
\label{alphabeta2}
\end{figure}

Since the CP asymmetries all depend on $\rho$ and $\eta$, the ranges for
$\sin 2\alpha$, $\sin 2\beta$ and $\sin^2 \gamma$ shown in Table
\ref{cpasym1} are correlated. That is, not all values in the ranges are
allowed simultaneously. We illustrate this in Fig.~\ref{alphabeta1},
corresponding to the ``experimental fit" (Fit 1), by showing the region in
$\sin 2\alpha$-$\sin 2\beta$ space allowed by the data, for various values
of $\fbd\sqrt{\hat{B}_{B_d}}$. Given a value for
$\fbd\sqrt{\hat{B}_{B_d}}$, the CP asymmetries are fairly constrained.
The parameters used in the central figure in Fig. ~\ref{alphabeta1},
namely $\fbd\sqrt{\hat{B}_{B_d}}=180$ MeV and $\hat{B}_K =0.8$, are
the best theoretical estimates at present. This figure and the third row in
Table \ref{cpasym1} then yield: $\sin 2 \beta > 0.42$ and $\sin^2 \gamma >
0.68$.
However, since there is still considerable uncertainty in the values of the
coupling constants, a more reliable profile of the CP asymmetries at
present is given by the ``combined fit" (Fit 2) of \cite{alpisa95}, where 
present theoretical and experimental values
have been convoluted  in their allowed ranges. The
resulting correlation is shown in Fig.~\ref{alphabeta2}. From this figure
one sees that the smallest value of $\sin 2\beta$ occurs in a small region
of parameter space around $\sin 2\alpha\simeq 0.4$-0.6. Excluding this
small tail, one expects the CP-asymmetry in $\bdbarp\to J/\Psi K_S$ to be
at least 30\%.

%
%
%
%

It may be difficult to extract $\gamma$ using the techniques described
above. First, since $\bsbarp\to D_s^\pm K^\mp$ involves the decay of $B_s$
mesons, such measurements must be done at hadron colliders. At present, it
is still debatable whether this will be possible. Second, the method of
using $B^\pm\to\dcp K^\pm$ to obtain $\gamma$ requires measuring the rate
for $B^+ \to D^0 K^+$. This latter process has an expected branching ratio
of $\lsim O(10^{-6})$, so this too will be hard. 

Recently, a new method to measure $\gamma$ was proposed \cite{PRL}. Using a
flavour SU(3) symmetry, along with the neglect of exchange- and
annihilation-type diagrams, it was suggested that $\gamma$ could be found
by measuring rates for the decays $B^+ \to \pi^0 K^+$, $B^+ \to \pi^+ K^0$,
$B^+ \to \pi^+ \pi^0$, and their charge-conjugate processes. The $\pi K$
final states have both $I=1/2$ and $I=3/2$ components. The crucial
ingredient is that the gluon-mediated penguin diagram contributes only to
the $I=1/2$ final state. Thus, a linear combination of the $B^+ \to \pi^0
K^+$ and $B^+ \to \pi^+ K^0$ amplitudes, corresponding to $I = 3/2$ in the
$\pi K$ system, can be related via flavour SU(3) to the purely $I = 2$
amplitude in $B^+ \to \pi^+ \pi^0$, permitting the construction of an
amplitude triangle. The difference in the phase of the $B^+ \to \pi^+
\pi^0$ side and that of the corresponding triangle for $B^-$ decays was
found to be $2 \gamma$. SU(3) breaking can be taken into account by
including a factor $f_K/f_\pi$ in relating $B\to\pi\pi$ decays to the
$B\to\pi K$ decays \cite{GHLR}. 

The key assumption is that the penguin is mediated by gluon exchange.
However, there are also electroweak contributions to the penguins
\cite{EWPs}. These electroweak penguins (EWP's) are not constrained to be
isosinglets. Thus, in the presence of EWP's, there is no longer a triangle
relation $B\to \pi K$ and $B\to\pi\pi$ amplitudes \cite{DH}. Indeed,
electroweak penguins can, in principle, even invalidate the isospin
analysis in $B\to \pi\pi$, since the $I=2$ amplitude will include a
contribution from EWP's, and hence will no longer have a well-defined weak
CKM phase. However, theoretical estimates \cite{DH,GHLREWP} show that
electroweak penguins are expected to be relatively unimportant for
$B\to\pi\pi$.

The question of the size of EWP's has therefore become a rather interesting
question, and a number of papers have recently appeared discussing this
issue \cite{EWPsize}. These include both theoretical predictions, as well
as ways of isolating EWP's experimentally. The general consensus is that
EWP's are large enough to invalidate the method of Ref.~\cite{PRL} for
obtaining $\gamma$. However, two new methods making use of the flavour
SU(3) symmetry, and which do not have any problems with electroweak
penguins, have been suggested. Both are rather complicated, making use of
the isospin quadrangle relation among $B\to\pi K$ decays, as well as
$B^+\to\pi^+\pi^0$ plus an additional decay: $B_s \to \eta\pi^0$ in one
case \cite{GHLREWP}, $B^+ \to \eta K^+$ with $\eta=\eta_8$ in the other 
\cite{DH2}. Although
electroweak penguins do not cause problems, SU(3)-breaking effects which
cannot be parametrized simply as a ratio of decay constants are likely to
introduce errors of about 25\% into both methods. It is clear that this is
a subject of great interest at the moment, and work will no doubt continue.
Recently, another quadrangle relation involving charged
$B$ decays $B^+ \to \eta K^+$ and $B^+ \to \eta^\prime K^+$ along with the
decays $B^+ \to \pi^+ K^0$ and $B^+ \to \pi^0 K^+$ has been proposed to
 determine the weak phase $\gamma$ \cite{GR96}.

\subsection{Summary of the CKM fits and CP asymmetries in $B$ Decays}

We summarize the results of this section:

\smallskip

(i) We have presented an update of the CKM unitarity triangle using the
theoretical and experimental improvements in the following quantities: 
$\Vcbabs$, $\vert V_{ub}/V_{cb} \vert$, $\delmd$, $\tau(B_d)$,
${\overline{\mt}}$,
$\hat{\eta}_{cc}$, $\hat{\eta}_{ct}$. The fits can be used to exclude
extreme values of the pseudoscalar coupling constants, with the range
$130~\mbox{MeV} \leq f_{B_d} \sqrt{\hat{B}_{B_d}} \leq 270~\mbox{MeV}$
still allowed for $\hat{B}_K=1$. The lower limit of this range is quite
$\hat{B}_K$-independent, but the upper limit is strongly correlated with
the value chosen for $\hat{B}_K$. For example, for $\hat{B}_K=0.8$ and
$0.6$, $f_{B_d} \sqrt{\hat{B}_{B_d}} \leq 240$ and 210 MeV, respectively,
is required for a good fit. The solutions for $\hat{B}_K = 0.8 \pm 0.2$
are slightly favoured by the data as compared to the lower values. These
numbers are in very comfortable agreement with QCD-based estimates from sum
rules and lattice techniques, and recently also with estimates from improved
chiral perturbation theory. 
 The statistical significance of the fit is,
however, not good enough to determine the coupling constant more precisely.
The fits in \cite{alpisa95} show that $\hat{B}_K \leq 0.4$ is strongly 
disfavoured by the data, since the quality of fit for such values is very poor.
This value of $\hat{B}_K$ is now of very little theoretical interest, 
anyway.

\smallskip

(ii) The newest experimental and theoretical numbers restrict the allowed
CKM unitarity triangle in the $(\rho,\eta)$-space somewhat more than
before. However, the present uncertainties are still enormous -- despite
the new, more accurate experimental data, our knowledge of the unitarity
triangle is still poor. This underscores the importance of measuring
CP-violating rate asymmetries in the $B$ system. Such asymmetries are
largely independent of theoretical hadronic uncertainties, so that their
measurement will allow us to accurately pin down the parameters of the CKM
matrix. Furthermore, unless our knowledge of the pseudoscalar coupling
constants improves considerably, better measurements of such quantities as
$\xd$ will not help much in constraining the unitarity triangle. On this
point, help may come from the experimental front. It may be possible to
measure the parameter $\fbd$, using isospin symmetry, via the
charged-current decay $\bu\to\tau^\pm \nu_\tau$. With $\vert V_{ub}/V_{cb}
\vert =0.08 \pm 0.02$ and $\fbd=180\pm 50~{\rm MeV}$, one gets a branching
ratio $BR(\bu\to\tau^\pm\nu_\tau)=(1.5$--$14.0)\times 10^{-5}$, with
a central value of $5.2\times 10^{-5}$. This
lies in the range of the future LEP and asymmetric $B$-factory experiments,
though at LEP the rate $Z \to B_c X \to \tau^\pm \nu_\tau X$ could be just
as large as $Z \to B^\pm X \to \tau^\pm \nu_\tau X$. Along the same lines,
the prospects for measuring $(\fbd,\fbs)$ in the FCNC leptonic and photonic
decays of $\bd $ and $\bs$ hadrons, $(\bd,\bs)\to\mu^+\mu^-, (\bd,\bs) \to
\gamma\gamma$ in future $B$ physics facilities are not entirely dismal.

\smallskip

(iii) We have determined bounds on the ratio $\vert V_{td}/V_{ts} \vert$
from our fits. For $130~\mbox{MeV} \leq f_{B_d} \sqrt{\hat{B}_{B_d}} \leq
270~\mbox{MeV}$, i.e.\ in the entire allowed domain, at 95 \% C.L. we find
\be
0.13 \leq \left\vert {V_{td} \over V_{ts}} \right\vert \leq 0.35~.
\ee
The upper bound from our analysis is more restrictive than the current
experimental upper limit following from the CKM-suppressed radiative
penguin decays $BR(B \to \omega + \gamma )$ and $BR(B \to \rho + \gamma )$,
which at present yield at 90\% C.L. \cite{Tomasz95}
\be
\left\vert {V_{td} \over V_{ts}} \right\vert \leq 0.64 - 0.75~,
\ee
depending on the model used for the SU(3)-breaking in the relevant form
factors \cite{abs93,bksnsr}. Long-distance effects in the decay $B^\pm \to
\rho^\pm + \gamma$ may introduce theoretical uncertainties comparable to
those in the SU(3)-breaking part but the corresponding effects in the
decays $B^0 \to (\rho^0,\omega) +\gamma$ are expected to be very small
\cite{ab95}. Furthermore, the upper bound is now as good as that obtained
from unitarity, which gives $0.08 \leq \vert V_{td}/V_{ts} \vert \leq
0.36$, but the lower bound from our fit is more restrictive.

\smallskip

(iv) Using the measured value of $\mt$, we find
\begin{equation}
\xs = \left(20.3 \pm 4.5\right)\frac{\fbbs}{(230 ~\mbox{MeV})^2}~.
\end{equation}
Taking $f_{B_s}\sqrt{\hat{B}_{B_s}}= 230$ (the central value of lattice-QCD
estimates), and allowing the coefficient to vary by $\pm 2\sigma$, this
gives
\begin{equation}
11.4 \leq \xs \leq 29.4~.
\end{equation}
No reliable confidence level can be assigned to this range -- all that one
can conclude is that the SM predicts large values for $\xs$, most of which
lie above the ALEPH 95\% C.L. lower limit of $\xs > 8.8$.

\smallskip

(v) The ranges for the CP-violating rate asymmetries parametrized by $\sin
2\alpha$, $\sin 2\beta$ and and $\sin^2 \gamma$ are determined at 95\% C.L.
to be
\begin{eqnarray}
&~& -1.0 \leq \sin 2\alpha \le 1.0~, \nonumber \\
&~& 0.21 \leq \sin 2\beta \le 0.93~, \\
&~& 0.12 \leq \sin^2 \gamma \le 1.0~. \nonumber
\end{eqnarray}
(For $\sin 2\alpha < 0.4$, we find $\sin 2\beta \ge 0.3$.) Electroweak
penguins may play a significant role in some methods of extracting
$\gamma$. Their magnitude, relative to the tree contribution, is therefore
of some importance. One factor in determining this relative size is the
ratio of CKM matrix elements $\vert V_{td}/V_{ub} \vert$. We find
\be
1.2 \leq \left\vert {V_{td}\over V_{ub}} \right\vert \leq 5.8 ~.
\ee

\noindent
\section{Acknowledgements}
 I would like to thank Patricia Ball, Vladimir Braun,
Gian Giudice, Marco Ciuchini, Christoph Greub,
L. Handoko, Marek Jezabek, Boris Kayser, David London, Guido Martinelli,
Tak Morozumi, Jon Rosner, Hubert Simma, Amarjit Soni and Arkady
Vainshtein for helpful discussions. I also thank Vladimir Braun, Christoph 
Greub and Jon Rosner for reading the manuscript and for suggesting many
improvements. The help of Patricia Ball for providing
Figure 2 and Christoph Greub for providing Figures 3 and 4
is acknowledged. I am grateful to Gudrun Hiller,
L. Handoko and Tak Morozumi for allowing me to include Figures 9 and 10
based on our joint paper \cite{AHHM96} prior to its publication. 
Informative discussions on experimental issues with
Roy Aleksan, Klaus Honscheid, Tatsuya Nakada,  Henning Schr\"oder,
Tomasz Skwarnicki, Paris Sphicas and Ed 
Thorndike are also gratefully acknowledged. Finally, 
I would like to thank the organizers of the Nathiagali summer college,
in particular its patron, Dr. Ishfaq Ahmed, 
the Co-Director of the college, Professor
Riazuddin, and its scientific secretary, Dr. K.A. Shoaib, for their
warm hospitality, scientific coordination and a great stay in the Murree 
hill tracts. These lectures are dedicated to Professor Abdus Salam on his 70th
birthday.


\end{document}